\documentclass[11pt, a4paper]{article}
\pdfoutput=1
\usepackage{jheppub}
\usepackage{mathdots, amsmath, amsfonts, amssymb}
\usepackage[latin1]{inputenc}

\usepackage{caption}
\usepackage{mathrsfs}
\usepackage{bbm}
\usepackage{epsfig}
\usepackage{latexsym}
\usepackage{slashed}
\usepackage{color}
\usepackage{amsthm}
\usepackage{amscd}
\usepackage{amstext}
\usepackage{enumerate}
\usepackage[section]{placeins}

\usepackage{hyperref}

\usepackage{graphics}

\newcommand{\be}{\begin{equation}}
\newcommand{\ee}{\end{equation}}
\newcommand{\ba}{\begin{eqnarray}}
\newcommand{\ea}{\end{eqnarray}}
\newcommand{\bi}{\begin{itemize}}
\newcommand{\ei}{\end{itemize}}
\newcommand{\ben}{\begin{enumerate}}
\newcommand{\een}{\end{enumerate}}

\newcommand{\eps}{\varepsilon}
\newcommand{\dd}{\mathrm{d}}
\newcommand{\nn}{\nonumber}

\title{{\bf Phase diagram of 4D field theories with chiral anomaly from holography}}
\author{Martin Ammon,}
\author{Julian Leiber}
\author{and Rodrigo P. Macedo}

\affiliation{{\it 
Theoretisch-Physikalisches Institut, Friedrich-Schiller University of Jena,}\\
{\it Max-Wien-Platz 1, 07743 Jena, Germany.}\\}
\emailAdd{martin.ammon@uni-jena.de}
\emailAdd{julian.leiber@uni-jena.de}
\emailAdd{rodrigo.panosso-macedo@uni-jena.de}

\abstract{Within gauge/gravity duality, we study the class of four dimensional CFTs with chiral anomaly described by Einstein-Maxwell-Chern-Simons theory in five dimensions. In particular we determine the phase diagram at finite temperature, chemical potential and magnetic field.  At high temperatures the solution is given by an electrically and magnetically charged AdS Reissner-Nordstroem black brane. For sufficiently large Chern-Simons coupling and at sufficiently low temperatures and small magnetic fields, we find a new phase with helical order, breaking translational invariance spontaneously. For the Chern-Simons couplings studied, the phase transition is second order with mean field exponents. Since the entropy density vanishes in the limit of zero temperature we are confident that this is the true ground state which is the holographic version of a chiral magnetic spiral.}

\arxivnumber{1601.02125}

\begin{document}
\maketitle

\section{Introduction}
One of the amazing developments emerging from the research in string theory is the idea of
a gauge/gravity duality  \cite{Maldacena:1997re}. Remarkably, the duality relates the strongly coupled regime of gauge theories to the weakly coupled regime of the dual string theory or (super-)gravity. Consequently, it has become a powerful tool to study strongly interacting systems by using a conjectured dual weakly coupled gravitational theory. At present, holographic descriptions of non-perturbative phenomena include, among other applications, condensed matter physics, high energy physics and quark-gluon plasma.\footnote{For textbooks see \cite{Ammon:2015wua,Nastase,Schalm}, for reviews see \cite{Hartnoll:2009sz,Herzog:2009xv,McGreevy:2009xe,CasalderreySolana:2011us}.} 

Many of these real world systems of interest involve a finite chemical potential and strongly-coupled degrees of freedom. However, only a few reliable methods exist to compute physical observables for these systems, with real-time physics being particularly difficult to study. 

Using the framework of gauge/gravity duality it is possible to study strongly coupled conformal field theories at finite temperature and finite charge density. For example, the simplest bottom-up holographic model of strongly-interacting matter is Einstein--Maxwell theory with negative cosmological constant. The dual field theory is some CFT with a global U(1) symmetry. The asymptotically AdS Reissner-Nordstroem black brane solution describes thermal equilibrium states with a finite U(1) charge density. 

If the field theory spacetime is even, and the global current is anomalous, the dual gravitational theory also contains a Chern-Simons term for the U(1) gauge field. In a series of papers \cite{D'Hoker:2009mm,D'Hoker:2009bc,D'Hoker:2010rz,D'Hoker:2010ij,D'Hoker:2012ej,D'Hoker:2012ih}, D'Hoker and Kraus constructed the electrically and magnetically charged black brane solutions in five dimensional Einstein-Maxwell-Chern-Simons theory which are dual to strongly coupled four dimensional conformal field theories with chiral anomaly at finite temperature, chemical potential and magnetic field. The phase diagram of these field theories exhibits interesting features, such as a quantum critical point.

The instability of the asymptotically Reissner-Nordstroem black brane solution has attracted much attention due to its relevance to quantum phase transitions in the dual strongly interacting quantum field theory
at finite density. For example, in the presence of (charged) scalar fields or non-abelian vector fields new phases were studied which are reminiscent of s-wave and p-wave superfluids \cite{Hartnoll:2008vx,Hartnoll:2008kx,Gubser:2008wv,Ammon:2009xh}. Moreover, for large enough chiral anomaly, and for low temperatures new spatially modulated phases  were found \cite{Nakamura:2009tf,Ooguri:2010kt,Donos:2012wi} at zero magnetic field. 

In this paper, we consider the class of strongly coupled four dimensional CFTs with chiral anomaly whose gravitational dual description is given in terms of Einstein-Maxwell-Chern-Simons theory.\footnote{It would be interesting to add scalar fields along the lines of \cite{Erdmenger:2015qqa} and investigate the interplay between the quantum critical point as well as spatially modulated and s-wave superfluid phases.} We study thermal equilibrium states for finite temperature, chemical potential and magnetic field and determine the phase diagram. Depending on the coefficient of the chiral anomaly we find a spatially modulated phase\footnote{Spatially modulated phases in the presence of magnetic fields were also discussed in \cite{Domenech:2010nf,Bolognesi:2010nb,Ammon:2011je,Almuhairi:2011ws,Bu:2012mq,Cremonini:2012ir,Montull:2012fy,Salvio:2012at,Salvio:2013jia,Bao:2013fda, Jokela:2014dba,Donos:2015eew}.} extending the results of \cite{Donos:2012wi} to non-zero magnetic fields. In particular, the quantum critical point \cite{D'Hoker:2010rz,D'Hoker:2010ij} is hidden within this new phase. 

The new spatially modulated phase discussed in this paper may be viewed as a holographic version of a chiral spiral \cite{Basar:2010zd}\footnote{For other work on holographic chiral spirals see \cite{Kim:2010pu,BallonBayona:2012wx}.}, although technically speaking we only have one anomalous current in contrast\footnote{Hence, if we compare to QCD, the anomalous current may be identified with the axial current. Moreover, $\mu$ should be viewed as an axial chemical potential, and $B$ as an axial magnetic field.} to QCD. Moreover, the interplay between the quantum critical point and spatially modulated phases is also observed in certain meta-magnetic materials in condensed matter physics. In particular these materials may have a quantum critical point due to the meta-magnetic phase transition which is hidden behind a nematic phase (e.g. see \cite{2010ARCM}). 
 
The remainder of the paper is organised as follows. In section \ref{sec:holoset} we summarize the holographic setup used here. First, we present our coordinate ansatz in the gravitational theory which exhibits Bianchi $\textrm{VII}_\textrm{0}$ symmetry implying that the corresponding equations of motion are ordinary differential equations. Then, we briefly state the asymptotic expansions close to the horizon and conformal boundary and discuss how to extract the thermodynamic observables of the dual CFT from the gravitational theory. 

In section \ref{sec:magneticblackbrane} we numerically construct asymptotically $AdS_5$ black brane solutions with non-trivial electric charge density and magnetic field breaking translational invariance spontaneously. In particular, we determine the phase diagram at finite temperature, chemical potential and magnetic field. Moreover, we characterise the new phase by identifying the order parameters and critical exponents close the phase transition. Finally we explicitly show that the entropy density vanishes in the limit of zero temperature. 

In section \ref{sec:summary} we summarise the results of the paper and conclude with a few remarks on possible future directions. More details concerning the equations of motion, thermodynamics, special cases and numerics are given in the appendices to the paper. Note that in appendix \ref{sec:NumDetail} some techniques are presented to improve the numerical accuracy, specifically at low temperatures, which may be relevant also for other holographic setups. 

\section{Holographic setup}\label{sec:holoset}
To describe a strongly coupled four-dimensional field theory with chiral anomaly within the framework of gauge/gravity duality, we consider a simple toy model on the gravity side. Let us collect the minimal features of this toy model. First, it has to contain a metric (with components $g_{mn}$), and a U(1) gauge field $A=A_{m}\,\dd x^{m}$. Second, the spacetime should be asymptotically $AdS_5$, i.e. using coordinates $(x^m)=(x^\mu, z)$, where $x^\mu$ may be identified with the field theory coordinates\footnote{Moreover, $x^0=t$ and $x^i$ are the spatial coordinates of the field theory.} and $z$ is the radial coordinate of $AdS_5$. The line-element $\dd s^2$ reads
\begin{equation}
\dd s^{2} = \frac{L^2}{z^2} \left( \dd z^2 + \eta_{\mu\nu} \, \dd x^\mu \, \dd x^\nu \right) \, ,
\end{equation}
in the limit $z \rightarrow 0.$ Here, $L$ is the radius of $AdS_5$. The metric $g_{mn}$ is dual to the energy-momentum tensor $T_{\mu\nu}^{cft}$ of the corresponding four-dimensional CFT, while the gauge field $A=A_m \, \dd x^m$ is dual to the current $J_\mu^{cft}$ on the field theory side. Due to the chiral anomaly of the CFT the current $J_\mu^{cft}$ is not conserved, i.e.
\begin{equation}
\partial^\mu \left\langle J_\mu^{cft} \right\rangle = \frac{\gamma}{8} \, \epsilon^{\mu\nu\rho\sigma} \, \tilde{F}_{\mu\nu} \, \tilde{F}_{\rho\sigma} \, ,
\end{equation}
where $\epsilon^{\mu\nu\rho\sigma}$ denotes the totally antisymmetric tensor of four dimensional Minkowski spacetime which we normalise such that $\epsilon^{0123}=1$. Moreover, $\tilde{F}$ is the field strength tensor of an external field $\tilde{A}$ which may be viewed as a source term for the current $J^{cft}$. The chiral anomaly with its coefficient $\gamma$ is dual to a Chern-Simons term on the gravity side, which is another vital ingredient of the gravitational toy model. To summarize, the action of the gravitational toy model (with the features highlighted above) reads 
\begin{equation}\label{eq:actionS}
S_{grav}=\frac{1}{2\kappa^{2}}\left[\int_{\mathcal{M}}\!\dd^{5}x\, \sqrt{-g}\left(R+\frac{12}{L^{2}}-\frac{L^2}{4}F_{mn}F^{mn}\right)-\frac{\gamma}{6}\int_{\mathcal{M}} A\wedge F\wedge F\right]\,,
\end{equation}
where $g=\det g_{mn}$. Moreover, $F = \dd A$ is the field strength tensor of the U(1) gauge field\footnote{Note that the two gauge fields $A$ and $\tilde{A}$ are closely related but are not identical. In particular, the gauge field $A$ lives in the five-dimensional curved spacetime, while the external field $\tilde{A}$ is dual to the current $J^{cft}$ and hence is defined on the four-dimensional Minkowski spacetime of the dual field theory.}  $A$ and $2\kappa^{2}\equiv 16\pi G_{5}$ with the five-dimensional gravitational constant $G_5$. For $\gamma = 2/\sqrt{3}$, the action \eqref{eq:actionS} coincides with the bosonic part of minimal gauged supergravity in five dimensions and hence is a consistent truncation of the most general class of type IIB supergravity in ten dimensions or supergravity in eleven dimensions which are dual to $\mathcal{N}=1$ superconformal field theories, see e.g. \cite{Buchel:2006gb,Gauntlett:2006ai,Gauntlett:2007ma,Colgain:2014pha}. However, in this paper $\gamma$ is treated as a free parameter and we study the phase diagram as a function of $\gamma$. This action has to be supplemented by boundary terms \cite{Henningson:1998gx, Balasubramanian:1999re,Taylor:2000xw} of the form
\begin{equation}\label{eq:actionSbdy}
S_{bdy}= \frac{1}{\kappa^2} \int_{\partial\mathcal{M}}\! \dd^4x \, \sqrt{-h} \left( K - \frac{3}{L} + \frac{L}{4} R(h) + \frac{L}{8} \ln\left( \frac{z}{L} \right) F_{\mu\nu} F^{\mu\nu} \right) \, .
\end{equation}
Here, $h_{\mu\nu}$ is the metric induced by $g_{mn}$ on the conformal boundary of $AdS_5$. The extrinsic curvature $K_{mn}$ is given by 
\begin{equation}
K_{mn} = \mathcal{P}_m^{\ \, o} \,  \mathcal{P}_n^{\ \, p} \, \nabla_o n_p \, ,\qquad \mbox{with} \quad \mathcal{P}_m^{\ \, o} = \delta_m^{\ o} - n_m n^o \, ,
\end{equation}
where $\nabla$ is the covariant derivative and $n_m$ are the components of the outward pointing normal vector of the boundary $\partial\mathcal{M}.$ Moreover, $K$ is the trace of the extrinsic curvature with respect to the metric at the boundary. 

The first term in \eqref{eq:actionSbdy} is the standard Gibbons-Hawking term which is needed for the well-posedness of the variational principle. The other terms remove divergencies and hence are required for the proper renormalisation of various physical quantities \cite{Henningson:1998gx, Balasubramanian:1999re,Taylor:2000xw}. In our case, the conformal boundary of $AdS_5$ is flat Minkowski space and hence the Ricci scalar associated with the boundary metric, $R(h)$, vanishes. Note that the last term in \eqref{eq:actionSbdy} is not invariant under diffeomorphisms. As we will see explicitly, this term in the boundary action is needed to remove the divergence associated with the trace anomaly
\begin{equation}
\eta^{\mu\nu} \left\langle T^{cft}_{\mu\nu} \right\rangle = -\frac{1}{4} \tilde{F}_{\mu\nu} \tilde{F}^{\mu\nu}
\end{equation}
on the field theory side. To keep notations compact we set $2\kappa^{2}=1$ as well as $L=1$ from now on. The equations of motion associated with the action \eqref{eq:actionS} read 
\begin{equation}\label{eq:EOM1}
R_{mn}=-4g_{mn}+\frac{1}{2}\left(F_{mo}F_{n}{}^{o}-\frac{1}{6}g_{mn}F_{op}F^{op}\right)
\end{equation}
for the metric, as well as 
\begin{equation}\label{eq:EOM2}
\dd\star F+\frac{\gamma}{2}F\wedge F=0
\end{equation}
for the gauge field. Equivalently, we can rewrite \eqref{eq:EOM2} as 
\begin{equation}\label{eq:EOM2a}
\nabla_m F^{m n} + \frac{\gamma}{8 \sqrt{-g}} \, \tilde{\epsilon}^{nmopq} F_{mo} F_{pq} = 0 \, ,
\end{equation}
where $\tilde{\epsilon}^{mnopq}$ is the totally antisymmetric Levi-Civita symbol in five spacetime dimensions with $\tilde{\epsilon}^{t123z}=1$.

Moreover, the field strength tensor has to satisfy the Bianchi identitiy $\dd F=0$. For example, a solution to the equations of motion \eqref{eq:EOM1} and \eqref{eq:EOM2} is given by the electrically charged, asymptotically $AdS_5$ Reissner-Nordstroem (RN) black brane with metric and gauge field given by
\begin{equation}
\label{eq:RN_Sol}
\dd s^{2}= \frac{L^2}{z^2} \left( \frac{\dd z^2}{u(z)} - u(z) \,  \dd t^2+ \sum\limits_{i=1}^{3} \dd x^i \, \dd x^i \right), \quad\qquad A = -E(z) \dd t .
\end{equation}
Herefrom we get the field strength tensor  
\be
\quad F= e(z)\,\dd t \wedge\dd z, \quad {\rm with} \quad e(z) = \frac{d}{dz}E(z).
\ee
The function $u(z)$ and $E(z)$ are given by 
\be
\label{eq:RN_u}
u(z)=1-z^{4}\left( 1 + \frac{1}{3}\left(\frac{\rho}{2}\right)^{2}(1-z^{2})\right), \, \qquad\quad \, A_t(z)=-E(z) = \mu (1-z^2)
\ee
leading to the electric field 
\be
\label{eq:RN_e}
e(z)=\rho \, z.
\ee 
The parameter $\rho = 2\, \mu$ is related to the density of the dual field theory, i.e. $\langle J^{cft}_t \rangle =- \rho.$ In a series of papers \cite{D'Hoker:2009mm,D'Hoker:2009bc,D'Hoker:2010rz,D'Hoker:2010ij,D'Hoker:2012ej} (see \cite{D'Hoker:2012ih} for a review) the electrically charged RN black brane in asymptotically $AdS_5$ spacetime was generalized to allow for a constant non-vanishing magnetic field $B$.\footnote{We sometimes refer to this solution as the charged magnetic brane solution.} Without loss of generality, we can assume that the constant magnetic field $B$ is aligned in $x_3$ direction. As reviewed in the introduction, and as explicitly reproduced in appendix \ref{sec:DhokerKraus}, the solution exhibits interesting features, such as a quantum critical point.

Here, we study particular instabilities against spatial modulation for the electrically and magnetically charged Reissner-Nordstroem black brane. As shown in \cite{Nakamura:2009tf,Ooguri:2010kt} for $\gamma > \gamma_c \approx 1.158$, the electrically charged AdS-RN black brane is unstable against spatial modulation below  a  critical  temperature,  suggesting  that  the  system  is in a spatially modulated phase in which the current acquires a helical order. The corresponding backreacted solution for zero magnetic field was presented in \cite{Donos:2012wi}. In this paper we find numerical evidence that the helical structure at zero magnetic field persists at finite magnetic field, at least in some part of the phase diagram. In particular we construct electrically and magnetically charged black branes with (reduced) Bianchi $\textrm{VII}_\textrm{0}$ symmetry, which give rise to the helical order in the currents and energy-momentum tensor.

\begin{figure}[t!]
\begin{center}
\includegraphics[width=8.7cm]{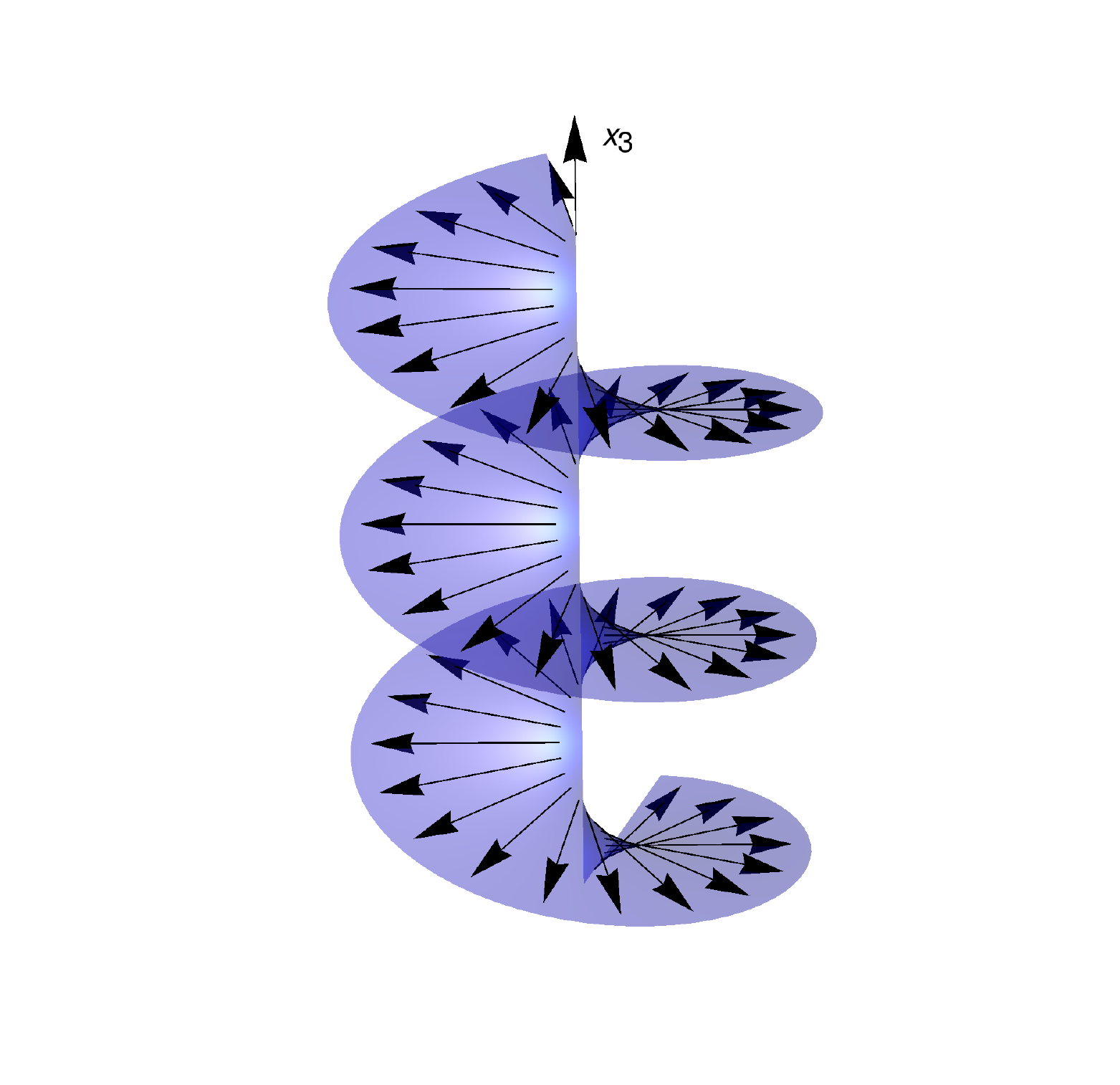}
\end{center}
\caption{Helical structure displaying the tangent vectors dual to $\omega_k.$}
\label{fig:Helix}
\end{figure}

The Bianchi $\textrm{VII}_\textrm{0}$ symmetry is manifest using the one-forms $\omega_i$ defined by 
\begin{eqnarray}\label{eq:omegaforms}
\omega_{1} & = & \cos(k \,x_{3})\,\dd x_{1}-\sin(k \,x_{3})\,\dd x_{2}\,,\\ \nn
\omega_{2} & = & \sin(k \, x_{3})\,\dd x_{1}+\cos(k \,x_{3})\,\dd x_{2}\, , \\ \nn
\omega_{3} & = & \dd x_{3}\,.
\end{eqnarray}
Note that $\omega_{1}\wedge\omega_{2}=\dd x_{1}\wedge\dd x_{2}$ as well as $\dd \omega_1 = k \, \omega_2 \wedge \omega_3 $, $\dd \omega_2 =- k \, \omega_1 \wedge \omega_3 $ and  $\dd \omega_3=0$. The meaning of the differential forms $\omega_i$, or to be precise their dual tangent vectors, is apparent from figure \ref{fig:Helix}. Using the differential forms \eqref{eq:omegaforms} we assume\footnote{This assumption is justified a posteriori since the new black brane solution will have zero entropy density for $T\rightarrow 0$ and hence we speculate that this is the true ground state of the system.} that the helical structure is parallel to the magnetic field $B$, i.e. $\omega_3$ is aligned along the magnetic field. $\omega_{1}$ and $\omega_{2}$ span the plane of the spatial directions perpendicular to the magnetic fields, i.e. the ($x_1$,$x_2$)-plane. 
In this paper, we determine the phase diagram at finite magnetic field, chemical potential and temperature. In particular, we use the following ansatz
\begin{eqnarray}
\label{eq:ansatzmetric}
\dd s^{2} & = & \frac{1}{z^{2}} \, \Bigg(\frac{\dd z^{2}}{u(z)}-u(z)\, \dd t^{2}+v(z)^{2}\, \alpha(z)^{-2}\, \omega_{2}^{2}   + w(z)^{2}\left(\omega_{3}+c(z) \, \dd t\right)^{2}   \notag\\
 &  &   +v(z)^{2}\, \alpha(z)^{2}\, \left(\omega_{1}+g(z)\, \omega_{3}+q(z)\, \dd t\right)^{2}  \Bigg)
\end{eqnarray}
 for the metric and
\begin{equation}\label{eq:ansatzF} 
F  =  e(z)\,\dd t\wedge\dd z+B\,\omega_{1}\wedge\omega_{2}+p(z)\,\dd z \wedge \omega_{3} +b'(z)\,\dd z\wedge\omega_{1}+b(z)\, \dd \omega_{1}
\end{equation}
for the field strength tensor $F = \dd A$. Note that $B$ has to be independent of $z$ in order to satisfy the Bianchi identitiy $dF=0$. 

The ansatz specified by \eqref{eq:ansatzmetric} and \eqref{eq:ansatzF} respects the Bianchi $\text{VII}_\text{0}$ symmetry mentioned above. The field strength tensor may be obtained from a gauge field $A$ of the form
\begin{equation}\label{eq:ansatzA}
A=-E(z)\, \dd t \, - B \, x_2 \, \dd x_1 \, +  b(z) \, \omega_1 + P(z) \, \omega_3 \, ,
\end{equation}
where $A_t(z)$ is related to the electric field $e(z)$ by $E'(z)=e(z)$. Moreover, $P'(z)=p(z)$, where $^\prime$ denotes the derivative with respect to $z.$

Note that the ansatz \eqref{eq:ansatzmetric}, and \eqref{eq:ansatzF} (or \eqref{eq:ansatzA} repsectively) generalises \cite{D'Hoker:2010rz} and \cite{Donos:2012wi}: for $\alpha(z)=1,b(z)=Q(z)=0$ as well as $g(z)=0$ and $k=0$ we obtain the original D'Hoker and Kraus background \cite{D'Hoker:2010rz}, while in the limiting case $c(z)=0$ as well as $B=0$ and $g(z)=p(z)=0$ we obtain a setup equivalent to Donos and Gauntlett \cite{Donos:2012wi} as discussed in appendix \ref{sec:DonosGauntlett}.

Inserting the ansatz \eqref{eq:ansatzmetric} and \eqref{eq:ansatzF} into the Einstein equations \eqref{eq:EOM1}, we obtain nine differential equations, i.e. seven second order differential equations for the metric functions $u(z)$, $v(z)$, $w(z)$, $q(z)$, $c(z)$, $\alpha(z)$ and $g(z)$ as well as two constraints which we denote by $\text{CON}_1$ and $\text{CON}_2$ and contain only first derivatives of the metric fields. Moreover, there are three independent equations of motion \eqref{eq:EOM2} for the gauge fields. While $b(z)$ has to satisfy a second order differential equation, the fields $p(z)$ and $e(z)$ satisfy first order equations of motion which can be recasted in the form
\begin{eqnarray}
 B \gamma e(z) + \partial_z E^*(z) = 0 \quad {\rm and} \quad B \gamma  p(z) + \partial_z P^*(z) = 0,  \label{eq:EOMEP}
\end{eqnarray}
with \ba
E^*(z) &=&  \frac{v(z)^2}{z \, w(z)} \, \bigg(c(z) w(z)^2 \left(q(z) b'(z)+e(z)\right) +(g(z) b'(z)-p(z))\left(u(z)-c(z)^2 w(z)^2\right)\bigg),  \nn \\
P^*(z) &=& \frac{v(z)^2 w(z)}{z} \, \bigg((c(z) g(z)-q(z)) \, b'(z)-c(z) p(z)-e(z) \bigg) + \frac{1}{2} \gamma \,  k \, b(z)^2 . \label{eq:DefEP}
\ea
The full form of the remaining equations of motion is not very enlightening. For these reasons we do not display them here (see appendix \ref{sec:fulleom} for more details).  Moreover, we explicitly checked that the two constraints are consistent. To be precise, using the equations of motion of the metric and gauge fields we showed that the constraints satisfy the following differential equations
\begin{equation}
\partial_z \left( \text{CON}_1(z) \right) + f(z) \, \text{CON}_1(z) = 0 \, \qquad \quad
\partial_z \left(\text{CON}_2(z) \right) + \tilde{f}(z) \, \text{CON}_2(z) = 0 \,
\end{equation}
for some functions $f(z)$ and $\tilde{f}(z).$ These differential equations for $\text{CON}_i(z)$ can be formally solved in terms of exponential functions. Hence, solving the constraints $\text{CON}_i(z_0)=0$ for some $z=z_0$ guarantees that the constraints are satisfied for all $z$. 

\subsection{Asymptotic Expansions}
In order to solve the equations of motion, we first consider the asymptotic expansion of the metric and gauge fields close to the horizon and the conformal boundary of the spacetime. For instance, imposing asymptotically AdS (see discussion in appendix \ref{sec:fulleom})
\[ 
u'(0)=0, \quad 
v(0)=1,  \quad
w(0)=1, \quad 
\alpha(0) = 1, \quad 
c(0) = 0, \quad 
q(0) = 0, \quad 
g(0) = 0,
\]
 \[
A_t(0)= - E(0) = \mu,  \quad 
e'(0) = \rho,  \quad 
P(0) = 0, \quad 
b(0) = 0,
\]
we obtain for the metric functions
\ba
&& u(z) = 1 + z^4\left[ {\mathbf u_4} +{\cal O}(z^2) \right]+ z^4\ln(z)\left[ \frac{B^2}{6}+ {\cal O}(z^2) \right]  \nn \\
&& v(z) = 1 + z^4\left[- \frac{{\mathbf w_4}}{2} +{\cal O}(z^2)\right] + z^4\ln(z)\left[- \frac{B^2}{24}+ {\cal O}(z^2) \right] \nn\\
&& w(z) = 1 + z^4\left[ {\mathbf w_4} +{\cal O}(z^2)\right] + z^4\ln(z)\left[ \frac{B^2}{12}+ {\cal O}(z^2) \right] \nn\\
&& \alpha(z) = 1 + z^4\left( {\mathbf a_4} +{\cal O}(z^2) + z^4\ln(z)\left[ -\frac{B^2}{2304}\left(B^2 + 192 {\mathbf a_4}\right) + {\cal O}(z^2) \right]\right) \label{eq:Exp_metric_AdS} \\
&& c(z) =z^4\left( {\mathbf c_4} + {\cal O}(z^2) + z^4\ln{z} \left[ - \frac{B^2}{12}  {\mathbf c_4} + {\cal O}(z^2)\right] \right) \nn  \\	
&& g(z) =  z^4\left( \frac{B}{2k}{\mathbf b_2} + {\cal O}(z^2) + z^4\ln(z) \left[  {\cal O}(1)\right]\right) \nn \\
&& q(z) = z^4\left( {\mathbf q_4} + {\cal O}(z^2) + z^4\ln(z) \left[  \frac{B^2}{24}  {\mathbf q_4} + {\cal O}(z^2)\right] \right). \nn
\ea
while the gauge field functions read
\ba
E(z)  &=&  -{\boldsymbol \mu} +  \frac{{\boldsymbol \rho}}{2} z^2  +  \frac{\gamma B {\mathbf p_1}}{8} z^4  +  {\cal O}(z^6),  \quad \
e(z) =  z\left( {\boldsymbol \rho}  +\frac{B\gamma}{2}{\mathbf p_1}z^2 + {\cal O}(z^4)  \right) \nn \\
P(z) &=& z^2\left( \frac{{\mathbf p_1}}{2}  + \frac{\gamma B {\boldsymbol \rho}}{8}  z^2 + {\cal O}(z^4 ) \right), \qquad 
p(z) =  z\left({\mathbf p_1}  +\frac{B\gamma}{2}{\boldsymbol \rho}z^2 + {\cal O}(z^4 ) \right) \label{eq:Exp_P} \\
b(z) &=& z^2 \left( {\mathbf b_2}  + {\cal O}(z^2) + z^4\ln(z)\left[ -\frac{\mathbf b_2}{12}B^2  + {\cal O}(z^2)\right] \right) \, . \nn
\ea
Using diffeomorphisms we can shift the horizon to $z=1.$ The event horizon condition imposes that $u(1)=0$. Then, it  follows from the regularity conditions that $c(1)=q(1)=0$. Therefore, the expansion around $z=1$ assumes the following structure for the metric functions
\ba
&&u(z) = (1-z)\left[ \mathbf{ \bar{u}_1} + {\cal O}(1-z) \right], \qquad\quad 
c(z) = (1-z)\left[ \mathbf{ \bar{c}_1}  + {\cal O}(1-z)\right], \nn \\ 
&&q(z) = (1-z)\left[ \mathbf{ \bar{q}_1} + {\cal O}(1-z)\right], \qquad\quad 
w(z) = \mathbf{ \bar{w}_0} + {\cal O}(1-z),\label{eq:expMetric_Hrz} \\
&&g(z) =  \mathbf{ \bar{g}_0} + {\cal O}(1-z),  \qquad\qquad\qquad\quad
v(z) =\mathbf{ \bar{v}_0} + {\cal O}(1-z),  \nn \\ 
&&\alpha(z) = \mathbf{ \bar{a}_0} + {\cal O}(1-z). \nn
\ea
Furthermore, we also impose regulartiy for $A_t$ at the horizon, i.e. $E(1)=0$. In turn, we obtain for the gauge field functions 
\ba
&&E(z) = (1-z)\left[ \mathbf{ -\bar{e}_0} + {\cal O}(1-z) \right] , \qquad\qquad
e(z) =  \mathbf{ \bar{e}_0} + {\cal O}(1-z) , \nn \\
&&P(z) = \mathbf{ \bar{P}_0} + {\cal O}(1-z) , \qquad\qquad\qquad\qquad \
p(z) =   \bar{p}_0 + {\cal O}(1-z) , \label{eq:expGaugeField_Hrz} \\
&&b(z) = \mathbf{ \bar{b}_0} + {\cal O}(1-z) \nn .
\ea
The boldface letters in \eqref{eq:Exp_metric_AdS}--\eqref{eq:expGaugeField_Hrz} denote quantities that are not determined by the expansion, i.e., their values are obtained only after a global solution is found.

In the ansatz for the gauge field \eqref{eq:ansatzA} the functions $E(z)$ and $P(z)$ appear. By integrating \eqref{eq:EOMEP} we can determine $P(z)$ and $E(z)$. For $\gamma B \neq 0$, the function $E(z)$ is given by
\be
\label{eq:Efunc}
E(z) = -  \frac{E^*(z)}{B \gamma}.
\ee
Since we identify $E(0)$ with the chemical potential $\mu$, i.e. $A_t(0)=-E(0)=\mu$, we can use eqs.~\eqref{eq:DefEP} and \eqref{eq:Efunc} as well as the asymptotic expansion \eqref{eq:Exp_P} to obtain 
\begin{equation}\label{eq:Efuncasymp}
p_1 = -B \, \gamma \, \mu \, .
\end{equation}
Similarly, we can solve the equation \eqref{eq:EOMEP} to determine $P(z)$. Note that in this case, we only have to demand regularity at the horizon and hence we cannot fix $P(1).$ However, we only want to study systems where we do not allow for a source term of the operator dual to $P(z)$ and hence we have to demand $P(0)=0.$  


\subsection{Thermodynamics}\label{sec:Thermo}
Next we describe how to extract thermodynamic information from our solutions which describe thermal equilibrium states in the dual CFT. We will work in the grand canonical ensemble in which the chemical potential $\mu$ is fixed.

In order to analyse the thermodynamical properties of the black brane solution we have to analytically continue to a Euclidean time $\tau = t_{(E)}$ by a Wick rotation of the form $\tau = i \, t$. Since the metric and the vector field should be real in Euclidean signature, we also have to introduce $q_{(E)} = -i  \,  q, c_{(E)} = -i  \,  c$ as well as $e_{(E)} = -i  \,  e.$ The leading order coefficients for $q_{(E)}$, $c_{(E)}$ and $e_{(E)}$ at the horizon, see \eqref{eq:expMetric_Hrz} and \eqref{eq:expGaugeField_Hrz}, are denoted by $\bar{q}_{1(E)}, \bar{c}_{1(E)}$ and $\bar{e}_{0(E)}$ respectively. Hence the Euclidean metric and the field strength tensor of the gauge field near the horizon $z=1$ are given by
\begin{eqnarray}
\dd s_{(E)}^{2} &= & \frac{1}{z^2}\left(\frac{\dd z^{2}}{u(z)}+u(z) \, \dd \tau^{2}+v(z)^{2} \, \alpha(z)^{-2} \, (\omega_{2})^{2} +w(z)^{2} \, \left(\omega_{3}+c_{E}(z) \, \dd\tau\right)^{2}\right.\notag\\ \label{eq:hor1exp}
 &  & \quad\quad + \left.
 v(z)^{2}\alpha(z)^{2} \, \left(\omega_{1}+g(z)\omega_{3}+q_{(E)}(z) \,\dd\tau\right)^{2}\right)\nonumber \\
&\approx & \frac{\dd z^{2}}{\bar{u}_1 \, (1-z)}+\bar{u}_1 \, (1-z) \, \dd \tau^{2}+\bar{v}_0^2 \, \bar{a}_0^{-2} \, (\omega_{2})^{2} +\bar{w}_0^{2} \,\left(\omega_{3}-\bar{c}_{1(E)} \, (1-z) \, \dd\tau\right)^{2} \notag\\
 & & \quad + \bar{v}_0^{2} \, \bar{a}_0^{2} \, \left(\omega_{1}+\bar{g}_0 \, \, \omega_{3}- \bar{q}_{1(E)}\, (1-z)  \, \dd\tau\right)^{2}
\end{eqnarray}
as well as
\begin{eqnarray} \notag
F_{(E)} & = & e_{(E)}(z) \,\dd \tau\wedge\dd z+B\,\omega_{1}\wedge\omega_{2}+p(z)\,\dd z\wedge\omega_{3}+b'(z)\,\dd z\wedge\omega_{1}+b(z)\, d\omega_{1}\, , \\ \label{eq:hor2exp}
F_{(E)}& \approx & \bar{e}_{0(E)}  \,\dd \tau\wedge\dd z+B\,\omega_{1}\wedge\omega_{2}+\bar{p}_{0}\,\dd z\wedge\omega_{3} +\bar{b}_1\,\dd z\wedge\omega_{1}+\bar{b}_0\, d\omega_{1}\, .
\end{eqnarray}
In the last lines of \eqref{eq:hor1exp} and of \eqref{eq:hor2exp}  we kept only the leading terms in the near horizon limit. The temperature of the black brane solution can be deduced by demanding regularity of the Euclidean metric \eqref{eq:hor1exp}. In particular, we find that the temperature is given by
\begin{equation}
T =  \frac{|\bar{u}_1|}{4 \pi} \, .
\end{equation}
The entropy $S$ is given by the Bekenstein-Hawking entropy of the black brane. Due to the infinite area of the event horizon it is more convenient to work with the entropy density $s$. Since we work in units where $2 \kappa^2 = 16 \pi G_5 = 1$, the entropy density is given by
\begin{equation}
s =  4 \pi \, \bar{v}_0^2 \, \bar{w}_0 \, .
\end{equation}
This entropy density can be also deduced from the grand canonical potential $\Omega$. In AdS/CFT, the grand canonical potential is identified with $T$ times the on-shell bulk action in Euclidean signature.  
We thus analytically continue to Euclidean signature and compactify the Euclidean time direction $\tau$ with period $1/T$.

In order to determine the total Euclidean action $S_{(E) tot}$,
\begin{equation}
S_{(E) tot} = S_{(E) grav} + S_{(E) bdy} \, ,
\end{equation}
we first perform the Wick-rotation on the action $S$ as given by \eqref{eq:actionS} (including its boundary terms \eqref{eq:actionSbdy}), and denote the result by $\bar{S}_{grav}$ and $\bar{S}_{bdy}$ respectively. Then the corresponding Euclidean actions $S_{(E) grav}$ and $ S_{(E) bdy}$ are given by $S_{(E) grav} = -i \bar{S}_{grav}$ and  $S_{(E) bdy} = -i \bar{S}_{bdy}$. Note that $S_{(E) tot} = - S_{tot}$. We will use the action $S_{tot}$ in Minkowski signature from now on. Hence the grand-canonical potential is given by
\begin{equation}
\Omega = T \, S_{(E) tot}^{o.s.} = - T \,  S_{tot}^{o.s.} \, ,
\end{equation}
where $o.s.$ indicates that we have to evaluate the total action on-shell. The total Euclidean on-shell action is displayed in appendix \ref{sec:MoreThermo}. Using the boundary and horizon expansions \eqref{eq:Exp_metric_AdS}--\eqref{eq:expGaugeField_Hrz}, we obtain for the density of the grand canonical potential (which we also denote by $\Omega$ to keep the notation simple)
\begin{equation}
\Omega =  - \bar{u}_1 \, \bar{v}_0^2\,\bar{w}_0-3\,u_4- \mu\,\rho +\frac{1}{3}\, B\,\gamma \, \int\limits_0^1 \dd z \, E(z)\, p(z) \, ,
\end{equation}
Due to the standard recipes of AdS/CFT, in particular the formula\footnote{From now on, we drop the superscript cft of the energy momentum tensor $T_{\mu\nu}^{cft}$  and of the current $J_{\mu}^{cft}$ to keep the notation simple. Moreover, recall that we set $2\kappa^2 \equiv 1$ from the beginning.} (see \cite{Balasubramanian:1999re})
\begin{equation}
\left\langle T_{\mu\nu} \right\rangle=\lim\limits_{z\rightarrow 0}\frac{1}{z^{2}}\left(-2K_{\mu\nu}+2(K-3) \, h_{\mu\nu}+\textrm{log}(z)\left(F_{\mu}^{\ \alpha}F_{\nu\alpha}-\frac{1}{4} \, h_{\mu\nu}F^{\alpha\beta}F_{\alpha\beta}\right)\right)
\end{equation}
we can extract the energy-momentum tensor of the dual conformal field theory. The non-vanishing components of the energy-momentum tensor are given by
\begin{eqnarray}
\nonumber
\left\langle T_{tt} \right\rangle &=& -3 \, u_4 \, , \\ 
\nonumber
\left\langle T_{t \omega_1} \right\rangle = \left\langle T_{\omega_1 t} \right\rangle &=& 4 \, q_4 \, , \\
\label{eq:EnergyMomTensor}
\left\langle T_{t x_3} \right\rangle = \left\langle T_{x_3 t} \right\rangle &=& 4 \, c_4 \, , \\ 
\nonumber
\left\langle T_{\omega_1 \omega_1} \right\rangle &=& -\frac{B^{2}}{4}+8\,a_4-u_4-4\,w_4  \, , \\ 
\nonumber
\left\langle T_{\omega_2 \omega_2} \right\rangle &=& -\frac{B^{2}}{4}-8 \, a_4-u_4-4\, w_4 \, , \\ 
\nonumber
\left\langle T_{\omega_1 x_3} \right\rangle = \left\langle T_{x_3 \omega_1} \right\rangle &=& \frac{2\, B\, b_2}{k} \, , \\ 
\nonumber
\left\langle T_{x_3 x_3} \right\rangle &=& 8 \, w_4 - u_4 \, .
\end{eqnarray}
In particular, the trace of the energy momentum tensor is given by 
\begin{equation}
\left\langle T^\mu{}_\mu \right\rangle = - \frac{B^2}{2} = -\frac{1}{4} \, \tilde{F}_{\mu\nu} \, \tilde{F}^{\mu\nu} \, .
\end{equation}
Similarly, we can also read off the expectation value of the current in the dual field theory using the relation \cite{D'Hoker:2009bc}
\begin{equation}
\left\langle J^{\mu} \right\rangle= \lim\limits_{z\rightarrow 0}\frac{1}{z^{3}}\left(h^{\mu\alpha}\partial_{z}A_{\alpha}+\frac{\gamma}{6}\epsilon^{\alpha\beta\gamma\mu}A_{\alpha}F_{\beta\gamma}\right) \, .
\end{equation}
For our ansatz, the non-zero components of the current are given by
\begin{equation}\label{eq:CFTcurrents}
\left\langle J_{t} \right\rangle = - \rho \, , \qquad\quad
\left\langle J_{\omega_1} \right\rangle = - 2 \, b_2 \, , \qquad\quad
\left\langle J_{x_3} \right\rangle =  p_1 \, .
\end{equation}
Hence we can write $\Omega$ in the form 
\begin{equation}\label{eq:grandpotential2}
\Omega = U - s \, T - \mu \left\langle J^t \right \rangle +\frac{1}{3}\, B\,\gamma \, \int\limits_0^1 \dd z \, E(z)\, p(z) \, ,
\end{equation}
where we have assumed that $u'(0)<0$ and hence $\bar{u}_1 > 0$ which is the case for our numerical results. Moreover, the charge density reads  $\left\langle J^t \right\rangle = \rho$, while the energy density  is given by $U = \left\langle T^{tt} \right\rangle = -3 \, u_4.$

Note that for $\gamma=0$, the grand canonical potential \eqref{eq:grandpotential2} reduces to its standard form $\Omega = U - s T - \mu \left\langle J^t \right \rangle$. If both $\gamma$ and $B$ are non-vanishing we obtain an additional contribution to the grand canonical potential due to the chiral anomaly. Moreover, combining \eqref{eq:CFTcurrents} and \eqref{eq:Efuncasymp}, we obtain a relation between $\left\langle J_{x_3} \right\rangle$ and $\mu$ of the form 
\begin{equation}\label{eq:Efuncasymp2}
\left\langle J_{x_3} \right\rangle = - B \, \gamma \, \mu \, .
\end{equation}
This is precisely the chiral magnetic effect. Due to the chiral anomaly we also expect to find \cite{Amado:2011zx}
\begin{equation}\label{eq:Ttx3}
\left\langle T_{t x_3} \right\rangle = \frac{\gamma}{2} \, B \, \mu^ 2 \, .
\end{equation}

\section{The magnetic helical black brane}\label{sec:magneticblackbrane}
We present now the results describing our magnetic helical black brane solution. Appendix \ref{sec:NumDetail} provides more details on the numerical techniques employed here. Due to the invariance of the metric (\ref{eq:ansatzmetric}) under scale transformation $\tilde{x}^m = \lambda \, x^m$, we express the results in terms of dimensionless quantities normalised by $\mu$. Since under the scale transformation we have $\tilde{\mu} = \lambda \, \mu$, the relevant physical observables are 
\ba
\label{eq:PhysObserv}
&& \bar{k} = \frac{k}{\mu}, \qquad \bar{T} = \frac{T}{\mu}, \qquad \bar{B} = \frac{B}{\mu^2}, \qquad  \bar{s} = \frac{s}{\mu^3}, \notag \\ 
 && \left<\bar{J}_{\mu}\right> = \frac{\left<{J}_{\mu}\right>}{\mu^3}, \qquad \bar{\Omega} = \frac{\Omega}{\mu^4}, \qquad \left<\bar{T}_{\mu \nu}\right> = \frac{\left<{T}_{\mu \nu}\right>}{\mu^4}. 
\ea
As described in \cite{Donos:2012wi}, for $\bar{B}=0$ one expects to construct the spatially modulated black brane solutions provided the Chern-Simons coupling be $\gamma > 1.158$. As representative examples, we  focus ourselves on the results with $\gamma = 1.5$. Besides, as a generalisation of the particular results from \cite{Donos:2012wi}, we also comment on some specific features of the case $\gamma = 1.7$.

We first address the question in which region of the parameter space $\{\bar{k}, \bar{B}\}$ we expect new solutions. Then we construct these solutions and single out the thermodynamically preferred ones. Following this we discuss thermodynamic properties of these solutions, with particular emphasis on the behaviour of near the critical temperature and in the low temperature limit.

\subsection{The phase boundary}

The magnetic helical black brane solution is described by the existence of a function $b(z) \neq 0$. As described in appendix \ref{sec:DhokerKraus}, in the limiting case $b(z) = 0$, one obtains the electrically charged Reissner-Nordstroem black brane with $\bar{B}=0$ or the electrically and magnetically charged brane for $\bar{B} \neq 0$. The boundary between the two regimes is therefore naturally defined as the region for which $b(z) \approx 0$.\footnote{Numerically, the boundary is characterised by $b_2 \approx 10^{-9}$ as introduced in \eqref{eq:Exp_metric_AdS}, see discussion in appendix \ref{sec:NumDetail}.}

\begin{figure}[t!]
\begin{center}
\includegraphics[width=7.5cm]{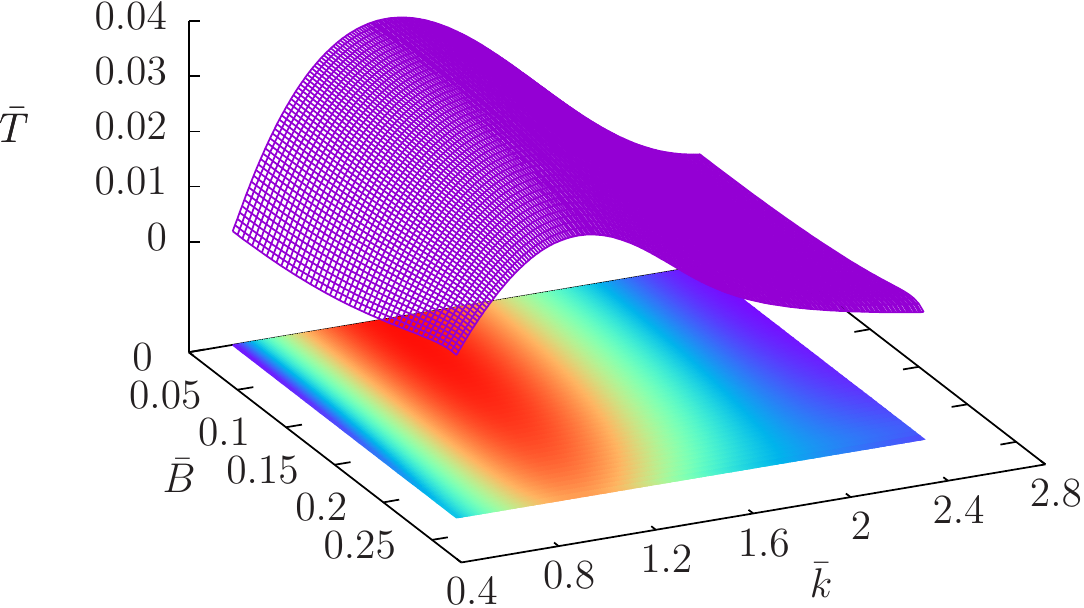}
\includegraphics[width=7.0cm]{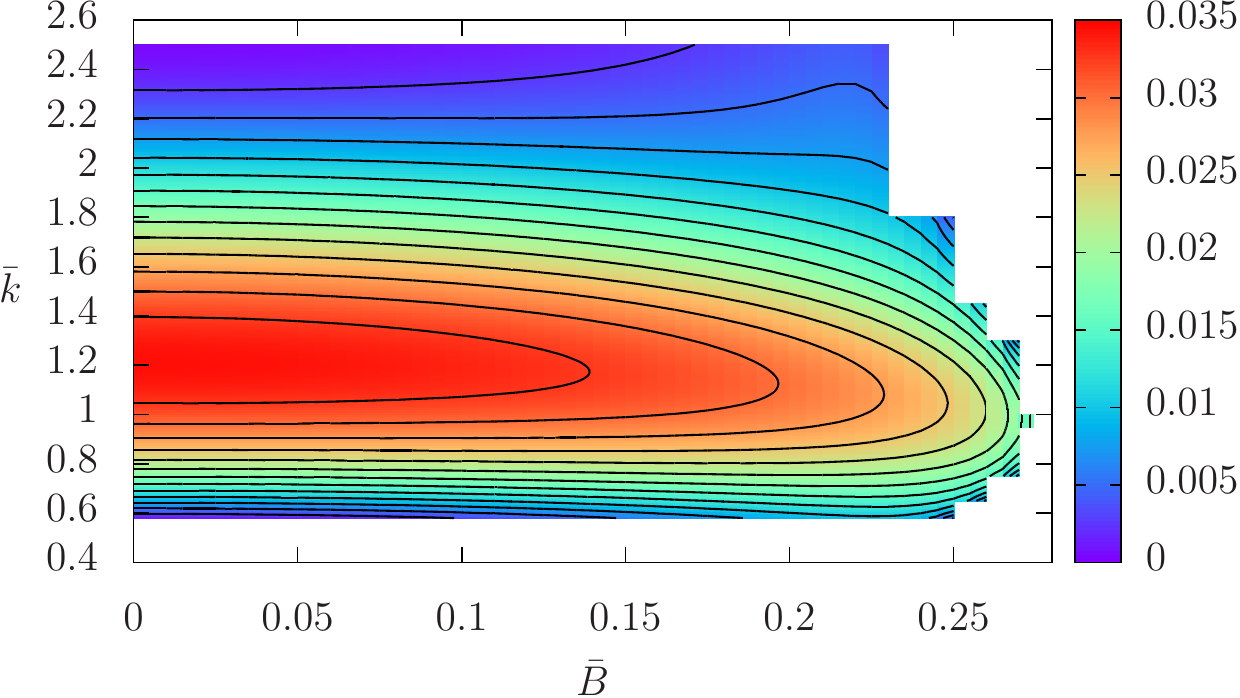}
\end{center}
\caption{Left panel: temperature $\bar{T}(\bar{k}, \bar{B})$ below which we find new magnetic helical black brane solution. Right panel: contour plot of isothermals in the $(\bar{k},\bar{B})$-plane. The highest temperature for which the magnetic helical black brane exists decreases as the magnetic field $\bar{B}$ is increased. The results are shown for $\gamma = 1.5$.}
\label{fig:T_versus_B_k}
\end{figure} 

\begin{figure}[t]
\begin{center}
\includegraphics[width=7.2cm]{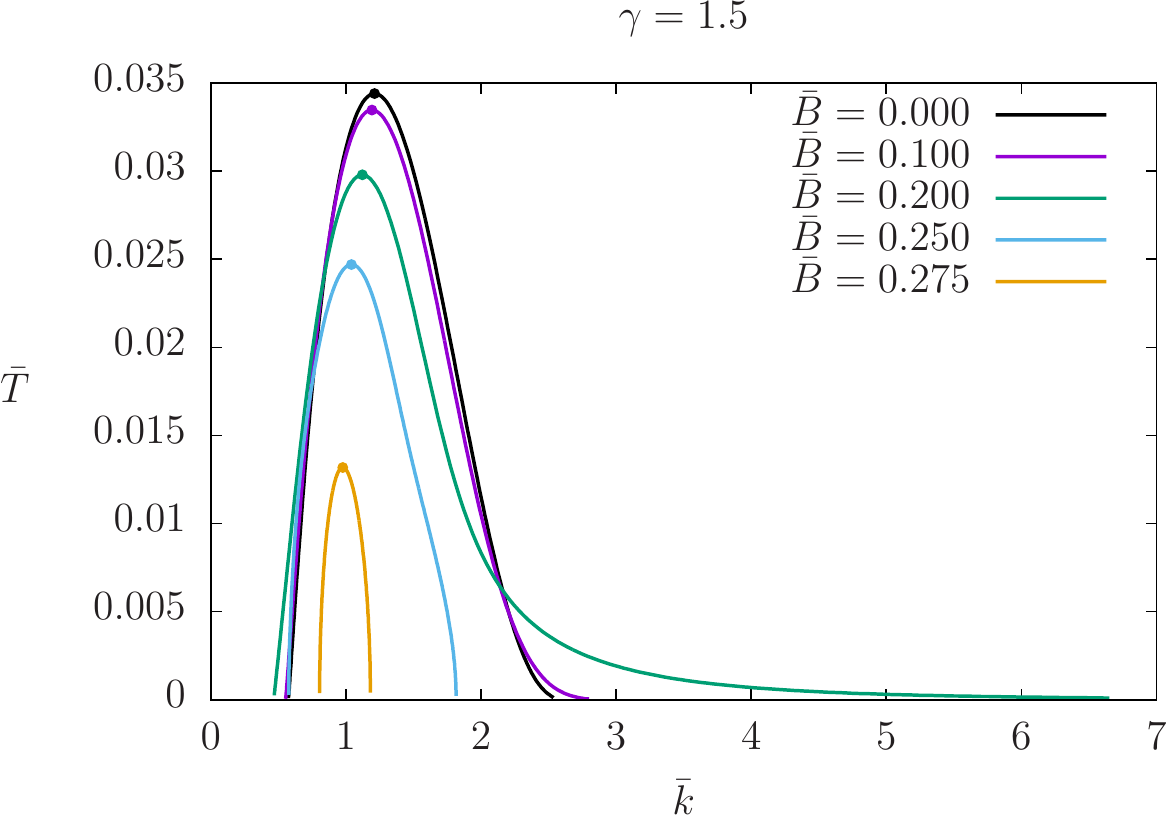}
\includegraphics[width=7.2cm]{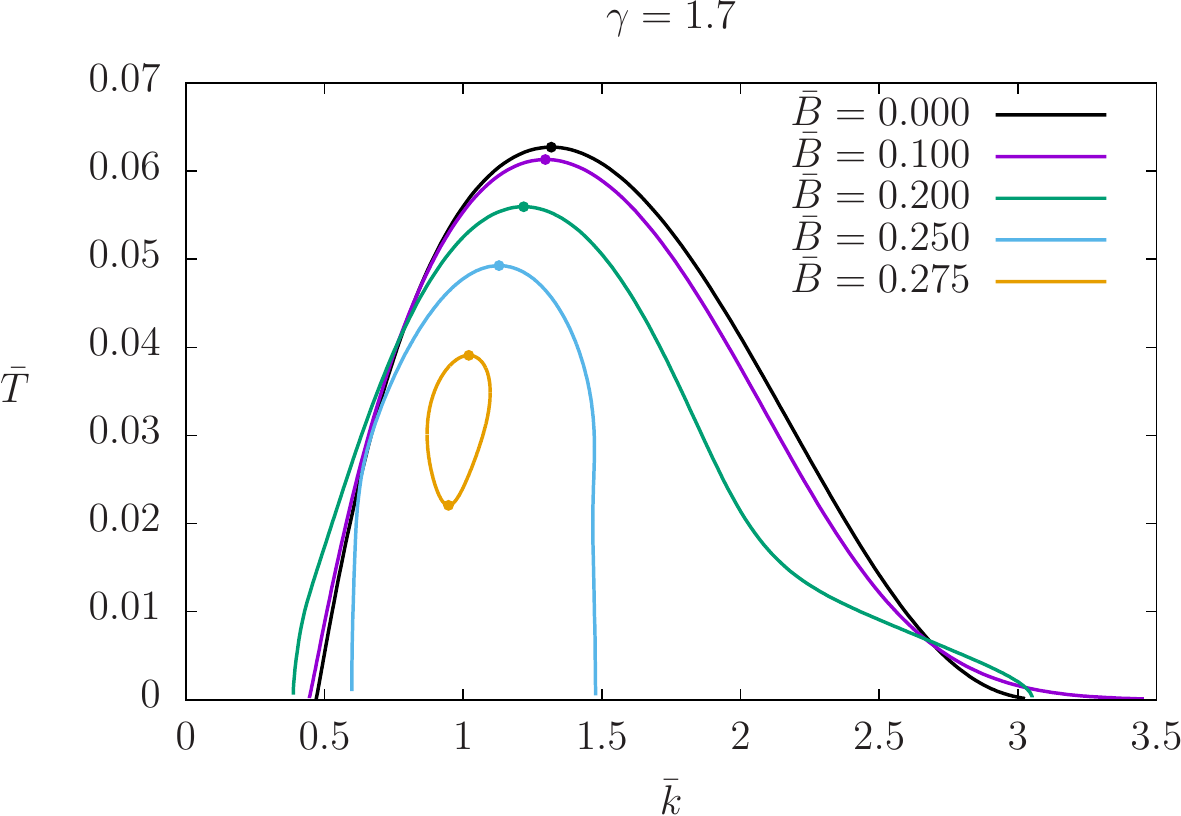}
\end{center}
\caption{Slices of $\bar{B}=$constant in the $(\bar{T}, \bar{k})$-plane. Left panel: results for $\gamma = 1.5$. The range of $\bar{k}$ for which magnetic helical black brane solutions exist first increases with increasing $\bar{B}$, reaching value as high as $\bar{k}\approx6.5$, and then shrinks again. Right panel: results for $\gamma = 1.7$. For $\bar{B} \gtrsim 0.274$ the new phase lies entirely within a closed curve.}
\label{fig:T_versus_k}
\end{figure}

In the left panel of figure ~\ref{fig:T_versus_B_k}, the highest temperature $\bar{T}$ for which the magnetic helical black brane exists is plotted as a function of $\bar{k}$ and $\bar{B}$. In other words, above the surface only the RN black brane exists while below the surface both the RN black brane as well as the magnetic helical black brane solution coexist. Projecting the surface onto the $\bar{B}=0$ plane we reproduce\footnote{The results shown in figure \ref{fig:T_versus_B_k} are for $\gamma=1.5$ while the results explicitly shown in \cite{Donos:2012wi} are for $\gamma=1.7$.}  the expected profile showed in \cite{Donos:2012wi} and the contour plot in the right panel of the same figure highlights the isothermal curves on the $(\bar{k},\bar{B})$-plane.

To appreciate this property, in figure \ref{fig:T_versus_k} we restrict ourselves to the $(\bar{T}, \bar{k})$-plane with $\bar{B}=$ constant. The left panel corresponds to the same results as in the previous figure, i.e. with $\gamma = 1.5$. It becomes evident that the critical temperature $\bar{T}_{\rm C} (\bar{B}) = \displaystyle \max_{\bar{k}} [ \bar{T}(\bar{k}, \bar{B})]$ decreases for increasing $\bar{B}$.

It is worth mentioning the equivalent results for $\gamma = 1.7$, depicted in the right panel of figure \ref{fig:T_versus_k}. For values $\bar{B} \gtrsim 0.274$, we observe that the magnetic helical phase lies entirely within a closed curve. In the particular example with $\bar{B} = 0.275$ displayed here, 
phase transitions occur at both $\bar{T}_{\rm C} \approx 0.03909$ and $\bar{T}_{\rm C} \approx 0.022061$.

\begin{figure}[t!]
\begin{center}
\includegraphics[width=10.0cm]{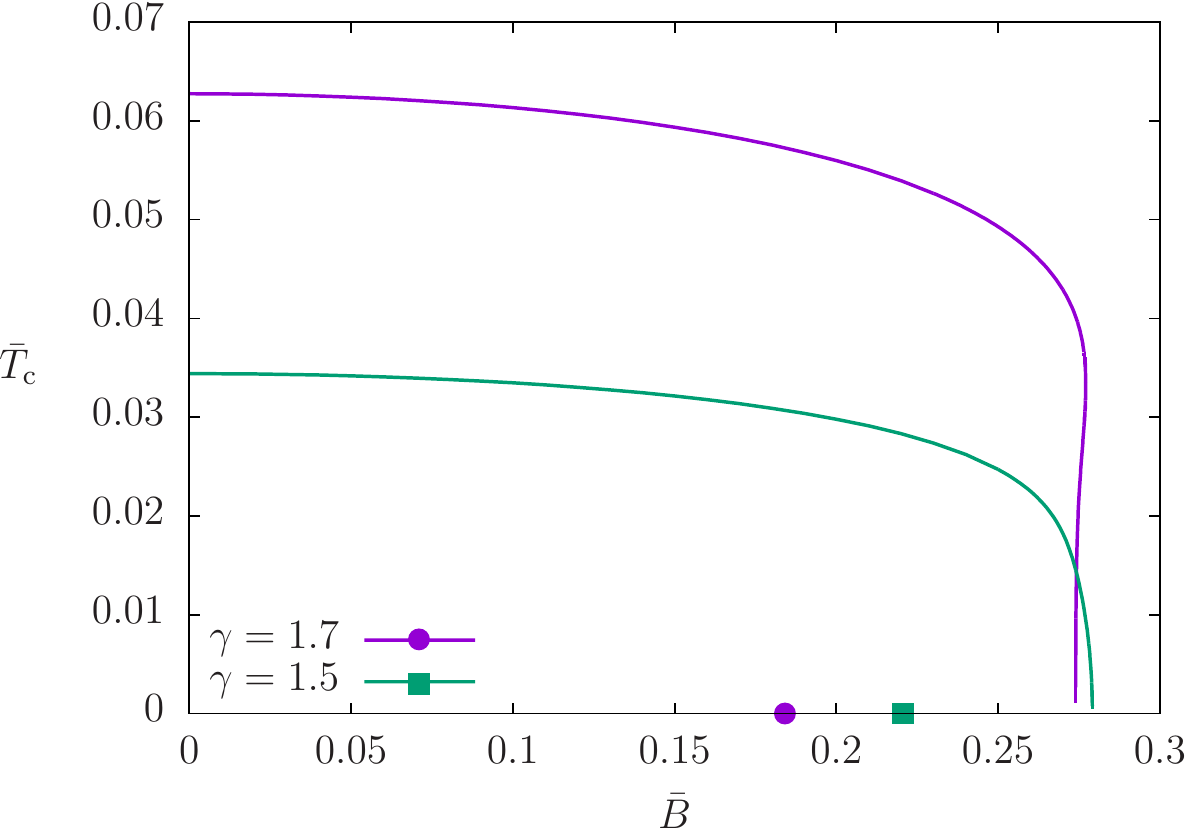}
\end{center}
\caption{Boundary of the magnetic helical phase in the $(\bar{T},\bar{B})$ phase diagram. There exist a maximum value $\bar{B}_0$ limiting the region where the new solution exists. For $\gamma = 1.5$ and $\gamma = 1.7$, we find $\bar{B}_0 \approx 0.279$ and $\bar{B}_0 \approx 0.274$, respectively. The corresponding quantum critical points are also displayed at $\bar{B}_{C} \approx 0.185$ ($\gamma = 1.7$) and $\bar{B}_{C} \approx 0.220$ ($\gamma = 1.5$). They lie within the new magnetic helical phase.}
\label{fig:T_versus_B}
\end{figure}

After identifying the critical temperature in the $(\bar{T}, \bar{k})$-plane, we study the dependence of $\bar{T}_{\rm C}$ on the magnetic field $\bar{B}$ and present the results in figure \ref{fig:T_versus_B}. Here again, it is evident that there exists a value $\bar{B}_0$ as $\bar{T}_{\rm C} \rightarrow 0$, which limits the region where the magnetic helical solution is expected to be found. For the particular examples treated here, these values are $\bar{B}_0 \approx 0.279$ ($\gamma = 1.5$) and $\bar{B}_0 \approx 0.274$ ($\gamma=1.7$). We also identify in the same figure the quantum critical point $\bar{B}_{\rm C}$ as found in \cite{{D'Hoker:2010rz}} (see also the discussion in appendix \ref{sec:DhokerKraus}). In particular, the critical values are $\bar{B}_{C} \approx 0.185$ and $\bar{B}_{C} \approx 0.220$ for $\gamma = 1.7$ and $\gamma =1.5$, respectively. It is interesting to notice that $\bar{B}_{\rm C}$ lies within the new phase region, meaning that the phase transition should occur before the system reaches the quantum critical point. 

The important question that arises now is what happens to the system as we lower the temperature and move inside the new phase along curves of constant $\bar{B}$. Of particular interests is the region $\bar{B}<\bar{B}_{\rm C}$ and the behaviour of the entropy $\bar{s}$ in the low temperature regime. We address this issue and discuss further details about the thermodynamics in the next section.

\subsection{Thermodynamic results}\label{sec:ThermoResults}

\begin{figure}[t!]
\begin{center}
\includegraphics[width=7.2cm]{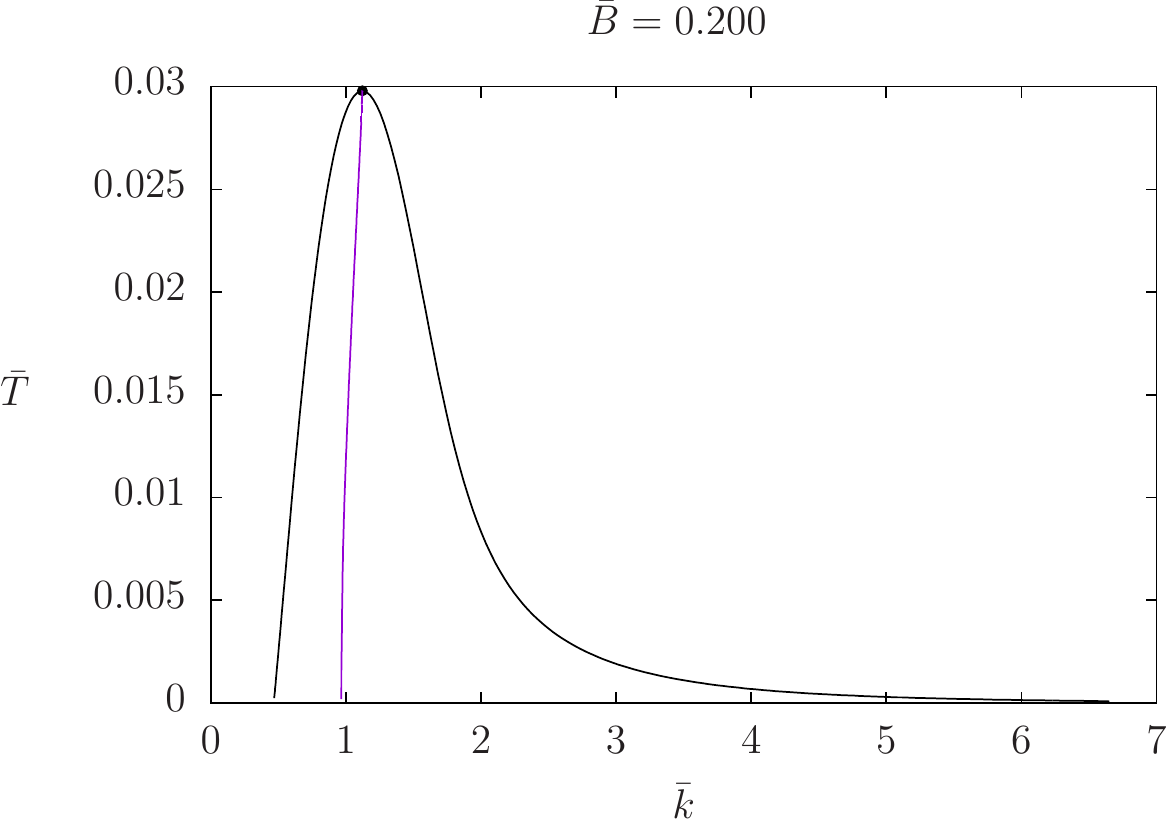}
\includegraphics[width=7.2cm]{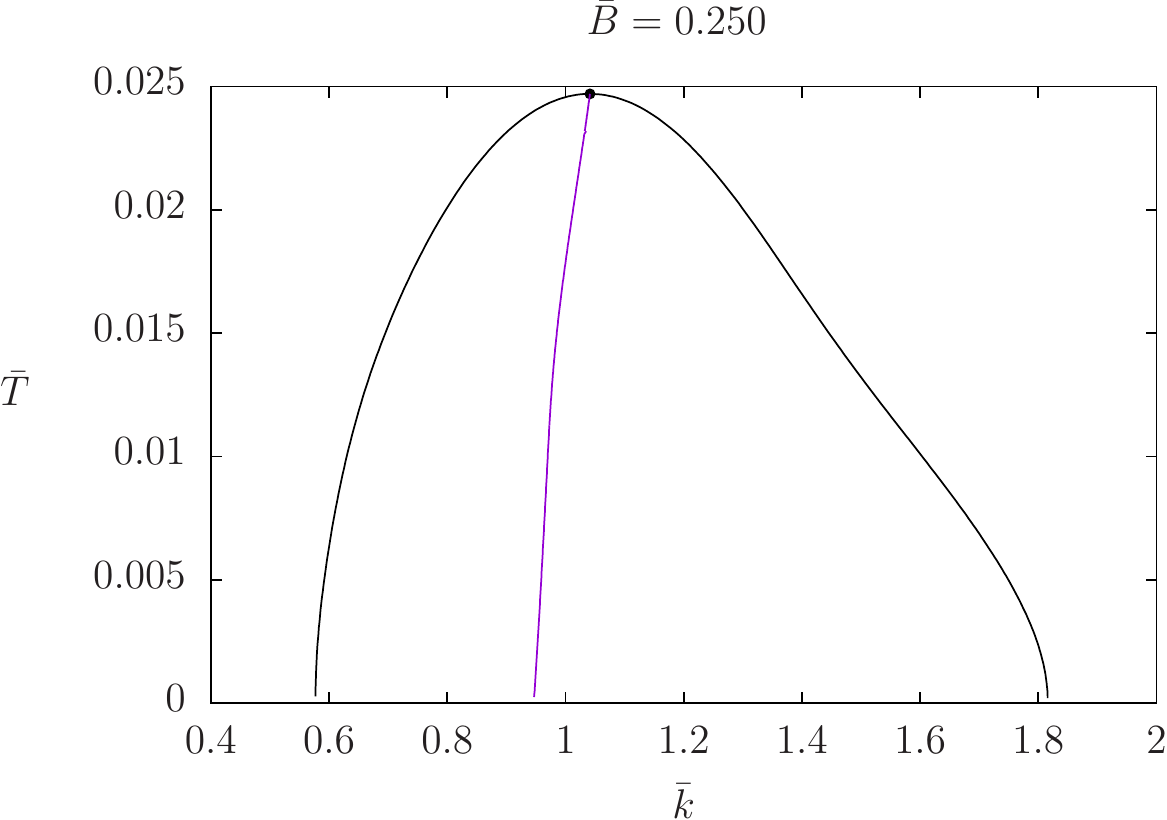}
\end{center}
\caption{Diagram of the $(\bar{T}, \bar{k})$-plane for fixed $\bar{B} = 0.200 < \bar{B}_{\rm C}$ (left panel) and $\bar{B} = 0.250> \bar{B}_{\rm C}$ (right panel). The thermodynamically most favourable physical states are found along the curve $\bar{k}_*(\bar{T})$ for which the grand canonical potential $\bar{\Omega}$ is minimised. }
\label{fig:Thermo_gamma15_Txk}
\end{figure}

For fixed $\bar{B}$ and for fixed temperature $\bar{T}$ we construct the solutions for different values of $\bar{k}.$ The solution corresponding to the physical state minimizes the grand canonical potential $\bar{\Omega}(\bar{k})$. The corresponding value for $\bar{k}$ minimising the grand canonical potential is denoted by $\bar{k}_*$. For fixed $\bar{B}$ we repeat this procedure for smaller temperatures $\bar{T}$ and hence obtain a trajectory $\bar{k}_*(\bar{T})$ in the  $(\bar{T},\bar{k})$-plane of thermodynamically preferred solutions. This trajectory is shown in figure \ref{fig:Thermo_gamma15_Txk} for the values $\bar{B} = 0.200 < \bar{B}_{\rm C}$ and $\bar{B} = 0.250> \bar{B}_{\rm C}$. In both cases, note that when lowering the temperature $\bar{T}$, the wave-number $\bar{k}_*(\bar{T})$ decreases and hence the pitch $\bar{p}_* = 2\pi / \bar{k}_*$ increases.

Along such trajectories of thermodynamically preferred solutions, we evaluate the observables derived in section \ref{sec:Thermo} and compare them to the corresponding values from the charged magnetic solution. In all the following figures, a continuous line represents a result within the new magnetic helical phase, whereas the dashed lines depict the results of \cite{D'Hoker:2010rz}  (see appendix \ref{sec:DhokerKraus} for more details).
\begin{figure}[b!]
\begin{center}
\includegraphics[width=9.0cm]{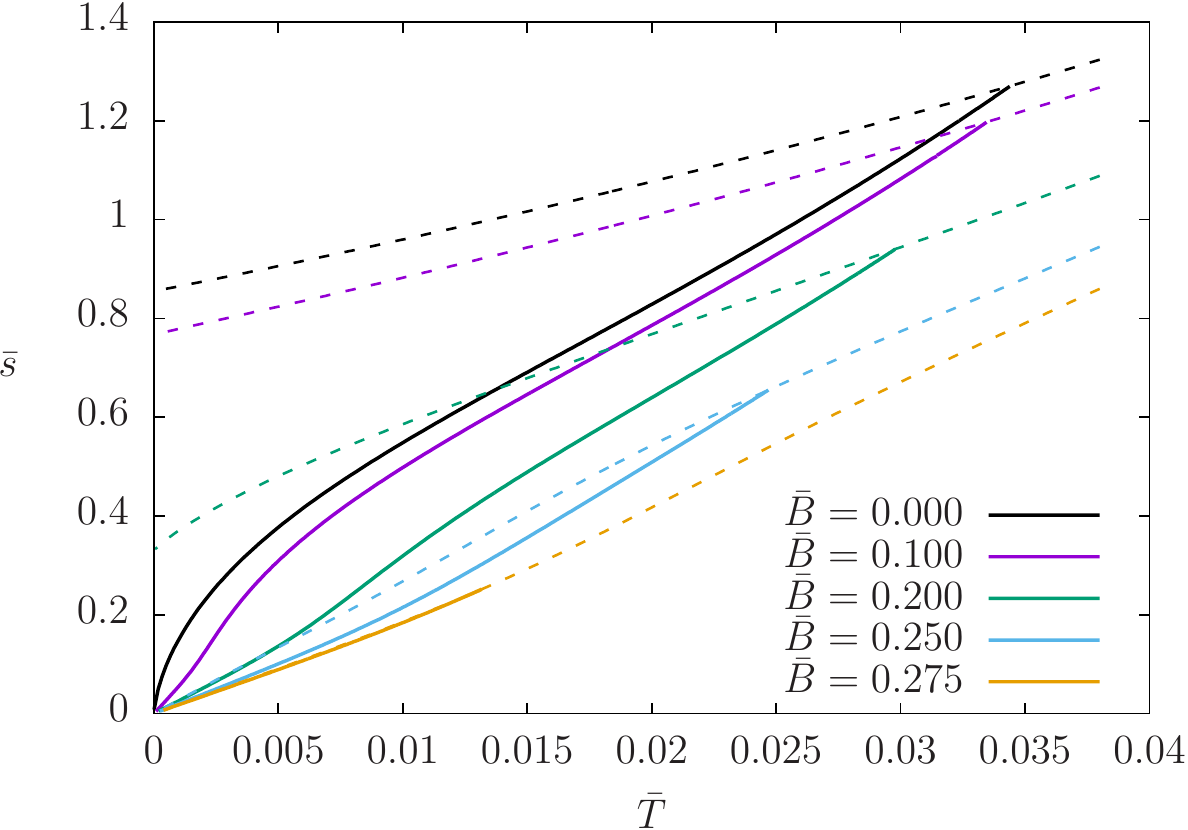}
\end{center}
\caption{Low temperature behaviour of entropy $\bar{s}$. The dashed lines are the charged magnetic results. For $\bar{B}< \bar{B}_{\rm C}$ the entropy does not vanish as $\bar{T} \rightarrow 0$. In the new magnetic helical phase (continuous line), $\bar{s}$ goes to zero for $\bar{T} \rightarrow 0$, regardless of the value of the magnetic field.}
\label{fig:Thermo_gamma15_s}
\end{figure}
First, in figure \ref{fig:Thermo_gamma15_s} we present the entropy density $\bar{s}$ as a function of $\bar{T}$. Let us first concentrate on the dahsed lines corresponding to the charged magnetic black brane. For $\bar{B}<\bar{B}_{\rm C}$ the entropy goes to a non-vanishing constant as $\bar{T}\rightarrow 0$ in agreement with the results of \cite{D'Hoker:2010rz}. However, for the new helical magnetic black brane construct in this paper, we observe that $\bar{s}\rightarrow 0$ as $\bar{T}\rightarrow 0$ regardless of the value of $\bar{B}$. Due to the vanishing entropy density we are confident that the magnetic helical black brane is dual to the true ground state of the CFT. Moreover, for fixed $\bar{B}$ the entropy is continuous close to the  phase transition, i.e. for $\bar{T} \lesssim \bar{T}_C(\bar{B}).$ Hence the phase transition is second order.

\begin{figure}[t!]
\begin{center}
\includegraphics[width=7.2cm]{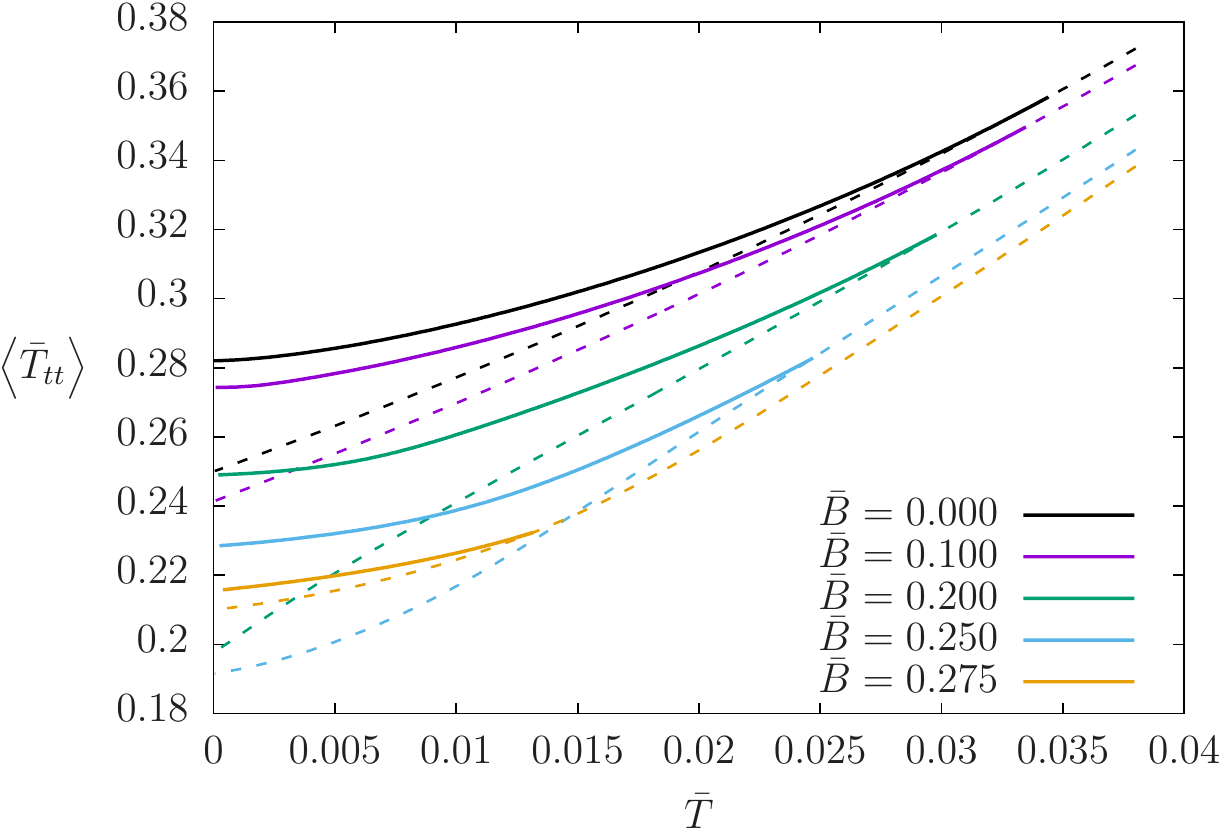}
\includegraphics[width=7.2cm]{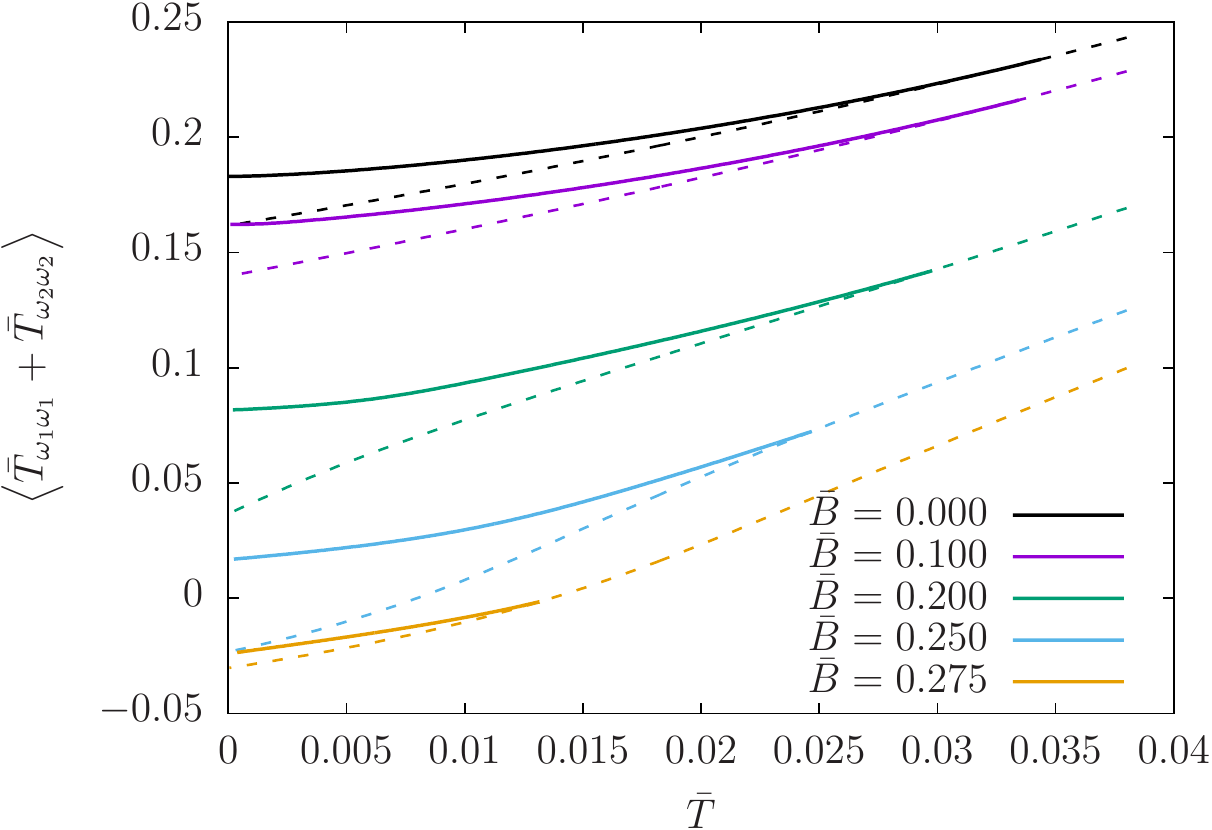}
\includegraphics[width=7.2cm]{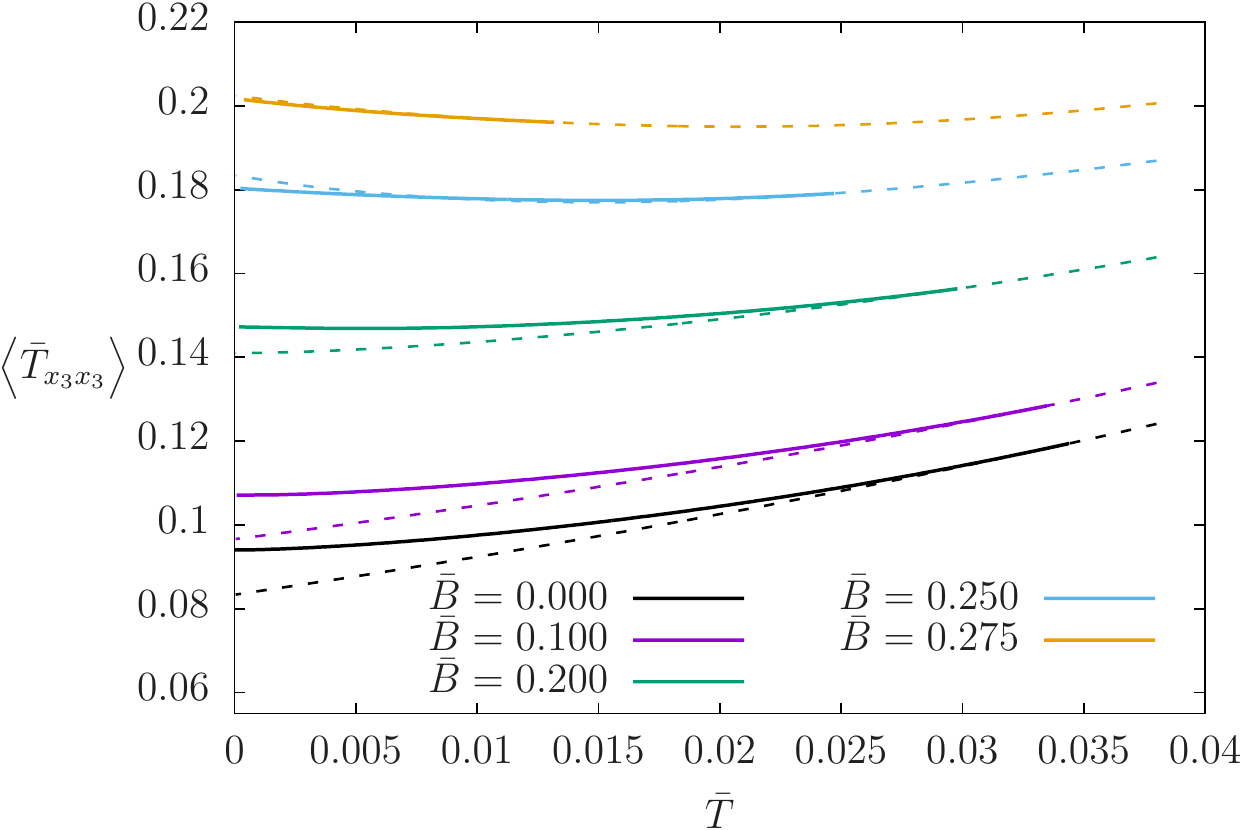}
\includegraphics[width=7.2cm]{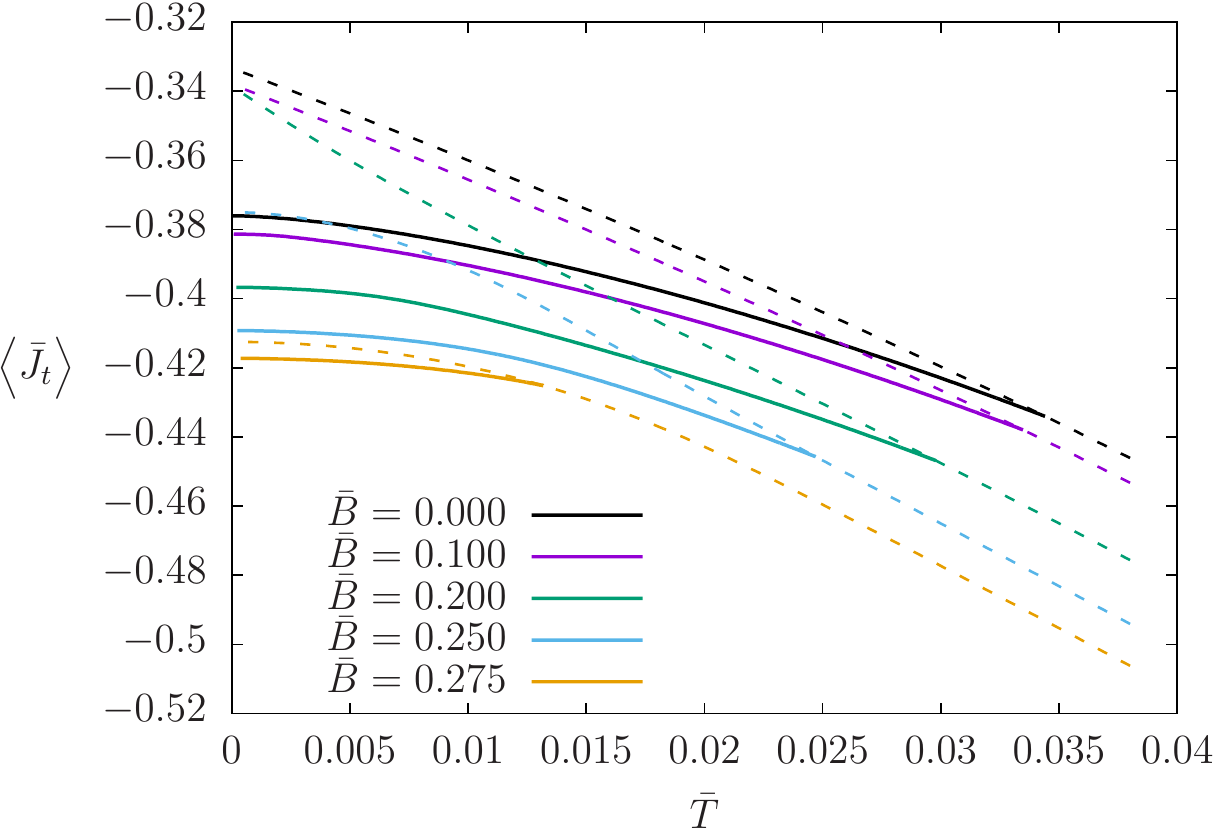}
\end{center}
\caption{Components $\left< \bar{T}_{tt} \right>$, $\left< \bar{T}_{\omega_1\omega_1} \right> + \left< \bar{T}_{\omega_2\omega_2} \right>$, $\left< \bar{T}_{x_3 x_3} \right>$ of the energy momentum tensor and $\left<\bar{J}_t\right>$ of the current. Note that $\left<J_t\right>=-\rho$ where $\rho$ is the charge density.  The dashed lines are the charged magnetic results, while the continuous lines correspond to the values in the magnetic helical phase. }
\label{fig:Thermo_gamma15_EnergyCurrents}
\end{figure}

Next, we turn to the non-vanishing components of the energy-momentum tensor $\left< \bar{T}_{\mu\nu} \right>$ and the current $\left<\bar{J}_\mu\right>$ of the dual field theory. Fig.~\ref{fig:Thermo_gamma15_EnergyCurrents} depicts the components $\left< \bar{T}_{tt} \right>$, $\left< \bar{T}_{\omega_1\omega_1} \right> + \left< \bar{T}_{\omega_2\omega_2} \right>$, $\left< \bar{T}_{x_3 x_3} \right>$ and $\left<\bar{J}_t\right>$. In all cases there are expected small deviation between the helical magnetic black brane and the charge magnetic black brane.
Moreover, from eq.~(\ref{eq:Efuncasymp2}) and the normalisation (\ref{eq:PhysObserv}) it is clear that $\left\langle \bar{J}_{x_3} \right\rangle = - \gamma \, \bar{B}$ is a constant. Furthermore, we also confirm that \eqref{eq:Ttx3} holds numerically, i.e. in terms of dimensionless quantities $ \displaystyle \left< \bar{T}_{t x_3} \right> = \frac{1}{2}\gamma \bar{B}$. Note that the relation \eqref{eq:Ttx3} is also satisfied for the charged magnetic black brane \cite{D'Hoker:2010rz} which we explicitly demonstrate in appendix \ref{sec:DhokerKraus}.

\begin{figure}[b!]
\begin{center}
\includegraphics[width=7.2cm]{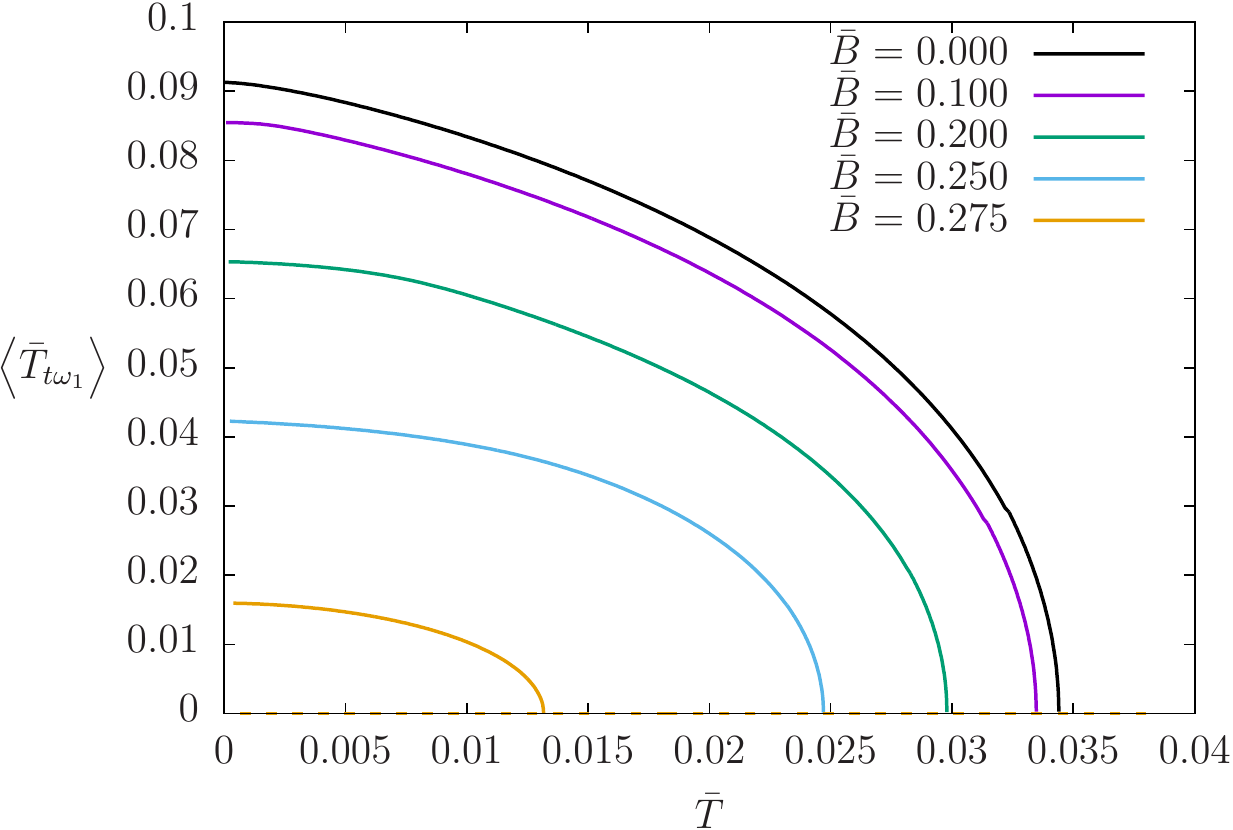}
\includegraphics[width=7.2cm]{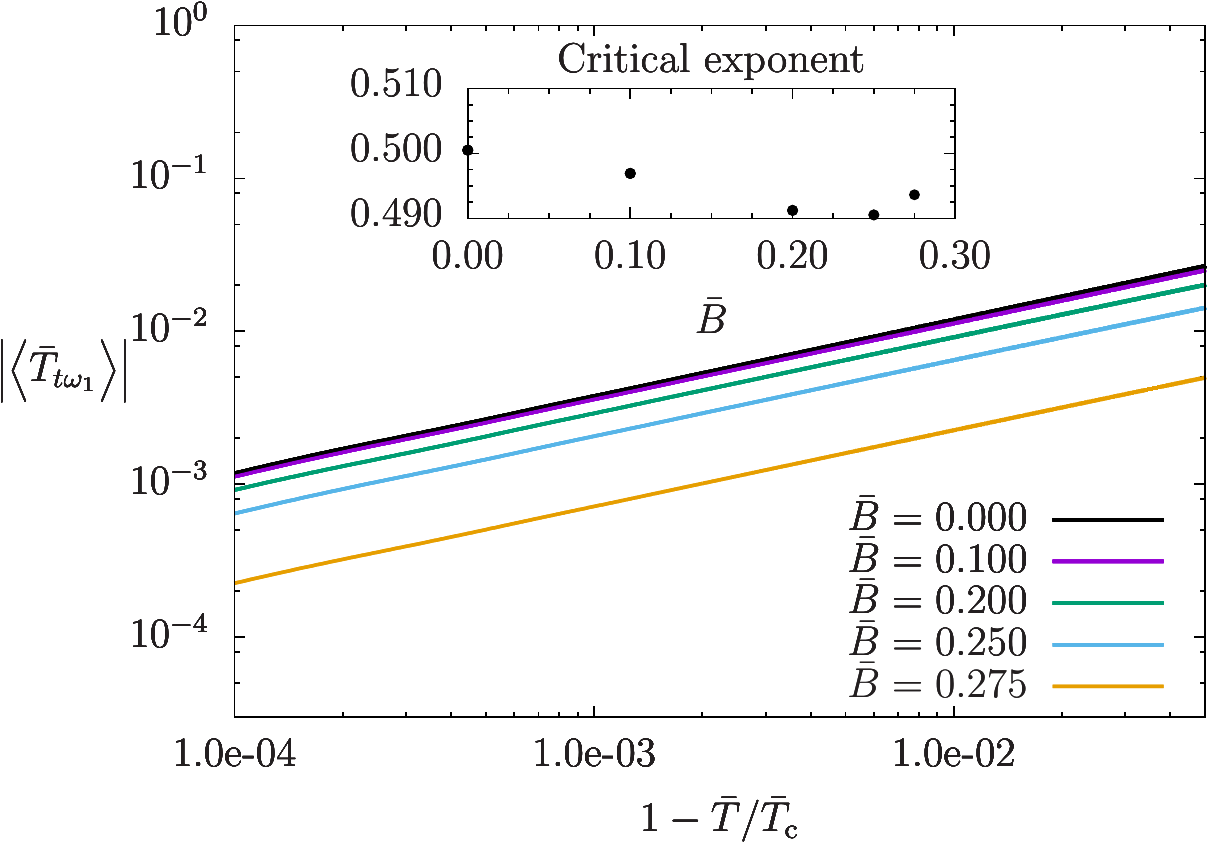}
\includegraphics[width=7.2cm]{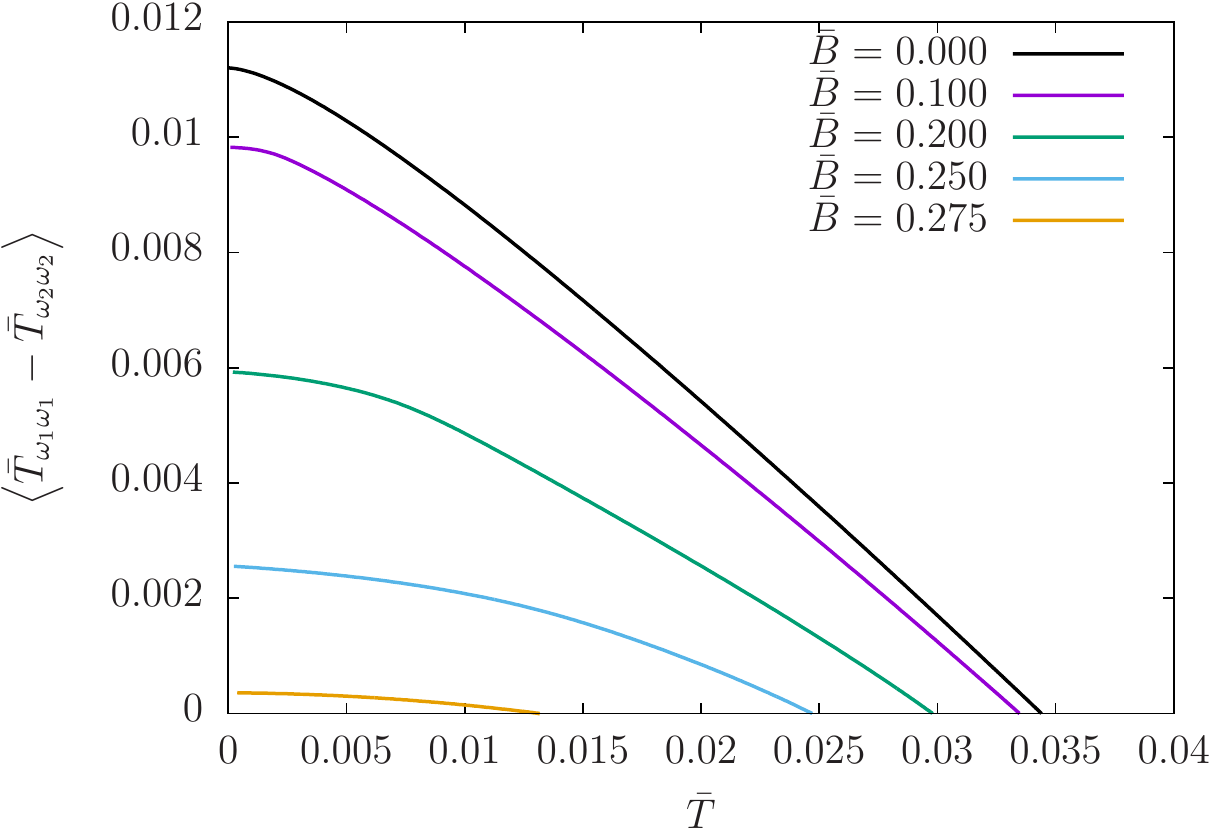}
\includegraphics[width=7.2cm]{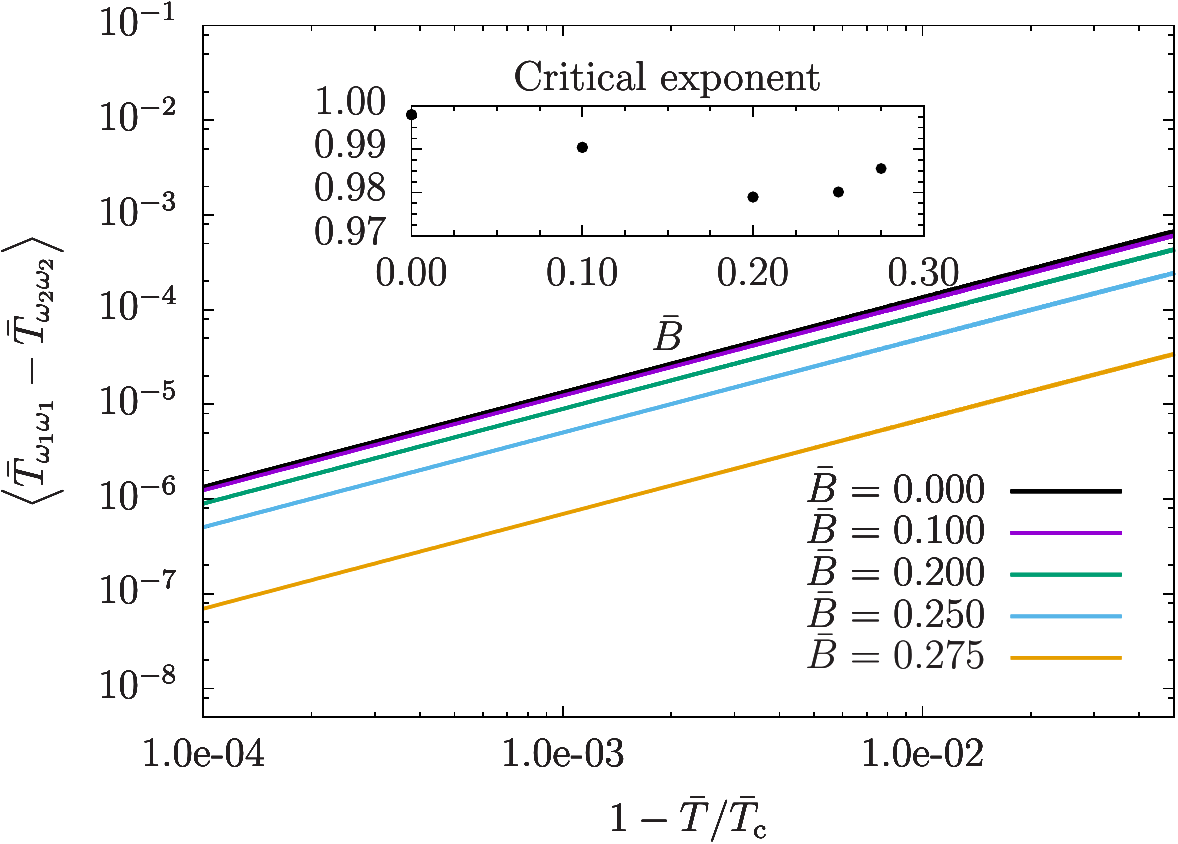}
\includegraphics[width=7.2cm]{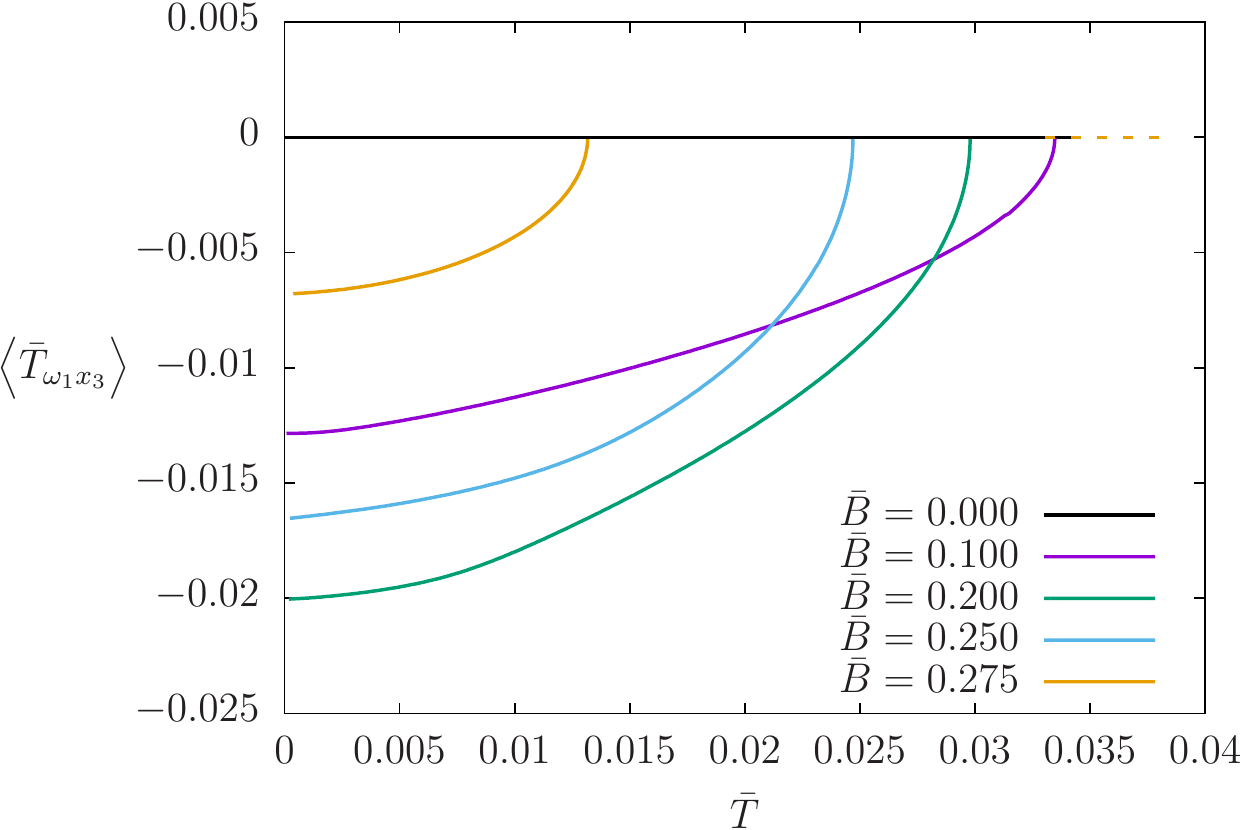}
\includegraphics[width=7.2cm]{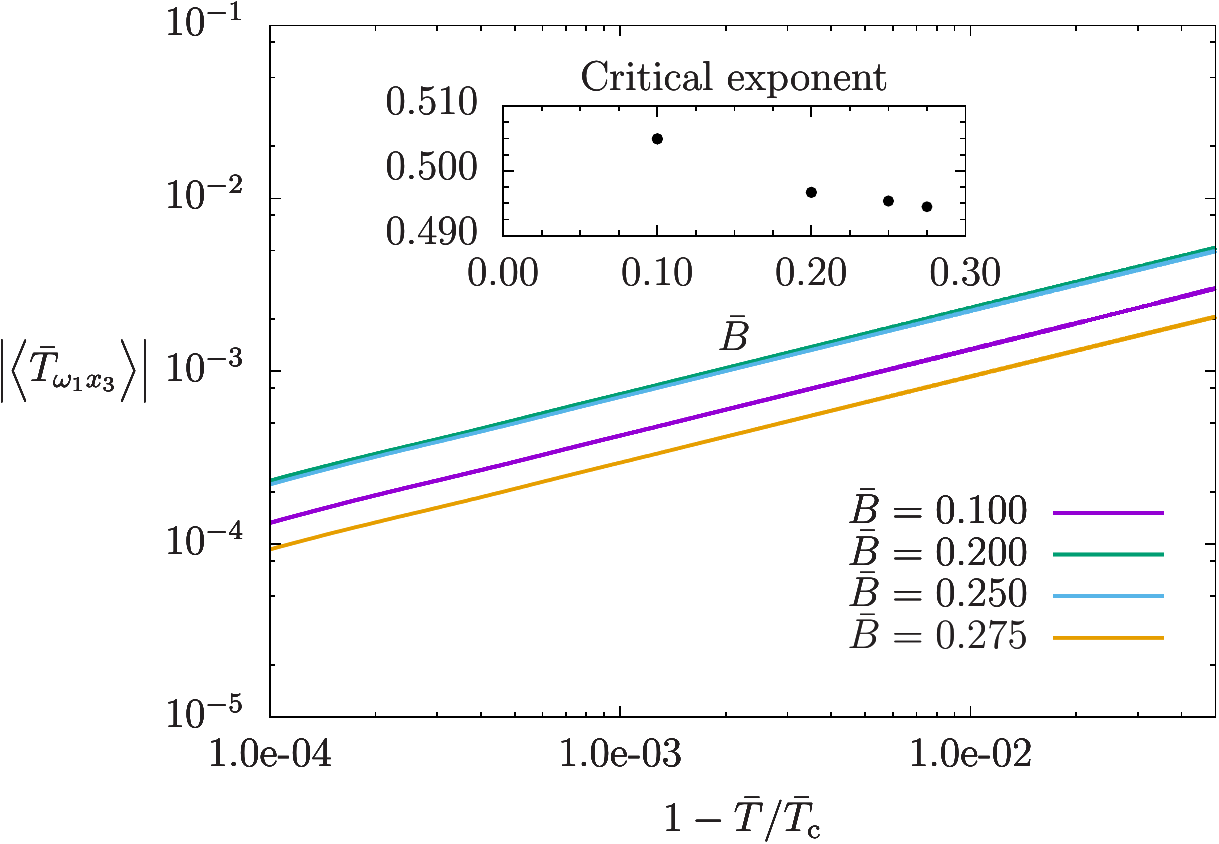}
\includegraphics[width=7.2cm]{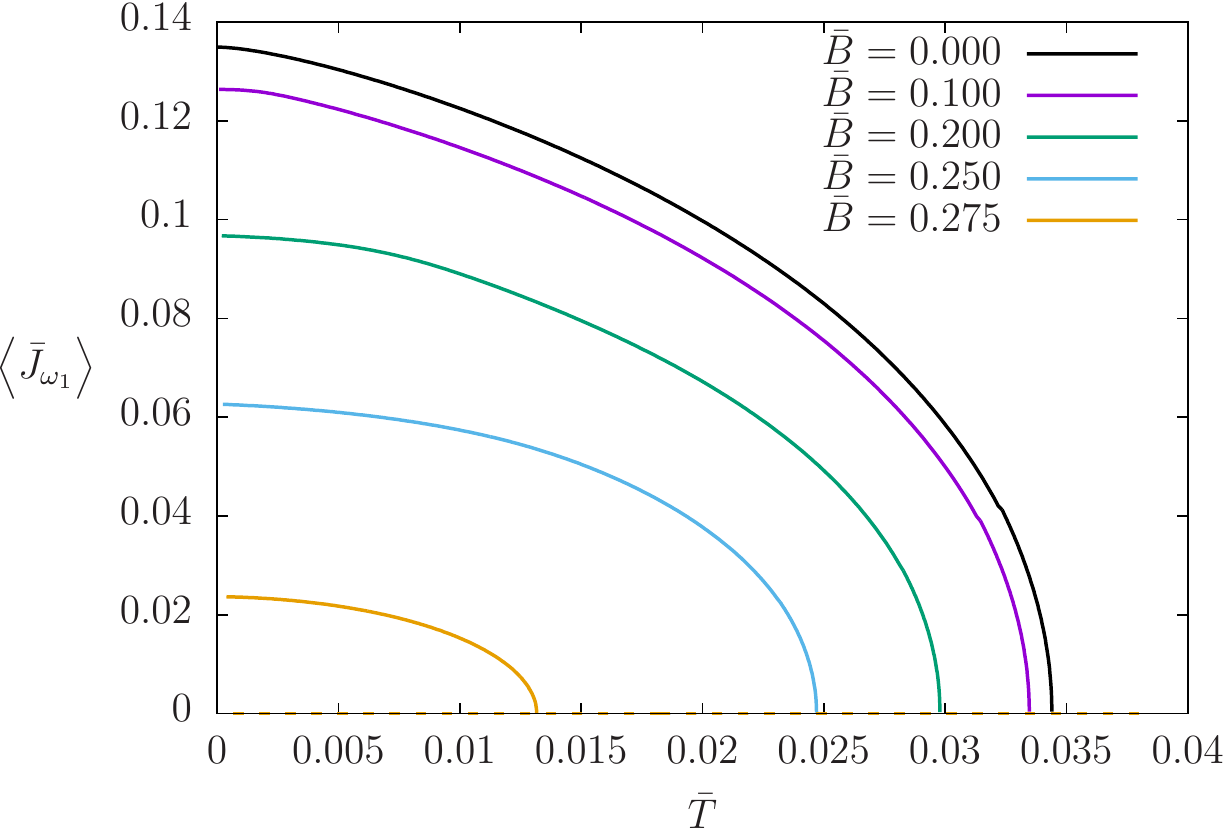}
\includegraphics[width=7.2cm]{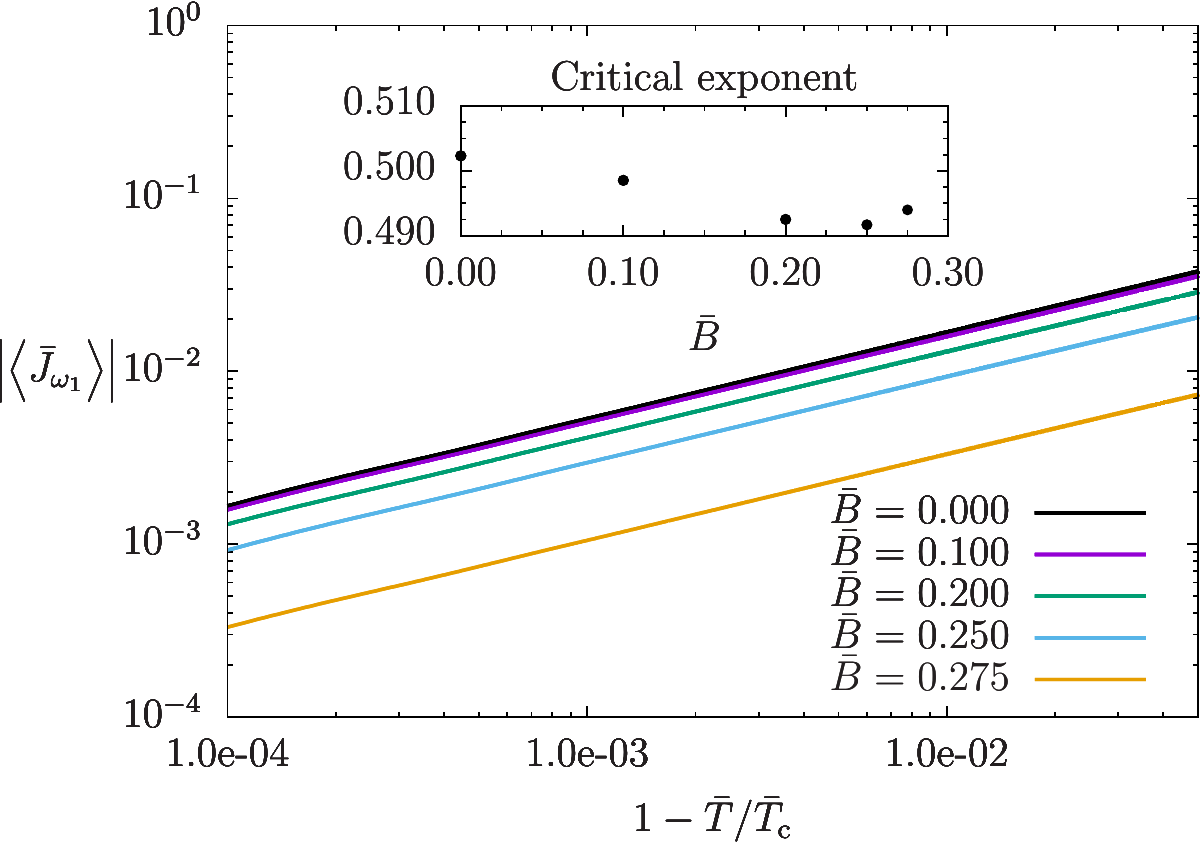}
\end{center}
\caption{Components $\left< \bar{T}_{t\omega_1} \right>$, $\left< \bar{T}_{\omega_1\omega_1} \right> - \left< \bar{T}_{\omega_2\omega_2} \right>$, $\left< \bar{T}_{\omega_1 x_3} \right>$ of the energy momentum tensor and $\left<\bar{J}_{\omega_1}\right>$ of the current. Left panel: these observables vanishes as $\bar{T}\rightarrow \bar{T}_{\rm C}$. Right panel: Behaviour in terms of $\left| 1 - \frac{\bar{T}}{\bar{T}_{\rm C}}\right|$. From the double logarithmic scale we infer the critical exponents depicted in the inset. We observe no systematic dependence on $\bar{B}$.}
\label{fig:Thermo_gamma15_EnergyCurrents_loglog}
\end{figure}
Finally, we display the results for the components  $\left< \bar{T}_{t\omega_1} \right>$, $\left< \bar{T}_{\omega_1\omega_1} \right> - \left< \bar{T}_{\omega_2\omega_2} \right>$, $\left< \bar{T}_{\omega_1 x_3} \right>$ and $\left<\bar{J}_{\omega_1}\right>$. As one can see in the left panels of figure \ref{fig:Thermo_gamma15_EnergyCurrents_loglog}, these components vanish as $\bar{T} \rightarrow \bar{T}_{\rm C}$. Hence these components are candidates for the order parameter. The right panels show the same observables against $\left| 1 - \frac{\bar{T}}{\bar{T}_{\rm C}}\right|$ in a double logarithmic scale.  The critical exponents are displayed in the inset of those figure. Within the interval range $\left| 1 - \frac{\bar{T}}{\bar{T}_{\rm C}}\right| \in [10^{-4}, 10^{-2} ]$, we do not observe any systematic dependency on the magnetic field $\bar{B}$ since the deviations are within the expected range of numerical errors. The critical exponents are those expected from mean field, i.e.
\ba
\left< \bar{T}_{t\omega_1} \right> \sim \left| 1 - \frac{\bar{T}}{\bar{T}_{\rm C}}\right|^{1/2}, \qquad\qquad\qquad\quad \  \left< \bar{T}_{\omega_1 x_3} \right> &\sim& \left| 1 - \frac{\bar{T}}{\bar{T}_{\rm C}}\right|^{1/2}, \nn \\ \label{eq:mfcritexp}
\left<\bar{J}_{\omega_1}\right> \sim \left| 1 - \frac{\bar{T}}{\bar{T}_{\rm C}}\right|^{1/2} \, , \qquad \left[ \left< \bar{T}_{\omega_1\omega_1} \right> - \left< \bar{T}_{\omega_2\omega_2} \right> \right] &\sim& \left| 1 - \frac{\bar{T}}{\bar{T}_{\rm C}}\right|^{1}. 
\ea

\section{Summary and Outlook}\label{sec:summary}
We studied strongly coupled four dimensional CFTs with chiral anomaly whose gravitational description is given in terms of Einstein-Maxwell-Chern-Simons theory in asymptotically AdS spacetime. In particular, we investigated the phase diagram at finite temperature, chemical potential and magnetic field and found a new spatially modulated phase for low temperatures and small magnetic fields. This is dual to asymptotically $\textrm{AdS}_5$ black brane solutions with non-trivial electric charge density and magnetic field spontaneously acquiring a helical current which we construct numerically. Note that this helical phase is supposed to exist only for large enough coefficient $\gamma$ of the chiral anomaly. In this paper we presented results mainly for $\gamma = 1.5$.

The new spatially modulated phase has interesting features. First, the quantum critical point at $B = B_{\rm C}$ is hidden in this new phase, at least for the values of $\gamma$ studied here. Second, the phase transition is second order with mean field exponents. Third, our numerical results indicate that the entropy density vanishes in the limit of zero temperature supporting our speculations that this is the true ground state of the system. Fourth, we have extracted how the wave-number of the helical structure changes with the parameters of the phase diagram. Finally, during the course of this work, as a side product we developed new numerical techniques which may be also useful for other holographic systems.

It will be very interesting to further analyse the system by addressing questions such as: does the phase diagram change for other values of the chiral anomaly coefficient? Are the states with helical structure aligned to the magnetic field really thermodynamically favoured? Answering this question will require to solve partial differential equations on the gravity side. 

Moreover, it will be worthwile to explore if there exists a simple relation between the location of the quantum critical point, given by $B = B_{\rm C}$, and the location of the phase boundary $B=B_0$ at zero temperature. Note that both, the new phase and the quantum critical point, are controlled by the chiral anomaly coefficient and hence such a relation may exist although it is not obvious in terms of the dual field theory.  

Another future direction is to study transport coefficients and quasi-normal modes within the new phase extending the results of \cite{Janiszewski:2015ura}. Finally, the results presented here can be generalised to models with an anomaly structure closer to the one of QCD and Weyl semimetals. In particular, the interplay between a magnetic field and chemical potentials for the vector/axial charge may be interesting.

\acknowledgments{
We are very grateful to Marcus Ansorg, Amadeo Jimenez Alba and Sebastian M\"ockel for valuable discussions and for insightful comments on the draft. JL and RM acknowledge financial support by \textit{Deutsche Forschungsgemeinschaft (DFG)} GRK 1523/2. RM was supported by CNPq under the programme "Ci\^encia sem Fronteiras". MA would like to thank KITPC and the organizers of the workshop "Holographic duality for condensed matter systems", as well as University of Vienna and the organizers of the workshop "Vienna Central European Seminar" for their hospitality.}

\appendix

\section{Equations of motion}\label{sec:fulleom}
In this section we give some details on the equations of motions and, in particular, how we treat them as a boundary value problem. For convenience, let us reproduce here eqs.~(\ref{eq:EOM1}) and (\ref{eq:EOM2a}) and define them as
\ba
{\cal E}_{mn} &=& R_{mn} + 4g_{mn}- \frac{1}{2}\left(F_{mo}F_{n}{}^{o}-\frac{1}{6}g_{mn}F_{op}F^{op}\right) \nn \\
 {\cal M}  &=& \dd\star F+\frac{\gamma}{2}F\wedge F. \nn
\ea
Furthermore, we complete the one-forms (\ref{eq:omegaforms}) with $\omega_0  = \dd t$ and $\omega_4 = \dd z$ and express the equations of motion in terms of the tetrad basis $\omega^a = \omega^a{}_m dx^m$, i.e., we look specifically at\footnote{In the tetrad basis, the equations do not present any trigonometric term related to $\cos(k \, x_3)$ or $\sin(k \, x_3)$.} \ba
{\cal E}^{ab} &=& \omega^a{}_m\,\omega^b{}_n \,{\cal E}^{mn} \\ 
{\cal M}^{abcd} &=& \omega^a{}_m\, \omega^b{}_n\, \omega^c{}_o\, \omega^d{}_p \, {\cal M}^{mnop}.
\ea
The non-zeros components $ \{ {\cal E}^{00}$, ${\cal E}^{01}$, ${\cal E}^{03}$, ${\cal E}^{11}$, ${\cal E}^{13}$, ${\cal E}^{22}$, ${\cal E}^{24}$, ${\cal E}^{33}$, ${\cal E}^{44} \}$ and $\{ {\cal M}^{1234}$, ${\cal M}^{1245}$, ${\cal M}^{2345}  \}$ form a system of 12 ordinary differential equations (ODE) for our 10 field variables: the components of the metric (7 functions) and gauge field (3 functions). In spite of being overdetermined, this system of ODE is consistent as already shown in the main text. Therefore, we must solve 10 out of 12 equations and ensure that the remaining 2 are satisfied for at least one value of $z$. Yet, we must assure that the chosen equations are independent of each other. 

The first point to notice is that the second derivatives appearing in each of the Maxwell-Chern-Simons equations involve only one of each gauge field function. In other words, the equations ${\cal M}^{1234}$, ${\cal M}^{1245}$ and ${\cal M}^{2345}$ can be straightforward regarded individually as equations for $E(z)$, $P(z)$ and $b(z)$, respectively. Next, we observe that ${\cal E}^{24}$ contains only first order derivatives. Finally, one can work the second derivates out of the remaining ${\cal E}^{ab}$ and obtain equations for each one of the metric fields $u(z) $, $c(z)$, $v(z)$, $w(z)$, $\alpha(z)$, $g(z)$ or $q(z)$. This procedure leaves us with 7 second order ODEs and one additional first order ODE (apart from ${\cal E}^{24}$).

By sorting out the second derivatives, we can associate for each one of the fields its respective ODE. In this way, we need boundary values at both $z=0$ and $z=1$. The only exception is the function $g(z)$, for which we work with the first order ODE ${\cal E}^{24}$ and therefore we are only allowed to fix the value at one of the surfaces. To exemplify the structure of the system of equations, let us collect the field variables and the equations of motion into the vector notation
\be
\label{eq:EOM_vec}
\vec{x} = \left[ \begin{array}{c}
 u(z)\\ c(z)\\ w(z)\\ v(z)\\ \alpha(z)\\ g(z)\\ q(z)\\ E(z)\\ P(z)\\ b(z) 
 \end{array}
 \right] ,\quad
  \vec{f}(\vec{x}; z) = \left[ \begin{array}{c}
f_u (u'', \vec{x}' , \vec{x}; z)\\ 
 f_c (c'', \vec{x}' , \vec{x}; z)\\ 
 f_w (w'', \vec{x}' , \vec{x}; z)\\ 
 f_v(v'', \vec{x}' , \vec{x}; z)\\ 
 f_\alpha (\alpha'', \vec{x}' , \vec{x}; z)\\ 
 f_g (\vec{x}' , \vec{x}; z)\\ 
 f_q (q'', \vec{x}' , \vec{x}; z)\\ 
 f_E (E'', \vec{x}' , \vec{x}; z)\\ 
 f_P (P'', \vec{x}' , \vec{x}; z)\\ 
 f_b (b'', \vec{x}' , \vec{x}; z) 
 \end{array}
 \right] \, .
 \ee
The boundary values are a mixture of regularity conditions imposed by the equations of motion and physical assumptions  insuring the surfaces $z=0$ and $z=1$ to represent the AdS boundary and the event horizon, respectively.

For example, at $z=1$ the horizon condition tells us that $u(1)=0$. Moreover due to regularity, we have to impose $E(1)=0$. By imposing such conditions on the remaining equations, we are left with regularity conditions involving the value of fields and their first derivatives at $z=1$ (Robin boundary conditions). In some specific cases, the conditions are rather simple and reduce to $q(1)=0$, $c(1)=0$. 

The next step is to study the asymptotic expansion around the AdS boundary $z=0$. In its most generic form, the expansions read
\ba
u(z) &=& 1 + u_2 z^2 + {\mathbf u_4}z^4 + {\cal O}(z^6) + z^4\ln(z)\left[ \hat{u}_4  +  {\cal O}(z^2) \right]\notag \\
		&& + {\mathbf u_1} z\left( 1 +  u_5z^4 +  {\cal O}(z^6)  + z^4\ln(z)\left[ \hat{u}_5  +  {\cal O}(z^2) \right] \right), \nn \\
v(z) &=& {\mathbf v_0} + v_2 z^2 +  v_4z^4 + {\cal O}(z^6) + z^4\ln(z)\left[ \hat{v}_4  +  {\cal O}(z^2) \right]\notag \\
		&& + {\mathbf u_1} z\left( v_1 + v_3 z^2 +  {\cal O}(z^2)  + z^4\ln(z)\left[ \hat{v}_5  +  {\cal O}(z^2) \right] \right), \nn \\
w(z) &=& {\mathbf w_0} + w_2 z^2 + {\mathbf w_4} z^4 + {\cal O}(z^6) + z^4\ln(z)\left[ \hat{w}_4  +  {\cal O}(z^2) \right]\notag \\
		&& + {\mathbf u_1} z\left( w_1 + w_3 z^2 +  {\cal O}(z^2)  + z^4\ln(z)\left[ \hat{w}_5  +  {\cal O}(z^2) \right] \right), \nn \\
\alpha(z) &=& {\mathbf a_0} + a_2 z^2 + {\mathbf a_4}z^4 + {\cal O}(z^6) + z^4\ln(z)\left[ \hat{a}_4  +  {\cal O}(z^2) \right]\notag \\
		&& + {\mathbf u_1} z\left(  a_3 z^2 + a_5z^4 +  {\cal O}(z^6)  + z^4\ln(z)\left[ \hat{a}_5  +  {\cal O}(z^2) \right] \right),
\ea
as well as 
\ba
c(z) &=& {\mathbf c_0} + c_2 z^2 + {\mathbf c_4}z^4 + {\cal O}(z^6) + z^4\ln(z)\left[ \hat{a}_4  +  {\cal O}(z^2) \right]\notag \\
		&& + {\mathbf u_1} z\left(  c_3 z^2 + c_5z^4 +  {\cal O}(z^6)  + z^4\ln(z)\left[ \hat{a}_5  +  {\cal O}(z^2) \right] \right) \, , \\
g(z) &=&  {\mathbf g_0} + g_2 z^2 + g_4z^4 + {\cal O}(z^6) + z^4\ln(z)\left[ \hat{a}_4  +  {\cal O}(z^2) \right]\notag \\
		&& + {\mathbf u_1} z\left(  g_3 z^2 + g_5z^4 +  {\cal O}(z^6)  + z^4\ln(z)\left[ \hat{a}_5  +  {\cal O}(z^2) \right] \right) \, , \nn\\
q(z) &=& {\mathbf q_0} + q_2 z^2 + {\mathbf q_4}z^4 + {\cal O}(z^6) + z^4\ln(z)\left[ \hat{q}_4  +  {\cal O}(z^2) \right]\notag \\
		&& + {\mathbf u_1} z\left(  q_3 z^2 + q_5z^4 +  {\cal O}(z^6)  + z^4\ln(z)\left[ \hat{q}_5  +  {\cal O}(z^2) \right] \right) \, , \nn\\
 E(z)  &=& {\mathbf E_0}  + \left(  {\mathbf E_2} z^2  + E_3 z^3  + {\cal O}(z^4) + z^2\ln(z)\left[ \hat{E}_2 + \hat{E}_4 z^2 + {\cal O}(z^4)\right] \right) \, , \notag \\
&& +   {\mathbf u_1} z^3 \left(  E_3 + E_5 z^2 +{\cal O}(z^4) +\ln(z)\left[\hat{E}_3 +{\hat E}_4 z^2 +{\cal O}(z^4) \right]\right),	\nn\\
P(z)  &=& {\mathbf P_0}  + \left(  {\mathbf P_2} z^2  + P_3 z^3  + {\cal O}(z^4) + z^2\ln(z)\left[ \hat{P}_2 + \hat{P}_4 z^2 + {\cal O}(z^4)\right] \right) \, , \notag \\
&& +   {\mathbf u_1} z^3 \left(  P_3 + P_5 z^2 +{\cal O}(z^4) +\ln(z)\left[\hat{P}_3 +{\hat P}_4 z^2 +{\cal O}(z^4) \right]\right),\nn	\\
b(z) &=&  b_0 + {\mathbf b_2} z^2 + b_4z^4 + {\cal O}(z^6) + z^2\ln(z)\left[ \hat{b}_2  + \hat{b}_4z^2   {\cal O}(z^4) \right]\notag \\
		&& + {\mathbf u_1} z\left(  b_3 z^2 + b_5z^4 +  {\cal O}(z^6)  + z^2\ln(z)\left[ \hat{b}_3  + \hat{b}_5 z^2 +  {\cal O}(z^2) \right] \right) \, . \nn
\ea
The quantities in boldface are free parameters, which can not be determined by the series expansion. All the other terms are fixed by them if one considers {\em all} equations of motion (including here the two first order differential equations). For example, we find that 
\be
\label{eq:b_0}
(B \, g_0 + k \, b_0)(1-w_0^2 \, c_0^2) + B \, w_0^2 \, c_0 \, q_0 = 0 \, .
\ee
Asymptotically AdS solutions require 
\[
v_0=1, \quad w_0 =1, \quad \alpha_0 = 1, \quad c_0=0, \quad g_0=0, \quad q_0=0 \, ,
\] 
and thus $b_0 = 0$ from (\ref{eq:b_0}). In our numerics we demand $u_1 = 0$ in order to fix all remaining diffeomorphisms. For the gauge field functions, we fix the chemical potential\footnote{From the ODE point of view, we could also prescribe $E_2 = \rho/2$ instead of $E_0$.} $E_0=-\mu$. Moreover, we do not allow for source term for operators dual to $P(z)$. Hence we impose $P_0=0$.
With this conditions, the expansions around $z=0$ assume the much simpler form given by (\ref{eq:Exp_metric_AdS}).

Once the solution is available, the thermodynamic observables (see appendix \ref{sec:MoreThermo}) require the knowledge of some coefficients related to higher derivatives, such as $u_4$, $w_4$, $a_4$, $c_4$ and $q_4$. Not only do we lose accuracy by calculating them numerically, but there are also some cases in which the derivative might not even exist due to the presence of terms $z^4\ln(z)$. In order to get access to all the needed coefficients with a reliable high accuracy, we incorporate the boundary conditions into our variables and introduce auxiliary fields via
\ba
&& u(z) = 1 + \frac{B^2}{6}z^4\ln(z) + z^4\left[ -1 + (1-z)\tilde{u}(z) \right], \quad\qquad\qquad \alpha(z) = 1 + z^4\tilde{\alpha}(z),\nn \\
&& v(z) = 1 - \frac{B^2}{24}z^4\ln(z) + z^4\tilde{v}(z),  \quad\qquad\qquad\ 
w(z) = 1 + \frac{B^2}{12}z^4\ln(z) + z^4\tilde{w}(z), \nn\\
&& c(z) = z^4(1-z)\tilde{c}(z), \label{eq:Aux_c} \quad\qquad \ q(z) = z^4(1-z)\tilde{q}(z), \quad\qquad \ g(z) = z^4\tilde{g}(z),  \nn \\
&& E(z) = (1-z)\tilde{E}(z), \quad\qquad \
P(z) = z^2\tilde{P}(z), \qquad\qquad\qquad
b(z) = z^2\tilde{b}(z). \label{eq:AuxVar}
\ea 
After substituting eqs.~(\ref{eq:AuxVar}) into the equations of motion (\ref{eq:EOM_vec}), we factor out powers of $z$ and $(1-z)$ and impose the resulting equations in the entire interval $z\in[0,1]$, i.e., as $z\rightarrow 0$ and $z \rightarrow 1$, the boundary conditions follow automatically from the limiting values of the the equations of motion written in terms of the auxiliary variables. The equations are then solved numerically with a spectral method, described in appendix \ref{sec:NumDetail}.

\section{More on thermodynamics} \label{sec:MoreThermo}
Calculating thermodynamic properties requires the evaluation of the on-shell action, which in principle is an integral over the numerically determined solutions.
It is more enlightening to have an analytic expression in which only boundary values of the solution have to be inserted. In this Appendix we show how to rewrite (part of) the on-shell Lagrangian as a total derivative.
This is usually also used to get Smarr Type formulas, for example see \cite{Bhattacharya:2011eea,Donos:2013woa}. We present here systematically how to do this step by step. First by taking the trace of the Einstein equations,
\begin{align}
R_{mn}& = -4g_{mn}+T_{mn} \, ,   &&R = -20 +T\,, \qquad\qquad\qquad \nonumber\\ \label{EOMapp1}
T_{mn}&=\frac{1}{2}\left(F_{mo}F_{n}^{\ o}-\frac{1}{6}g_{mn}F_{op}F^{op}\right) \, ,   &&T =\frac{1}{12}F_{mn}F^{mn}\,,
\end{align}
we can reformulate the Einstein-Maxwell part of the Lagrangian as
\begin{eqnarray}\
\mathcal{L}_{EM}= R+12-\frac{1}{4}F_{mn}F^{mn} = -8 -\frac{1}{6} F_{mn}F^{mn} \,.
\end{eqnarray}
The equations of motion \eqref{EOMapp1} may be written in the following form
\begin{equation}
2\,R_{t}^{\ t}=-8+ 2\,T_{\ t}^{t} = -8+ 2\,\tilde{T}_{\ t}^{t}-\frac{1}{6}F_{mn}F^{mn} \, ,
\end{equation}
where $\tilde{T}_{m}^{\ n} = F_{mo} \, F^{no}/2$. Hence we obtain for $\mathcal{L}_{EM}$
\begin{equation}
 \mathcal{L}_{EM} =2\,R_{\ t}^{t}- 2\,\tilde{T}_{\ t}^{t}\,.
\end{equation}
In order to rewrite $\sqrt{-g} \, \mathcal{L}_{EM}$ as a total derivative we have to massage $R_{\ t}^{t}$ and $\tilde{T}_{\ t}^{t}$. 

Let us start with $R_{\ t}^{t}$ by using the identity $R_{\ n}^{m} \, \xi^{n}=\nabla_{m}\nabla_{n}\xi^{n}$ for an arbitrary Killing vector $\xi$ (with components $\xi^{m}$). 
If the metric depends only on the radial coordinate z the identity can be expressed as a total derivative
\begin{equation}\label{killingid}
 \sqrt{-g} \, R_{\ n}^{m} \, \xi^{n}=-\partial_{z}\left(\sqrt{-g}\nabla^{z}\xi^{m}\right) \,.
\end{equation}
In particular $\sqrt{-g}  R_{\ t}^t$ reads
\ba
\sqrt{-g}R_{t}^{t} &=& -\partial_{z}\left(\sqrt{-g}\nabla^{z}\xi^{t}\right) \notag \\
&=&  -\partial_{z} \left(\frac{v^2\, w \left(z \left(c \left(\alpha ^2\, q\, v^2\, g'-w^2\, c'\right)-\alpha ^2\, q\, v^2\, q'+u'\right)-2\, u\right)}{2 z^4} \right) \,.
\ea
In order to rewrite $\sqrt{-g} \, \tilde{T}_{\ t}^{t}$ as a total derivative we analyse the Maxwell equations
\begin{equation}
d*F+\frac{\gamma}{2}F\wedge F=0
\end{equation}
in the following systematic way: The $F\wedge F$ term has a rather simple structure
\begin{eqnarray} \nonumber
\frac{\gamma}{2}F\wedge F & =
-\gamma\,  \left(k\, b(z)\, b'(z)+B\, p(z)\right) 				& \dd x_{1} \wedge \dd x_{2} \wedge\dd x_{3} \wedge \dd z\\ \nonumber
&+\, \gamma\,  k\, b(z)\, e(z) \sin(k \,x_{3})\,				& \dd t \wedge \dd x_{1} \wedge\dd x_{3} \wedge \dd z \\  \nonumber
&+\, \gamma\,  k\, b(z)\, e(z) \cos(k \,x_{3})\,				&\dd t \wedge \dd x_{2} \wedge\dd x_{3} \wedge \dd z\\
&+\, B\, \gamma\,  e(z) 							& \dd t \wedge \dd x_{1} \wedge\dd x_{2} \wedge \dd z
\end{eqnarray}
while the term $d*F$ is a sum of total derivatives of $z$ and $x_{3}$
\begin{equation}
d*F=\left( \frac{\partial(*F)_{mnp}}{\partial z} \, \dd z +\frac{\partial (*F)_{mnp}}{\partial x_3} \, \dd x_{3} \right) \wedge \dd x^m \wedge \dd x^n \wedge \dd x^p\,.
\end{equation}
The relevant parts of $*F$ are those, without either $\dd z$ or $\dd x_{3}$. This gives three independent equations of motion. By comparing to the remainder $\tilde{T}_{\ t}^t$ given by
\be
\label{eq:rem2}
-2\sqrt{-g} \, \tilde{T}_{\ t}^t  = e(z) \, M_{1d}(z),
\ee
with $M_{1d}(z)$ defined by\footnote{$P^*(z)$ was first introduced in eq. \eqref{eq:DefEP}.}
\ba
M_{1d}(z) &=& \frac{\gamma \, k\,b(z)^2}{2} - P^*(z) \nn \\ 
&=& \frac{v(z)^2\, w(z)}{z} \left(c(z) \left(p(z)-g(z)\, b'(z)\right)+q(z)\, b'(z)+e(z)\right),
\ea
we see that the Maxwell equation with legs $\dd x_{1} \wedge \dd x_{2} \wedge\dd x_{3} \wedge \dd z$ corresponds to equation relating $p(z)$ and $P^*(z)$ in \eqref{eq:EOMEP}. For convenience, we re-write it here in the form
\begin{eqnarray}\label{eq:Maxwell1}
\frac{\partial }{\partial z} M_{1d}(z) = \gamma  \, R_1(z)  \quad {\rm with} \quad  R_1(z) = B\, p(z)+k\, b(z)\, b'(z).
\end{eqnarray}
Now, in order to rewrite $-2\sqrt{-g} \, \tilde{T}_{\ t}^t$ as a total derivative, we multiply $M_{1d}(z)$ by $E(z)$ and substract again the additional term of $\partial_z (E(z)\, M_{1d}(z))$.

Finally we also have to consider the Chern-Simons term 
\begin{equation}
-\frac{\gamma}{6} A\wedge F\wedge F = \mathcal{C}(z,x_3, x_2) \, \dd t \wedge \dd x_1 \wedge \dd x_2 \wedge \dd x_3 \wedge \dd z
\end{equation}
plus the remaining term $E(z)\,R_1(z)$,
\ba
\label{eq:rem3}
E(z)\,R_1(z) + \mathcal{C}(z,x_3, x_2) &=&  \gamma \, k\,b(z)\,\text{E}(z)\, b'(z)-B\, \gamma \, \text{E}(z)\,p(z) \nn \\
&& - \text{E}(z)\, \gamma  \left(B\, p(z)-k\, b(z)\, b'(z)\right) + \dots \, .
\ea
The absent terms in \eqref{eq:rem3} are proportional to $x_2 \cos(k x_3)$ and vanish when we integrate over any symmetric interval with respect to $x_2$. The expression \eqref{eq:rem3} can be reformulated in terms of the following two total derivatives
\begin{eqnarray}
  \frac{\partial }{\partial z}\left( \frac{1}{3} \gamma\,  k\, b(z)^2\, \text{E}(z)\right), \qquad\quad 
  \frac{\partial }{\partial z}\left(-\frac{1}{3} B\, \gamma\,  \text{E}(z)\, P(z)\right) \, ,
\end{eqnarray}
plus the remaining term  $\frac{1}{3}\ B\, \gamma\,  E(z)\, p(z)$.
Collecting everything we end up with the action density $s$, defined by $S = \mbox{vol}(\mathbbm{R}^{3,1}) \, s$,
\ba
s =\int dz && \left[  2\sqrt{-g} R_{\ t}^t +\frac{\partial}{\partial z} \bigg( M_{1d}(z) \, E(z)\bigg) \right. \nn \\
&&  \left.  -  \frac{ \gamma}{3}\, \frac{\partial }{\partial z} \bigg( k\, b(z)^2\, E(z)+B\, E(z)\, P(z) \bigg)
- \frac{\gamma}{3} \, B \, E(z)\, p(z) \right] \, .
\ea
Finally, we have to insert the boundary and horizon expansions \eqref{eq:Exp_metric_AdS} -- \eqref{eq:expGaugeField_Hrz} 
\begin{eqnarray}
s =  \bar{u}_1\,\bar{v}_0^{2}\, \bar{w}_0 + \frac{B^2}{6} + 2 \, u_4 - \frac{2}{z^4} + \mu \rho + \frac{1}{3} B^2 \, \log(z) - \frac{1}{3}\, B\,\gamma\int\text{E}(z)\, p(z)\, dz \,.
\end{eqnarray}
The divergent parts are canceled by appropriate counterterms given by \eqref{eq:actionSbdy}. Hence the final result for the on-shell action density reads
\begin{eqnarray}
s =  \bar{u}_1\,\bar{v}_0^{2}\, \bar{w}_0  + 3 \, u_4  + \mu \rho - \frac{1}{3}\, B\,\gamma\int_0^1 dz \, \text{E}(z)\, p(z)\,.
\end{eqnarray}
Note that the final result still contains an integral and hence the Lagrangian does not seem to reduce to a total derivative. However, the integral expression is actually a boundary term in $x_2$ direction. We checked this explicitly by computing the Noether charges along the lines of \cite{Rogatko:2007pv,Suryanarayana:2007rk}.


\section{Special cases }\label{sec:Reproduction}
In this section, we look at particular limits of our system of equations in order recover the results already available in the literature. We begin with the discussion of the special case $B = 0$ corresponding to the helical black brane solutions studied by Donos and Gauntlett in \cite{Donos:2012wi}. Afterwards, we comment on the case $B \neq 0$ and its quantum critical point observed in the studies of the charged magnetic branes solutions by D'Hoker and Kraus in \cite{D'Hoker:2010rz}.

\subsection{The special case $B= 0$}{\label{sec:DonosGauntlett}}
In our coordinates, the helical black brane solution \cite{Donos:2012wi} corresponds to the choice of
\be
\label{eq:DonosGauntlett}
B= P(z) = c(z) = g(z) = 0.
\ee
In this case, the line element (\ref{eq:ansatzmetric}), the field strength tensor (\ref{eq:ansatzF}) and the gauge field (\ref{eq:ansatzA}) read
\ba
\dd s^{2} & = & \frac{1}{z^{2}}\left(\frac{\dd z^{2}}{u(z)}-u(z)\, \dd t^{2}+v(z)^{2}\, \alpha(z)^{-2}\, \omega_{2}^{2} +v(z)^{2}\, \alpha(z)^{2}\, \left(\omega_{1}+q(z)\, \dd t\right)^{2}  + w(z)^{2} \, \omega_{3}^{2}\right) \nn \\
F  &=&  e(z)\,\dd t\wedge\dd z +b'(z)\,\dd z\wedge\omega_{1}+b(z)\, d\omega_{1}  \\
A &=& -E(z)\, \dd t  +  b(z) \, \omega_1. \nn
\ea
Note that our ansatz differs slightly from the one presented in \cite{Donos:2012wi}: first, we compactify $r$ by $r = 1/z$ with $z \in [ 0, 1]$, second we re-label the one-forms: $\omega_{1 \ \rm here} = \omega_{2 \ \rm DG}, \omega_{2 \ \rm here} = \omega_{3 \ \rm DG}$ and $\omega_{3 \ \rm here} = \omega_{1 \ \rm DG}$ and third we use a slightly different metric ansatz which is closer to the one used by D'Hoker and Kraus (see appendix \ref{sec:DhokerKraus}). In particular, while in \cite{Donos:2012wi} the metric components in terms of $z$ satisfy $z^4\,g_{tt}g_{zz} = -f(z)^2$, we choose an ansatz such that $z^4\,g_{tt}g_{zz} = -1$ and $z^4 g_{\omega_2 \omega_2} \, g_{\omega_1 \omega_1} = v(z)^4$. Finally, the authors of \cite{Donos:2012wi} use scaling freedom of the coordinates to set $\mu = 1$, while the coordinate location of the event horizon is not known a priori. In contrast we fix the horizon to be located at $z=1$, and hence we are not allowed to set $\mu$ to one.

\begin{figure}[h!]
\begin{center}
\includegraphics[width=8.2cm]{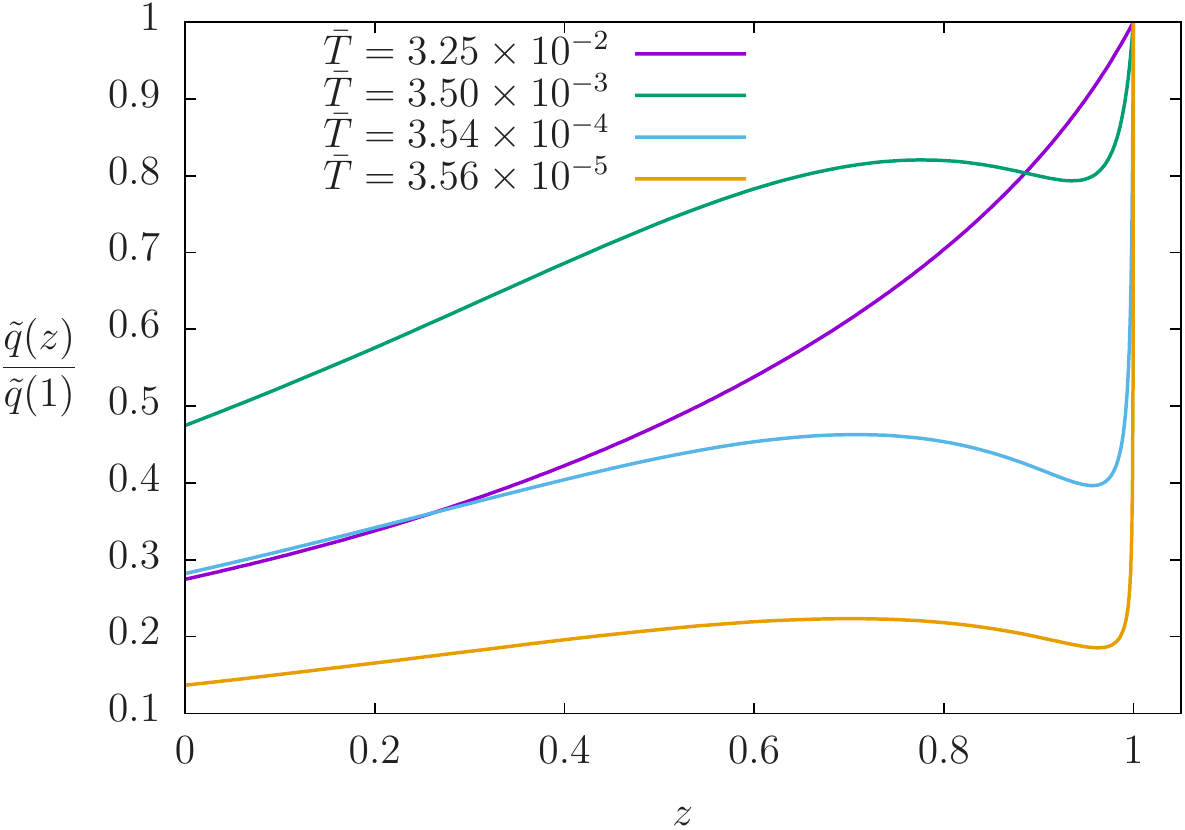}
\end{center}
\caption{Function $\tilde{q}(z)$ normalised by its value on $z=1$. In the regime $\bar{T}\rightarrow 0$, the function develops very strong gradients around the horizon. The numerical solution requires higher resolution and, eventually, specific techniques must be employed (such as the analytical mesh-refinement - see appendix \ref{sec:NumDetail}).  }
\label{fig:LowTemp_gamma17_funcQ}
\end{figure}

Our numerics pass an important check: We can reproduce all their results down to temperatures of order $\bar{T} \approx 10^{-5}$. The authors of \cite{Donos:2012wi} reported some difficulties in studying the behaviour of the solutions in the regime of very low temperatures, $\bar{T} \rightarrow 0$. In this regime, some functions develop strong gradients around the horizon. An example is depicted in  figure \ref{fig:LowTemp_gamma17_funcQ} for the results shown in section \ref{sec:ThermoResults} ($\gamma=1.5$). As we drop the temperature, the function $\tilde{q}(z)$ becomes steeper around $z=1$. Numerically, the solution must be obtained either by a massive increase in the number of grid points or by the development of specific techniques adapted to this drawback. In this paper, we use the so called analytical mesh-refinements \cite{Meinel:2008, Macedo2014}, described in appendix \ref{sec:NumDetail} to circumvent these problems. 

\subsection{Quantum critical point}{\label{sec:DhokerKraus}} 
We now turn our attention to the charged magnetic solution constructed by D'Hoker and Kraus \cite{D'Hoker:2010rz}. In our conventions, it corresponds to the choice of
\be 
\label{eq:DhokerKraus}
q(z) = g(z) = b(z) = 0 \quad {\rm and} \quad \alpha(z)=1,
\ee 
which leads to\footnote{One must be careful with the definition of $F$ and $A$ within the action. Our $S_{grav}$, given by eq.~(\ref{eq:actionS}), coincides with Donos and Gauntlett \cite{Donos:2012wi}, which in turn differs from the one used by D'Hoker and Kraus (DK) \cite{D'Hoker:2010rz}. It is crucial to keep in mind that $F_{\rm here} = 2 F_{\rm DK}$ and $A_{\rm here} = 2 A_{\rm DK}$. The relevant quantities $\mu$, $\rho$ and $B$ pick up this factor of $2$ accordingly.    }
\ba
\dd s^{2} & = & \frac{1}{z^{2}}\left(\frac{\dd z^{2}}{u(z)}-u(z)\, \dd t^{2}+v(z)^{2} \, \left( \omega_{1}^{2} + \omega_{2}^{2}\right)+ w(z)^{2} \, \left(\omega^{3}+c(z)\dd t\right)^{2}\right) \nn \\
F  &=&  e(z)\,\dd t\wedge\dd z+B\,\omega_{1}\wedge\omega_{2}+p(z)\,\dd z \wedge \omega_{3} \\
A &=& -E(z)\, \dd t \, - B \, x_2 \, \dd x_1 \, + P(z) \, \omega_3 \,. \nn
\ea 
Note that $ \omega_{1}^{2} + \omega_{2}^{2} = \dd x_1^2 + \dd x_2^2$ and $\omega_{1}\wedge\omega_{2}  = \dd x_1 \wedge \dd x_2$. Thus, the value of $k$ is actually irrelevant and the charged magnetic black brane solution is recovered just by eq.~(\ref{eq:DhokerKraus}).

Moreover, note that D'Hoker and Kraus normalise all their dimensionful quantities by the charge density $\rho = e'(0)$. For example, the dimensionless magnetic field $ \displaystyle \hat{B}$ is given by $ \displaystyle \hat{B} = \frac{B}{\rho^{2/3}}$, whereas we consider the dimensionless magnetic field $\bar{B}$ given by $\bar{B} = \displaystyle \frac{B}{\mu^2}$. As usual, the charge density $\rho$ and the chemical potential $\mu$ are thermodynamically conjugate. In other words, one may consider the chemical potential to be a function of the charge density or vice-versa. The choice of perspective, i.e. canonical ensemble versus grand canonical ensemble, does not affect the dynamics of the field theory.

Due to different normalisations used we have to be careful when using results from \cite{D'Hoker:2010ij}. For example, we first have to translate the location $\hat{B}_{\rm C}$ of the quantum critical point quoted in \cite{D'Hoker:2010ij}. The corresponding value of $\bar{B}_{\rm C}$ in our notation is related to $\hat{B}_{\rm C}$ by $\bar{B}_{\rm C} = \gamma^2 \hat{B}^3_{\rm C}.$

\begin{figure}[b!]
\begin{center}
\includegraphics[width=10.cm]{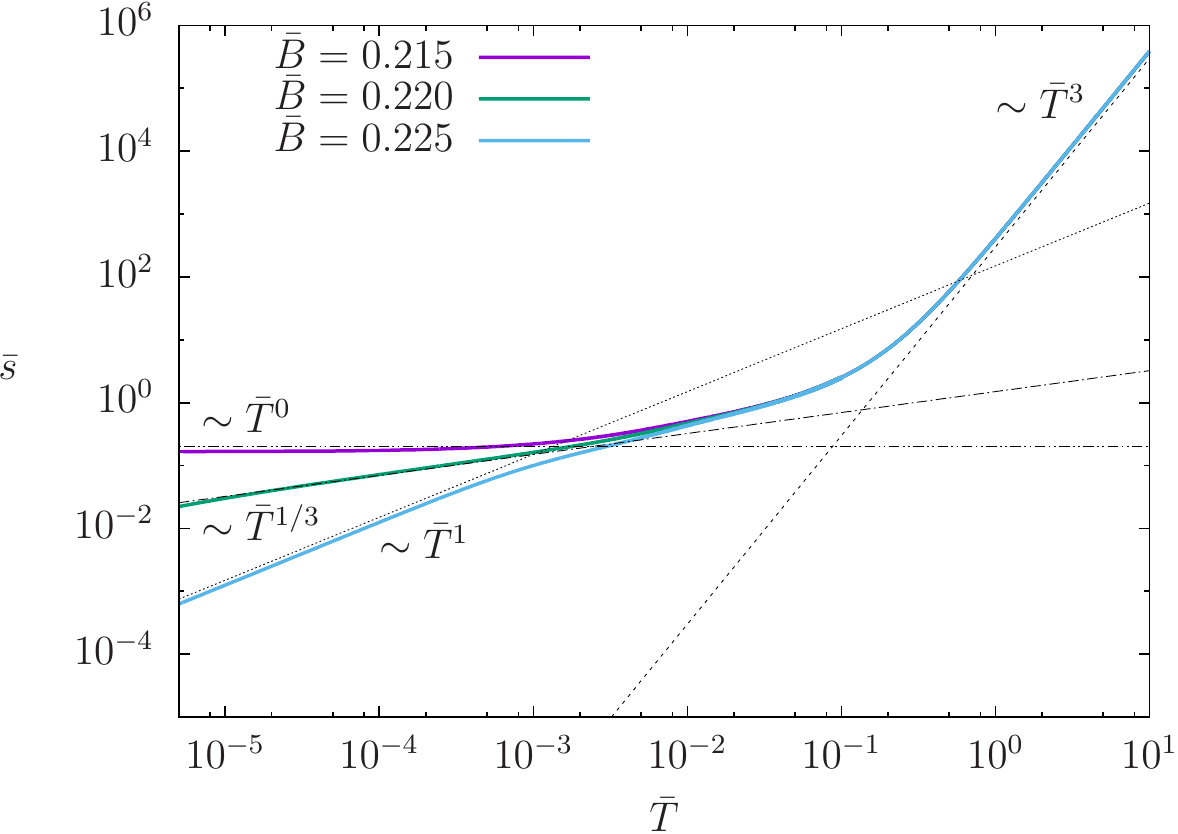}
\end{center}
\caption{Entropy $\bar{s}$ as a function of the temperature $\bar{T}$ for the charged magnetic brane. For $\gamma=1.5$ the quantum critical point is located at $\bar{B}_{C} \approx 0.220$ and the entropy decays as $\bar{s}\sim \bar{T}^{1/3}$. For $\bar{B}=0.215< \bar{B}_{C}$ the entropy goes to a constant in the low temperature regime, whereas for $\bar{B}=0.225 > \bar{B}_{C}$ one has $\bar{s}\sim \bar{T}$. For high temperatures we find the expected behaviour $\bar{s}\sim \bar{T}^{3}$. All results are in agreement with \cite{D'Hoker:2010rz}.}
\label{fig:gamma15_QtmCrtPoint}
\end{figure}

For instance, for $\gamma=1.7$ the quantum critical point is located at $\hat{B}_{C} \approx 0.400$ which corresponds to $\bar{B}_{C} \approx 0.185$ in our notation. For $\gamma = 1.5$ one obtains $\hat{B}_{C} \approx 0.461$ or $\bar{B}_{C} \approx 0.220$, respectively. The location of the quantum critical point is shown in the phase diagram figure \ref{fig:T_versus_B}. To check our numerics and the correct normalisation we determine the behaviour of the entropy density $\bar{s}$ as a function of $\bar{T}$ close to the quantum critical point $\bar{B} \approx \bar{B}_{C}$. We display in figure \ref{fig:gamma15_QtmCrtPoint} the entropy $\bar{s}$ as a function of $\bar{T}$ for $\gamma = 1.5$. Needless to say that we reproduce the results from \cite{D'Hoker:2010rz}.

\begin{figure}[b!]
\begin{center}
\includegraphics[width=7.2cm]{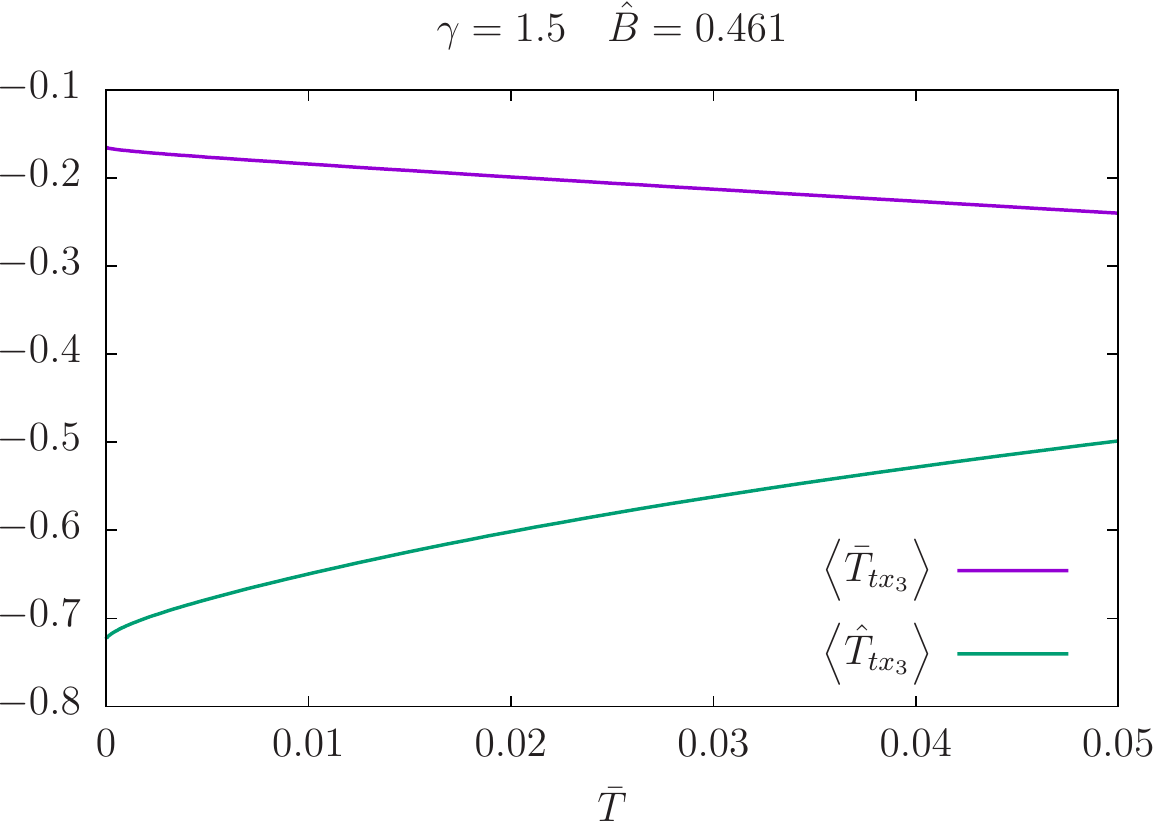}
\includegraphics[width=7.2cm]{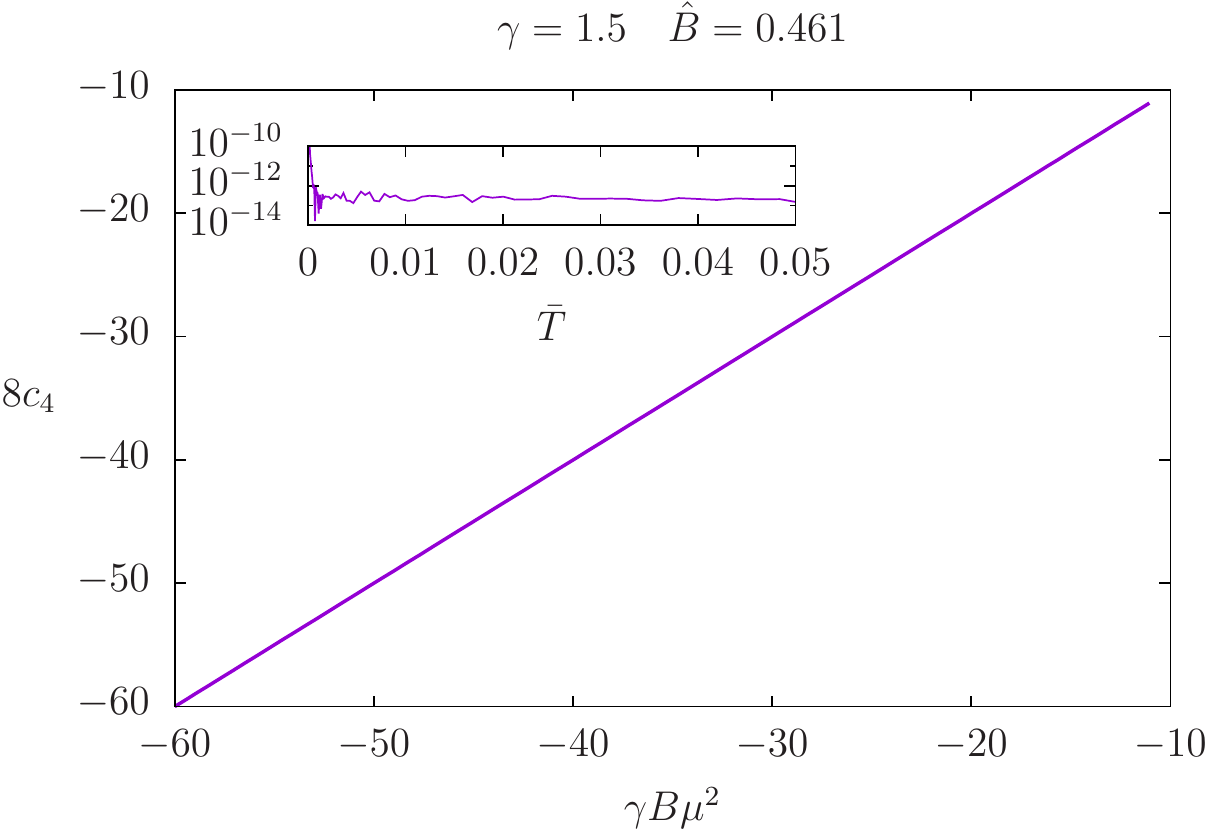}
\end{center}
\caption{
Component $\left<T_{tx_3}\right>$ of the energy momentum tensor for constant dimensionless magnetic field $\hat{B}$ normalised by $ \hat{B}= {B}/{\rho^{2/3}}$. Contrary to the results obtained for constant $\bar{B}=B/\mu^2$, neither $\left<\hat{T}_{tx_3}\right>$ nor  $\left<\bar{T}_{tx_3}\right>$ are constant (left panel). Yet, the relation $(\ref{eq:c4_gammaBmu})$ is still valid (right panel).
}
\label{fig:gamma15_Tt3_Bhat}
\end{figure}

A final comment concerning the behaviour of $\left<T_{tx_3}\right>$ given the different normalisations. As numerically checked along $\bar{B}=$ constant, $\displaystyle \left<\bar{T}_{tx_3}\right>= \frac{\left<T_{tx_3}\right>}{\mu^4} = \frac{\gamma \bar{B}}{2}$ is also a constant.  However, along $\hat{B}$, the left panel of figure \ref{fig:gamma15_Tt3_Bhat} shows that neither  $\displaystyle \left<\bar{T}_{tx_3}\right>$ nor $\displaystyle \left<\hat{T}_{tx_3}\right>= \frac{\left<T_{tx_3}\right>}{\rho^{4/3}}$ are constant. Yet, according to (\ref{eq:EnergyMomTensor}) and (\ref{eq:Ttx3}) the relation 
\be
\label{eq:c4_gammaBmu}
 \left<T_{tx_3}\right> = 4\,c_4 = \frac{\gamma B}{2}\,\mu^2
\ee
should still hold. We explicitly checked the validity of this expression in the right panel of figure \ref{fig:gamma15_Tt3_Bhat} for $\gamma = 1.5$ along the critical value $\hat{B} = 0.461$. In particular, the inset displays the difference $\displaystyle \left| 1 - \frac{8\,c_4}{\gamma \, B\, \mu^2 } \right|$, which is limited only by the numerical round-off error.

\section{Numerical details}\label{sec:NumDetail}
As mentioned before, the system of equations (\ref{eq:EOM_vec}) expressed in terms of the auxiliary variables (\ref{eq:AuxVar}) is solved numerically by means of a spectral method. In this section, we elaborate further on this topic and we give more details on the numerics involved. 

Spectral methods are best applied to differential equations whose solutions are known to be analytic. In such a case, the error coming from the numerics decays exponentially as one increases the grid resolution. On the other hand, the presence of logarithmic terms spoils this properties, rendering a merely algebraic convergence rate. Yet, the introduction of the auxiliary variables (\ref{eq:AuxVar}) removed the leading $z^4\ln(z)$ terms and we verify that our solutions show typically a rather efficient convergence rate, even for large values of $\bar{B}$.  

In addition to the high accuracy, spectral methods are also flexible enough to deal with other unknown parameter apart from the field functions. In order to fix these parameters, one needs to specify extra conditions together with the equations of motion. As a global scheme, the method makes no distinction between the unknown functions and parameters and solves the system of all variables at once. In the context of gauge/gravity dualities, it is possible to address the low temperature behaviour within spectral methods. In addition, spectral methods also allow to include singular points and hence the equations of motion can be solved in the whole domain. More details on spectral methods can be found in \cite{Boyd00,canuto_2006_smf} as well as in the specialised reviews \cite{Chesler:2013lia,Dias:2015nua}.

\subsection{Spectral Methods}
Let $n_{\rm fields}$ be the total number of unknown functions $X^I$  (with $I=1 ,\dots , n_{\rm fields}$) defined on the domain $z \in [0,1]$. Let us also assume the existence of $n_{\rm par}$ unknown real parameters $\chi^A$ (with $A=1, \dots, n_{\rm par}$). Moreover, we consider the following system of equations
\be
\label{eq:SysEq}
\left\{
\begin{array}{ll}
\vspace{0.1cm}
F_0^I(X'^I, X^I;\chi^A ) = 0  \qquad\quad\qquad &{\rm for} \quad z = 0 \, ,\\
\vspace{0.1cm}
F^I(X''^I, X'^I, X^I,z;\chi^A ) = 0 & {\rm for}\quad 0< z < 1 \, ,\\
F_1^I(X'^I, X^I;\chi^A ) = 0 & {\rm for}\quad z = 1 \, , \\
\Phi^A(X''^I, X'^I, X^I;\chi^A ) = 0 \, . &
\end{array}
\right.
\ee
Here, $X'^I$ and $X''^I$ are, respectively, the first and second derivative of the functions $X^I(z)$. $F_0^I$ and $F_1^I$ are the boundary conditions, whereas $F^I$ represent the equations of motions.  Finally, $\Phi^A$ stand for the extra conditions that fix the unknown parameters.

In order to solve numerically the system of equations (\ref{eq:SysEq}), we first provide a numerical resolution $N$ and  expand the functions $X^I(z)$ as
\be
\label{eq:SpecDecomp}
X^I (z) = \sum_{k=0}^{N} c^I{}_k \,T_k(2z-1) + R^I(z).
\ee
In the expansion above, the basis functions are the Chebyshev polynomials of first kind $T_k(\xi) = \cos[k\arccos(\xi)]$, $\xi \in[-1,1]$, while $R^I(z)$ are the residual functions. To determine the  Chebyshev coefficients $c^I{}_k$, we specify a set of grid points $z_i$ (with $i= 0, \dots, N$) and impose that the residual function vanishes exactly at the grid points, i.e., $R(z_i) = 0$. In other words, at the grid points, the unknown functions are given exactly by the spectral representation
\be
\label{eq:SpecDecomp_GridPoints}
X^I (z_i) = \sum_{k=0}^{N} c^I{}_k \,T_k(2z_i-1).
\ee
The Chebyshev coefficients are then obtained after the inversion of \eqref{eq:SpecDecomp_GridPoints}. From the $c^I{}_k$, we can obtain the coefficients $c'{}^I{}_k$ and $c''{}^I{}_k$ describing the spectral representation of the derivatives $X'{}^I(z)$ and $X''{}^I(z)$ as in eq.~\ref{eq:SpecDecomp}. In this paper, we work with the Chebyshev-Lobatto grid points
\be
\label{eq:LobattoGrid}
z_i =\frac{1}{2}\left[1+\cos\left(\pi\frac{i}{N}\right)\right], \quad\qquad i = 0, \dots, N.
\ee
Now we can combine the function values $X^I{}_i := X^I{}(z_i)$ and the parameters $\chi^A$ into the single vector $\vec{X}$ of length $n_{\rm total} = (N+1)n_{\rm Fields} + n_{\rm par}$
\be
\vec{X}^{\rm T} = \left( X^1_0, \dots, X^1_N, \dots X^{n_{\rm Fields}}_0, \dots X^{n_{\rm Fields}}_N | \chi^1, \dots, \chi^{n_{\rm par}}  \right)
\ee
from which we can also form the vectors $\vec{X}'$ and $\vec{X}''$ representing the discrete spectral derivatives with respect to z. These vectors are finally used to evaluate the system of equations \eqref{eq:SysEq} at the grid points (\ref{eq:LobattoGrid}), giving us a non-linear algebraic system
\be
\label{eq:SysAlgEq}
\vec{F}(\vec{X}) = 0
\ee
to be solved for the components of the vector $\vec{X}$.
The solution of the algebraic system (\ref{eq:SysAlgEq}) is obtained by a Newton-Raphson method, i.e. given an initial guess $\vec{X}_0$, the solution is iteratively approximated by
\be
\vec{X}_{n+1} = \vec{X}_{n} + \delta \vec{X}_n, \quad {\rm with} \quad \delta \vec{X}_n = - \left[\hat{J}(\vec{X}_n)\right]^{-1}\vec{F}(\vec{X}_n).
\ee
The inversion of the Jacobian matrix $\hat{J}(\vec{X})=\partial \vec{F}/\partial{\vec X}$ is performed with a LU decomposition. One can show that the Newton-Rapshon scheme always converges, providing the initial guess $\vec{X}_0$ is sufficiently close to a solution.

\subsection{Numerical solution: the initial guess}

In this work, we are looking for the numerical solution of the $n_{\rm Fields}=10$ metric and gauge field functions. If $\bar{B}=0$, the electrically charged Reissner-Nordstroem black brane (\ref{eq:RN_Sol}) is always a solution, regardless of $\bar{k}$ and $\gamma$. Similarly, for $\bar{B} \neq 0$ one always obtains charged magnetic black brane as trivial solution. The interesting cases are those values of $\gamma$ and $\bar{k}$ for which a non-trivial solution with $b(z)\neq 0$ and $q(z)\neq0$ also exists. 

Unfortunately, our first experiences showed that, for given boundary value $\tilde{E}(0) = -\mu$, the Newton-Raphson method always converged to a trivial solution of the non-linear algebraic system (\ref{eq:SysAlgEq}), regardless of the initial guess $\vec{X}_0$. In order to obtain the new solution describing the condensed phase, an extremely careful fine-tuning to the initial guess $\vec{X}_0$ seems to be needed. 

To solve this issue, we consider the parameter $\mu$ as a further unknown variable (thus $n_{\rm par}=1$) and impose the extra condition $\tilde{b}(0) = b_2 = \varepsilon$. By fixing a value $\varepsilon \neq 0$, we enforce that the Newton-Raphson scheme will necessarily converge to the non-trivial solution. 

For a given $\bar{k}$, $\gamma$, we start with $\bar{B}\sim 0$, and $\varepsilon \sim 0$. Then, the new solution should be just a small perturbation of the AdS Reissner-Nordstroem spacetime. Therefore, our initial-guess constitute of eqs.~(\ref{eq:RN_u})--(\ref{eq:RN_e}), with the slight modification $\tilde{b}(z) = \varepsilon$. Besides, we must also provide an initial-guess for the variable $\mu$. In all our experiments, the value $\mu = 2$ was sufficient for the convergence of the Newton-Raphson scheme. Once a solution is available, we can use it as initial-guess for a modified set parameters $\left\{ \gamma, \bar{B}, \bar{k},\varepsilon \right\}$. 

From the numerical point of view, fixing $\left\{ \gamma, \bar{B}, \bar{k},\varepsilon \right\}$ is an efficient method to find the non-trivial solution. However,
from the physical perspective, a system with a constant temperature $\bar{T}$, i.e. specified by $\left\{ \gamma, \bar{B}, \bar{k},\bar{T} \right\}$ is what one really wants to describe. Note that, as an alternative to the extra condition $\tilde{b}(0) = b_2 = \varepsilon$, we can indeed impose $\bar{T} = $constant. This corresponds to looking for the value of $\mu$ leading to a solution with a fixed value $\tilde{u}(1)$. Unfortunately, this approach does not guarantee that the method will give us the non-trivial solution. Depending on how far the initial-guess is from the trivial solution, the Newton-Rapshon scheme might converge to the charged magnetic black brane solution (with $\tilde{b}(z) = 0$). Therefore, for a fixed $\{ \gamma, \bar{B}\}$, our algorithm is a combination of both possibilities and can be divided into three stages:
\begin{enumerate}[I)]
\item Phase boundary: we set $\varepsilon = 10^{-9}$ and scan the values of $\bar{k} \in [\bar{k}_0, \bar{k}_1]$ to get the phase boundary\footnote{For some values of $\{ \gamma, \bar{B}\}$ one might find returning points, i.e., there might exist values of $\bar{k}$ with two different solutions $\bar{T}({\bar{k})}$. In such cases, we employed the methods described in \cite{Dias:2015nua} to scan the whole parameter range.}. With the knowledge of $\bar{T}(\bar{k})$ we find the point $\{\bar{k}_{\rm C}, \bar{T}_{\rm C} \}$ for which $\bar{T}_{\rm C} = \bar{T}(\bar{k}_{\rm C})$ is at a maximum.

\item Condensed phase: we then keep $\bar{k} = \bar{k}_{\rm C}$  fixed and find new solutions inside the phase by slowly increasing the values of $\eps$ until a given $0 < \bar{T}_0 < \bar{T}_{\rm C}$ is achieved. Typically, we set $\bar{T}_0 = 0.95 \, \bar{T}_{\rm C}$.

\item Constant temperature:~with the solution inside the phase provided by step II as initial guess, we no longer need to keep $\varepsilon$ fixed. We now solve at surfaces of constant $\bar{T} \in [\bar{T}_{\rm min}, \bar{T}_0] \cup [\bar{T}_{0}, \bar{T}_{\rm C}]$ in a given interval $\bar{k} \in [\bar{k}_0({\bar{T}}), \bar{k}_1({\bar{T}})]$ and find the physical state $\bar{k}_{*}(\bar{T})$ for which the grand canonical potential $\bar{\Omega}_{*} = \bar{\Omega}(\bar{k}_*)$ is at a minimum.  
\end{enumerate}

\begin{figure}[hp]
\begin{center} 
\includegraphics[width=6.cm]{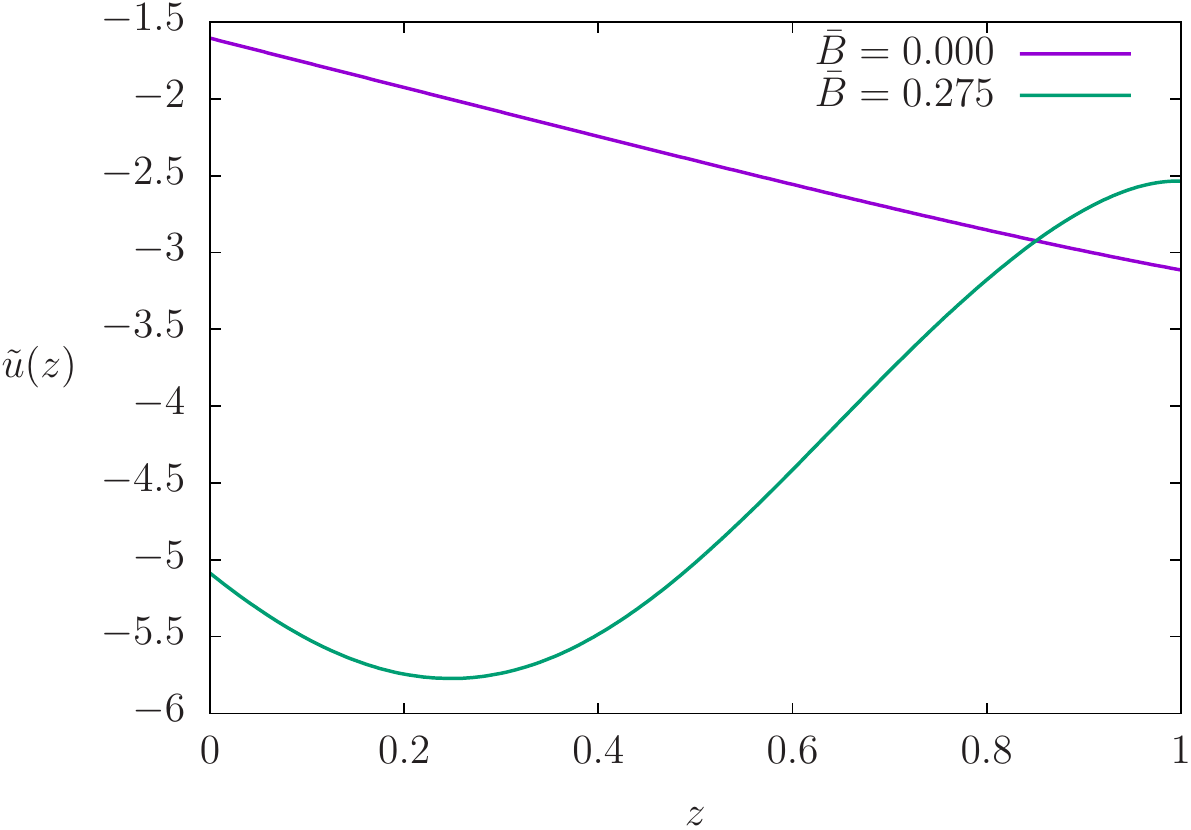}
\includegraphics[width=6.cm]{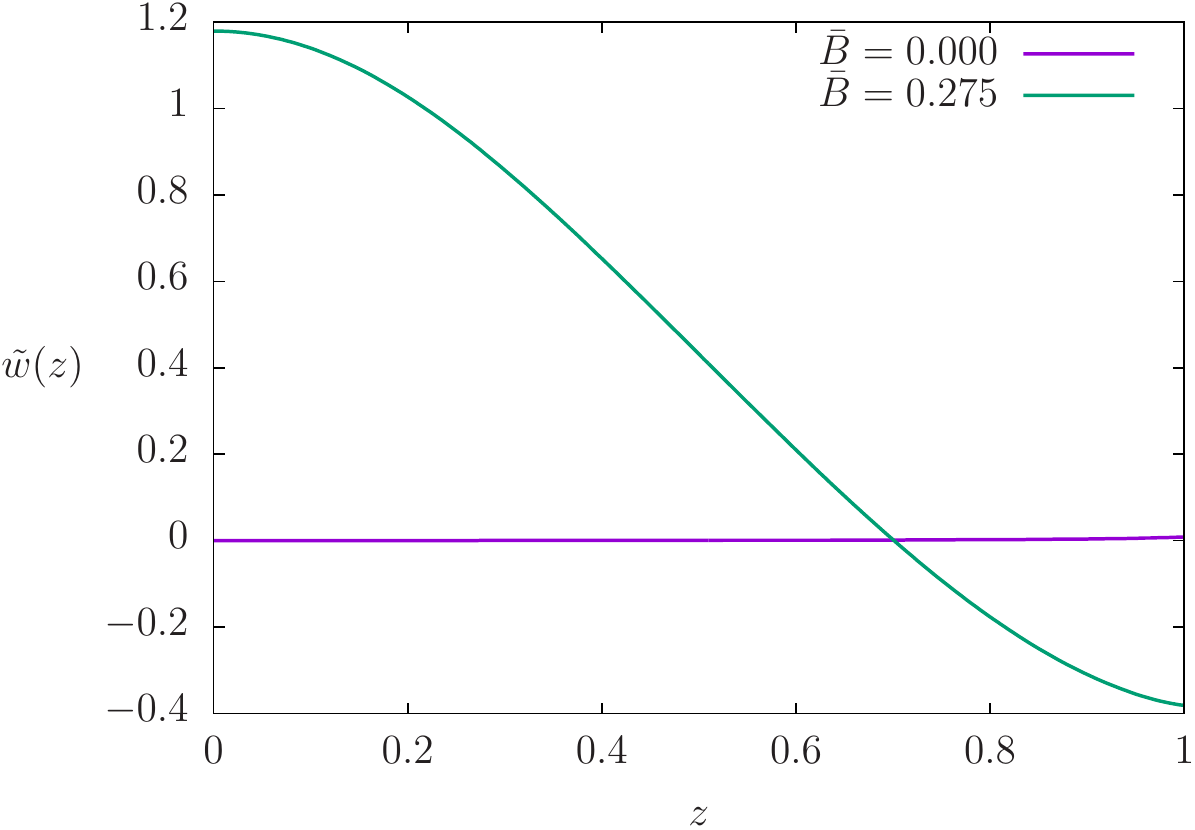}
\includegraphics[width=6.cm]{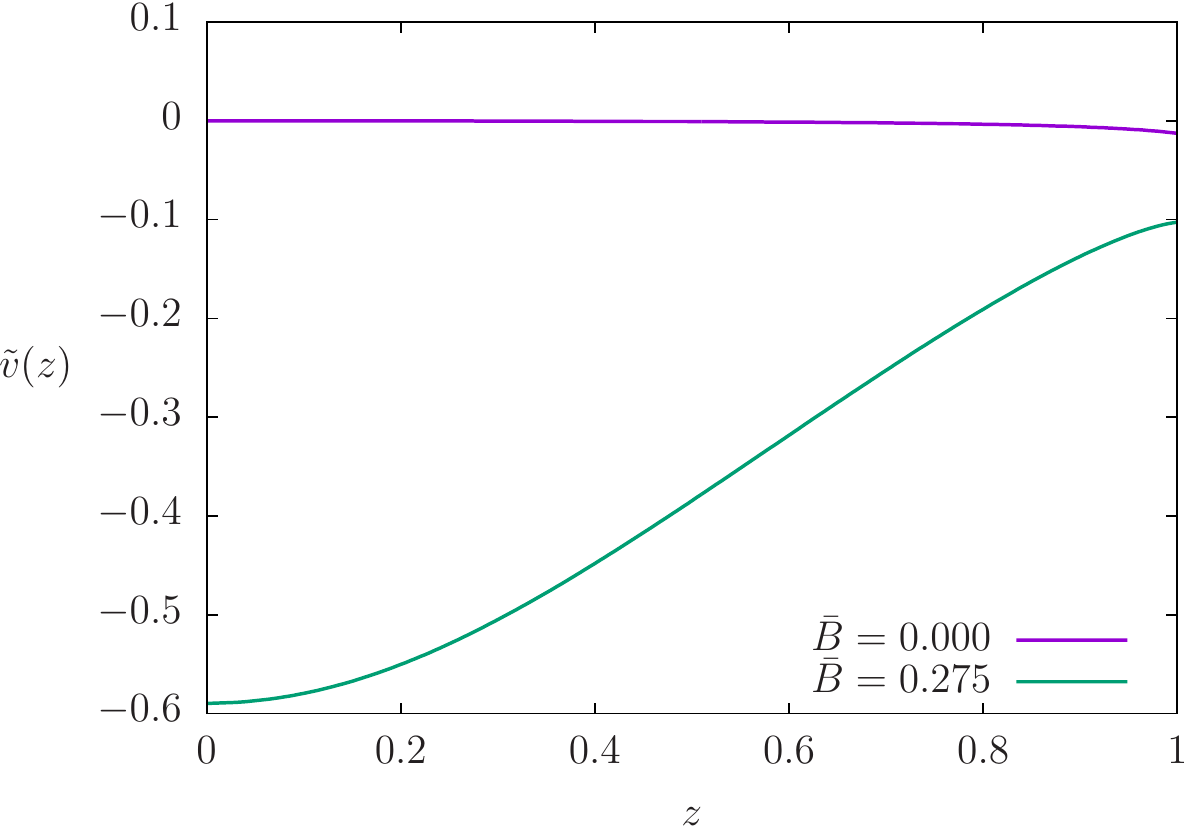}
\includegraphics[width=6.cm]{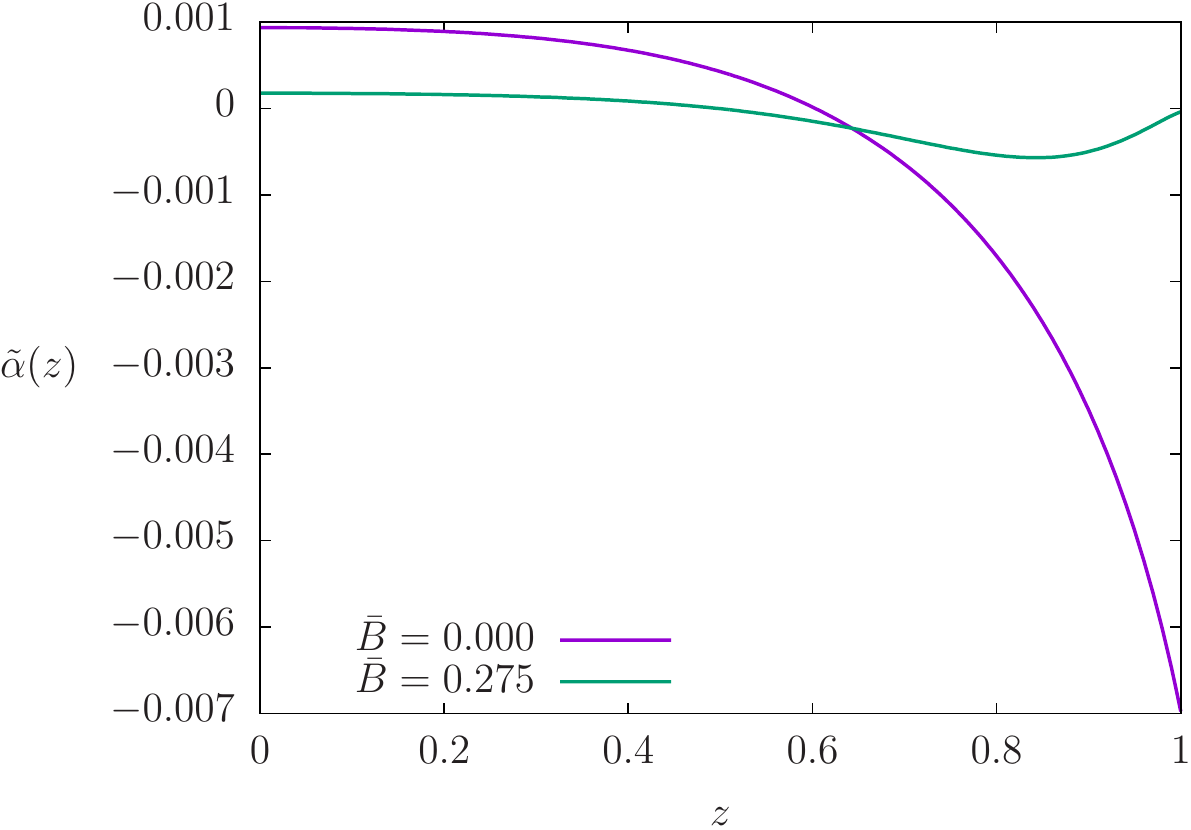}
\includegraphics[width=6.cm]{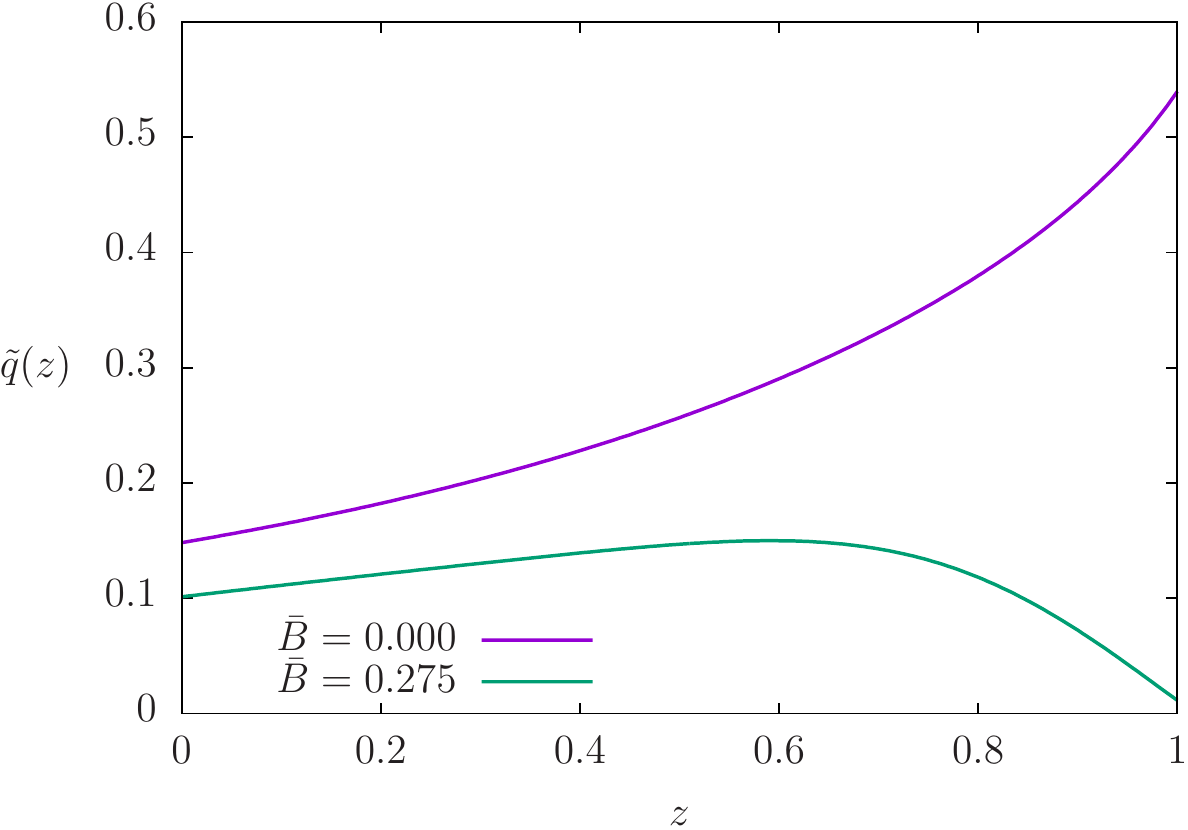}
\includegraphics[width=6.cm]{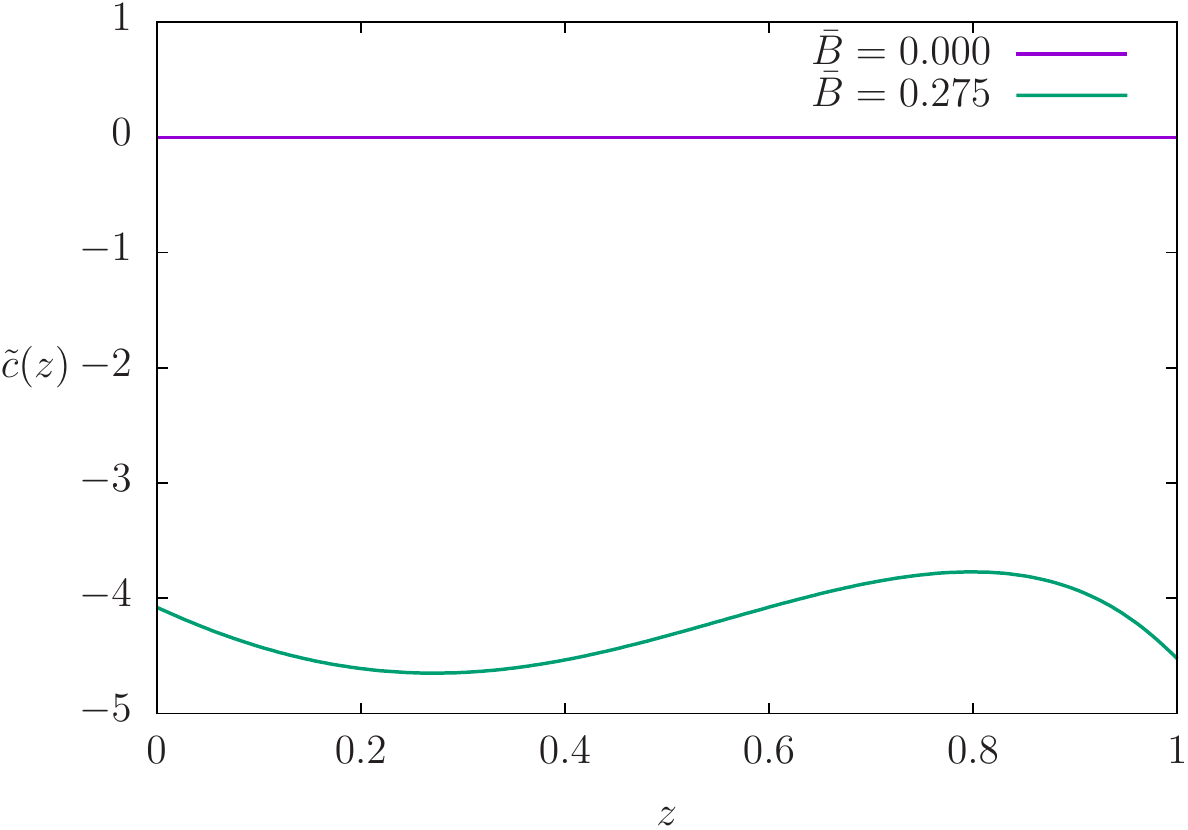}
\includegraphics[width=6.cm]{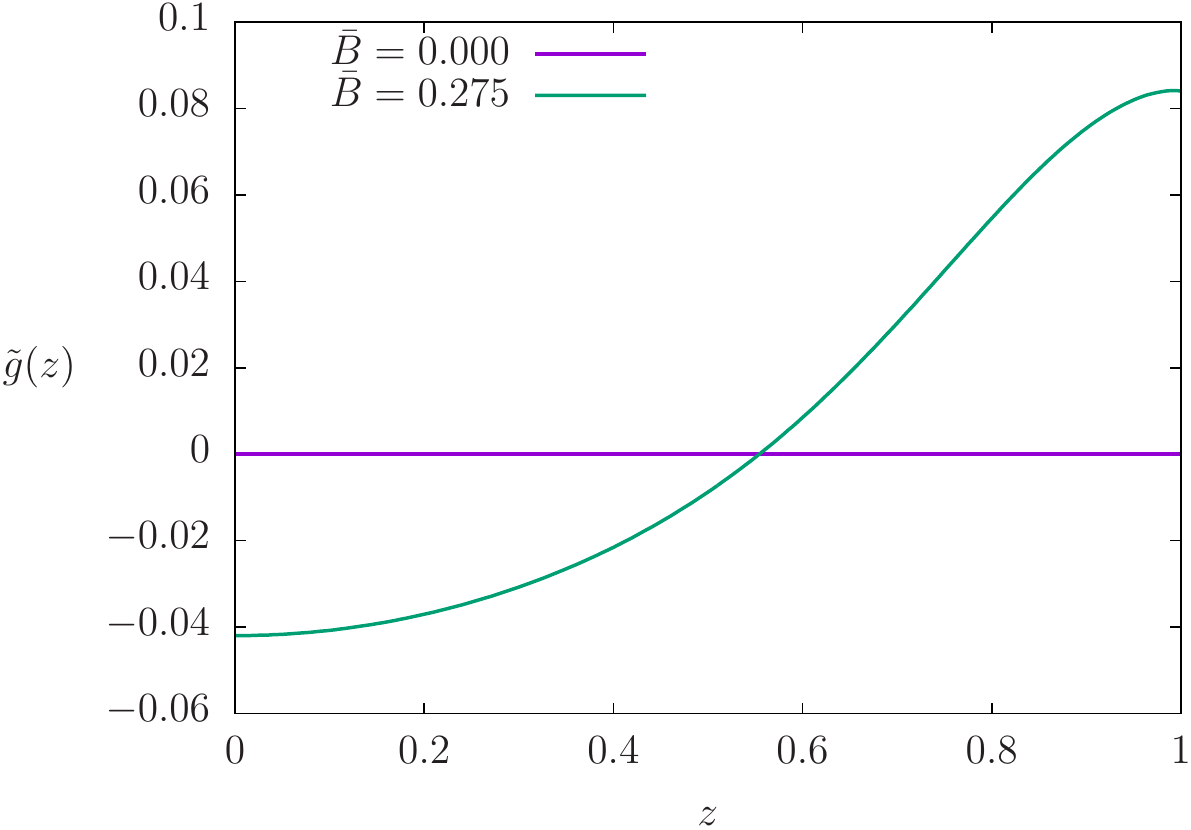}
\includegraphics[width=6.cm]{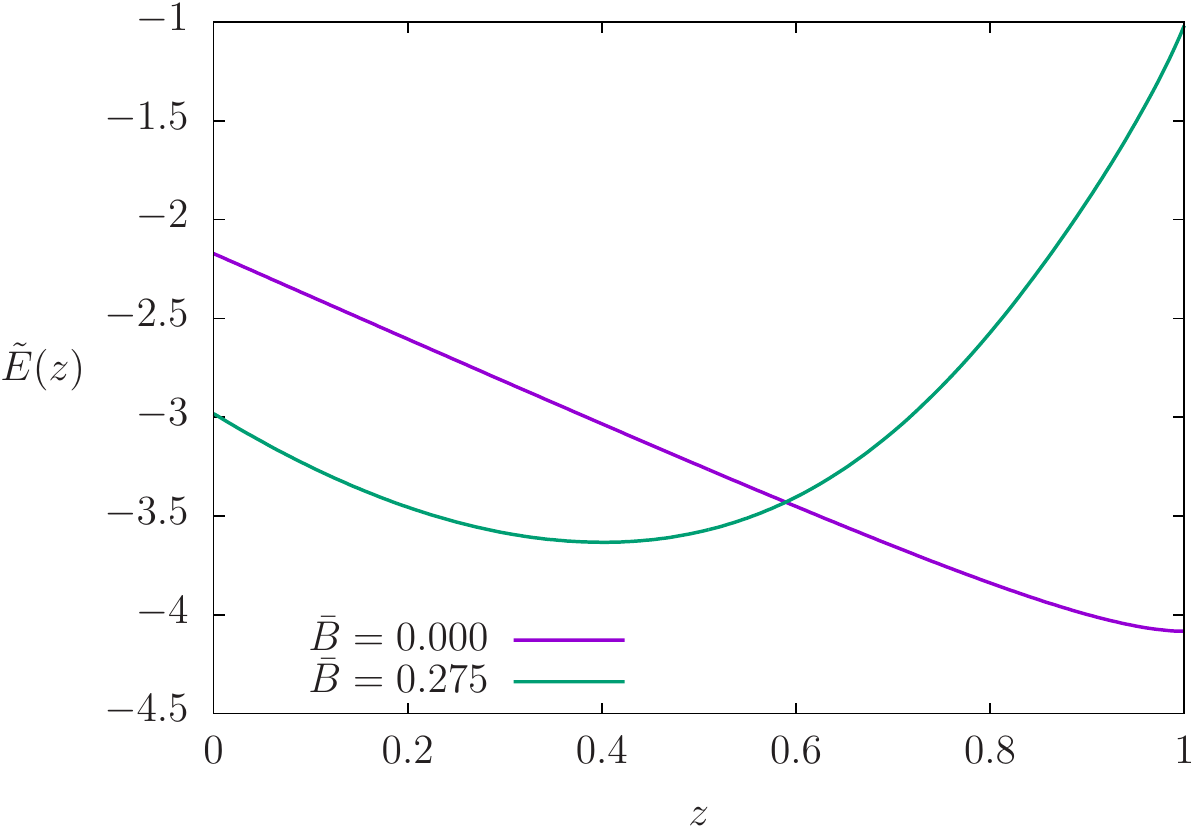}
\includegraphics[width=6.cm]{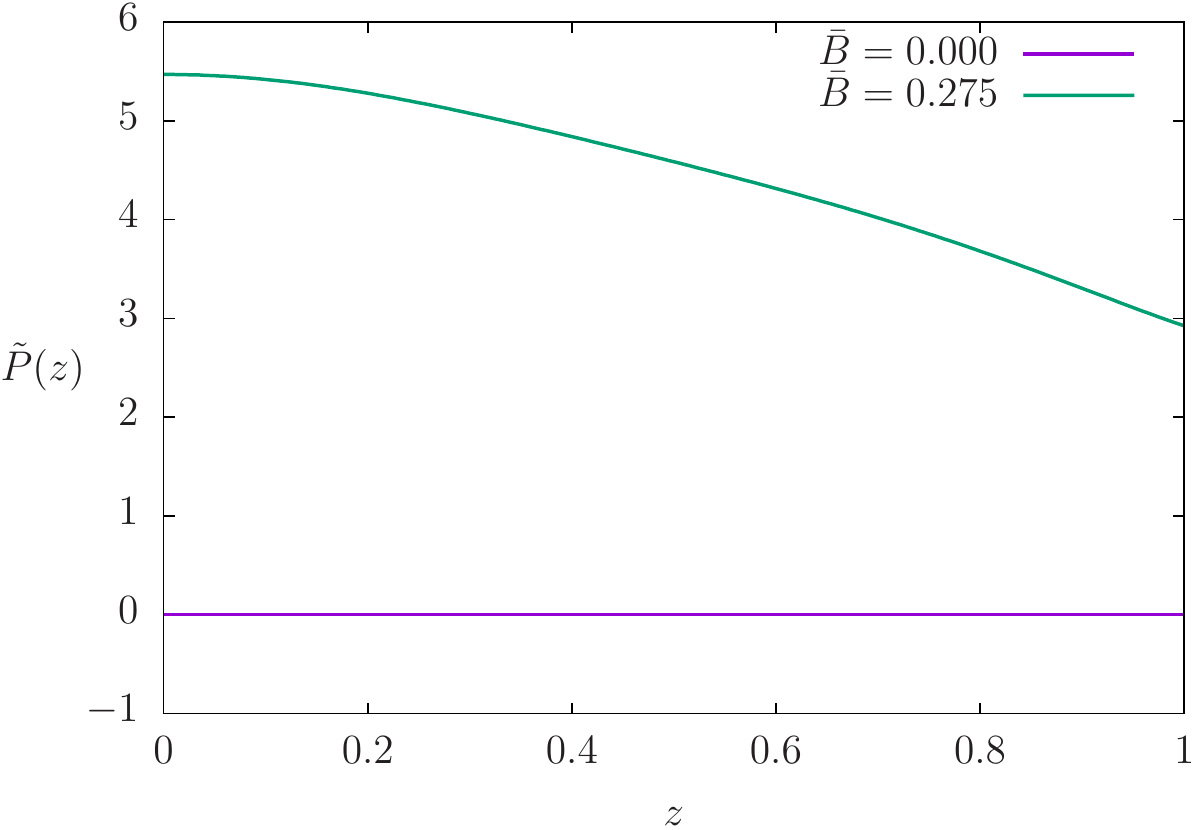}
\includegraphics[width=6.cm]{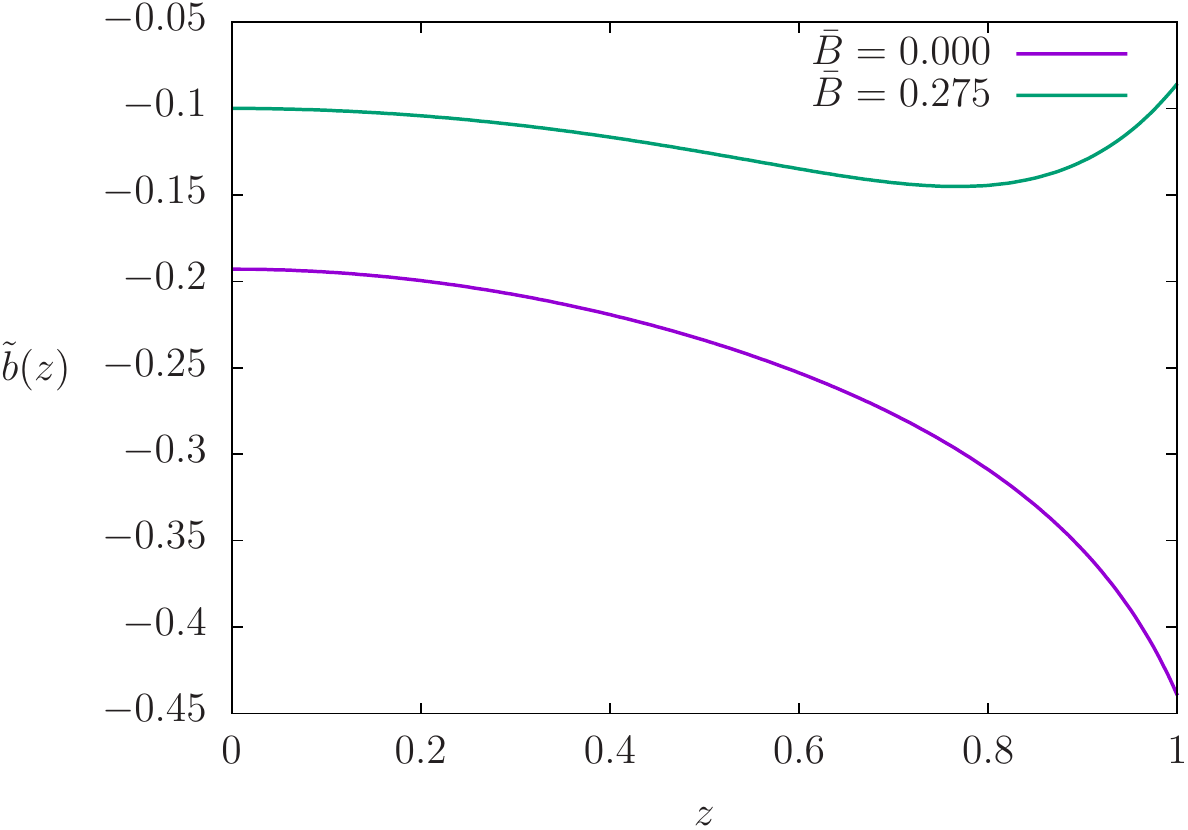}
\end{center}
\caption{Solutions for the auxiliary metric functions $\tilde{u}(z)$, $\tilde{w}(z)$, $\tilde{v}(z)$, $\tilde{\alpha}(z)$, $\tilde{q}(z)$, $\tilde{c}(z)$ and $\tilde{q}(z)$  and gauge field functions $\tilde{E}(z)$, $\tilde{P}(z)$ and $\tilde{b}(z)$ for the following two configurations: $\{\bar{B} = 0, \bar{k}=1.21, \bar{T}=3.25\times10^{-2}\}$ and $\{ \bar{B} = 0.275, \bar{k}=0.98, \bar{T}=1.25\times10^{-2}\}.$
}
\label{fig:gamma15_Sol}
\end{figure}

To illustrate the solutions, we show in figures \ref{fig:gamma15_Sol} the results for the metric and gauge field functions with $\gamma = 1.5$ for the following two configurations:
\[
 \{\bar{B} = 0, \bar{k}=1.21, \bar{T}=3.25\times10^{-2}\} \qquad {\rm and} \qquad \{ \bar{B} = 0.275, \bar{k}=0.98, \bar{T}=1.25\times10^{-2}\} \, ,
\] 
giving respectively $\mu = 2.172$ and $\mu = 2.982$ for the chemical potential.

\FloatBarrier
\FloatBarrier

\newpage
\subsection{Numerical error}

\begin{figure}[b!]
\begin{center}
\includegraphics[width=7.2cm]{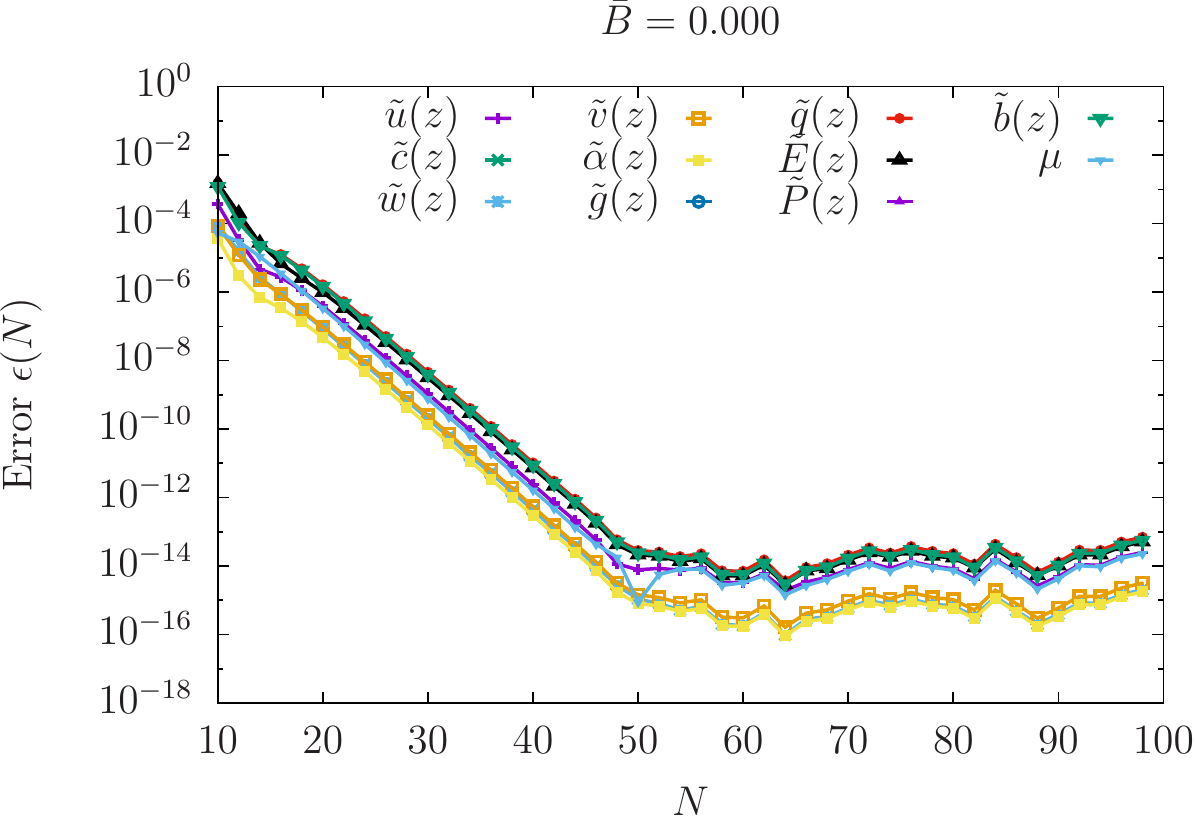}
\includegraphics[width=7.2cm]{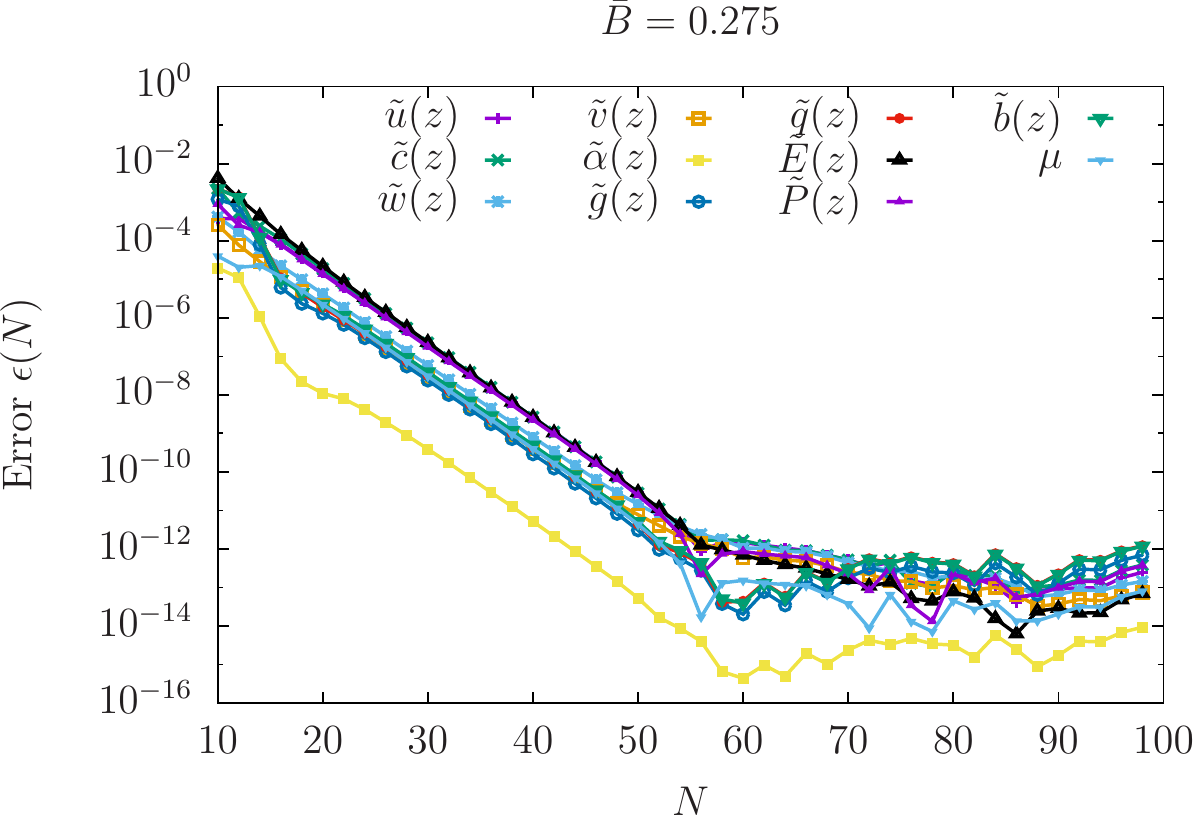}
\end{center}
\caption{Numerical errors for the two configurations $\{\bar{B} = 0, \bar{k}=1.21, \bar{T}=3.25\times10^{-2}\}$ (left panel) and $\{ \bar{B} = 0.275, \bar{k}=0.98, \bar{T}=1.25\times10^{-2}\}$ (right panel). For $\bar{B}=0$ we obtain the expected exponential convergence rate. For $\bar{B}\neq0$ one expects a merely algebraic decay due to the logarithmic terms. Its contribution, however, enters only within the machine precision.}
\label{fig:gamma15_Error}
\end{figure}

Finally, we discuss the accuracy of our method. Given a a high resolution $N_{\rm max}$, we consider a reference solution $\left\{X^I(z;N_{\rm max}),\rho(N_{\rm max})\right\}$ and define, for a lower resolution $N<N_{\rm max}$, the numerical error of the solution $\left\{X^I(z;N), \mu(N)\right\}$ by
\be
\epsilon^I(N) = \max_{z\in[0,1]}\left|X^I(z;N) - X^I(z;N_{\rm max})\right|, \qquad\quad \epsilon^\mu(N) = \left|\mu(N) - \mu(N_{\rm max})\right|.
\ee
Fig.~\ref{fig:gamma15_Error} displays the error for the configurations mentioned in the previous section. The left panel shows the case $\bar{B}=0$ and we see the typical exponential convergence provided by the spectral method. The right panel depicts the case $\bar{B}=0.275$. Despite the presence of logarithmic terms, the convergence rate is very efficient and we do not observe a significant influence of an algebraic decay within the machine limits imposed by round off errors. 

\begin{figure}[t!]
\begin{center}
\includegraphics[width=7.2cm]{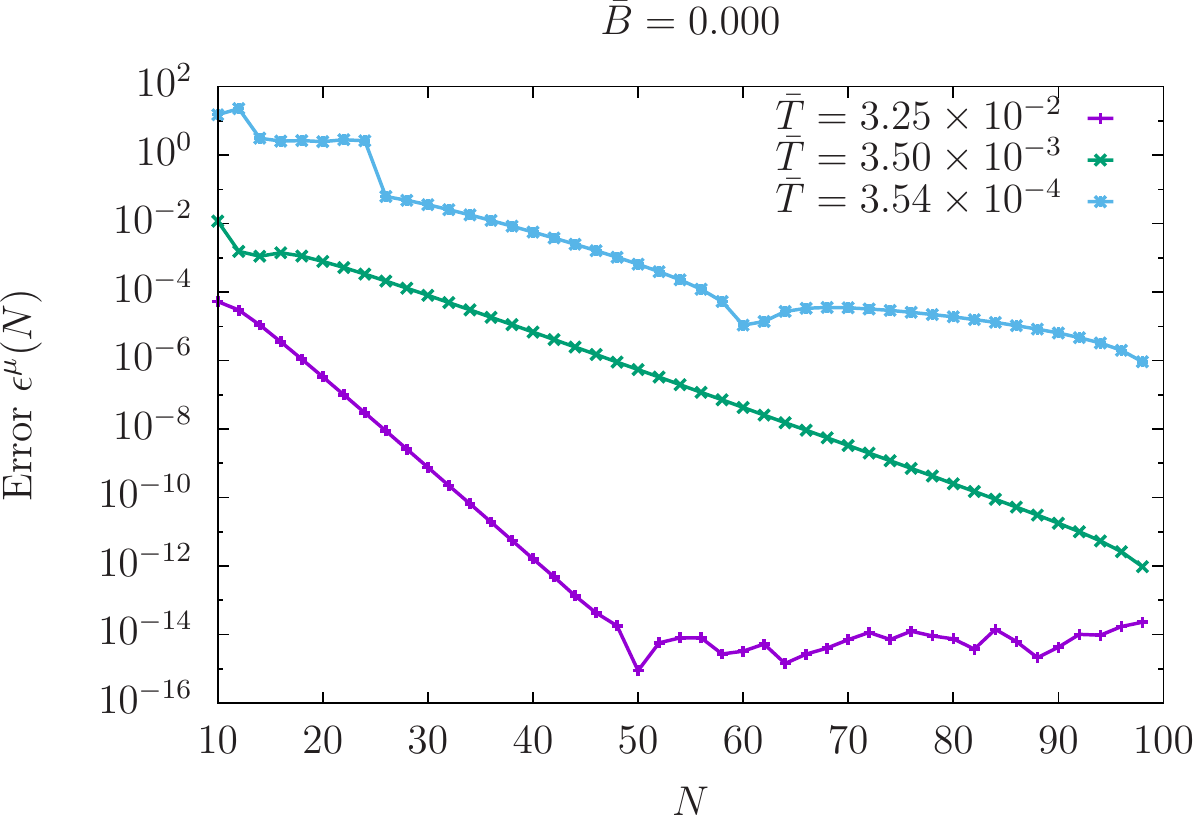}
\includegraphics[width=7.2cm]{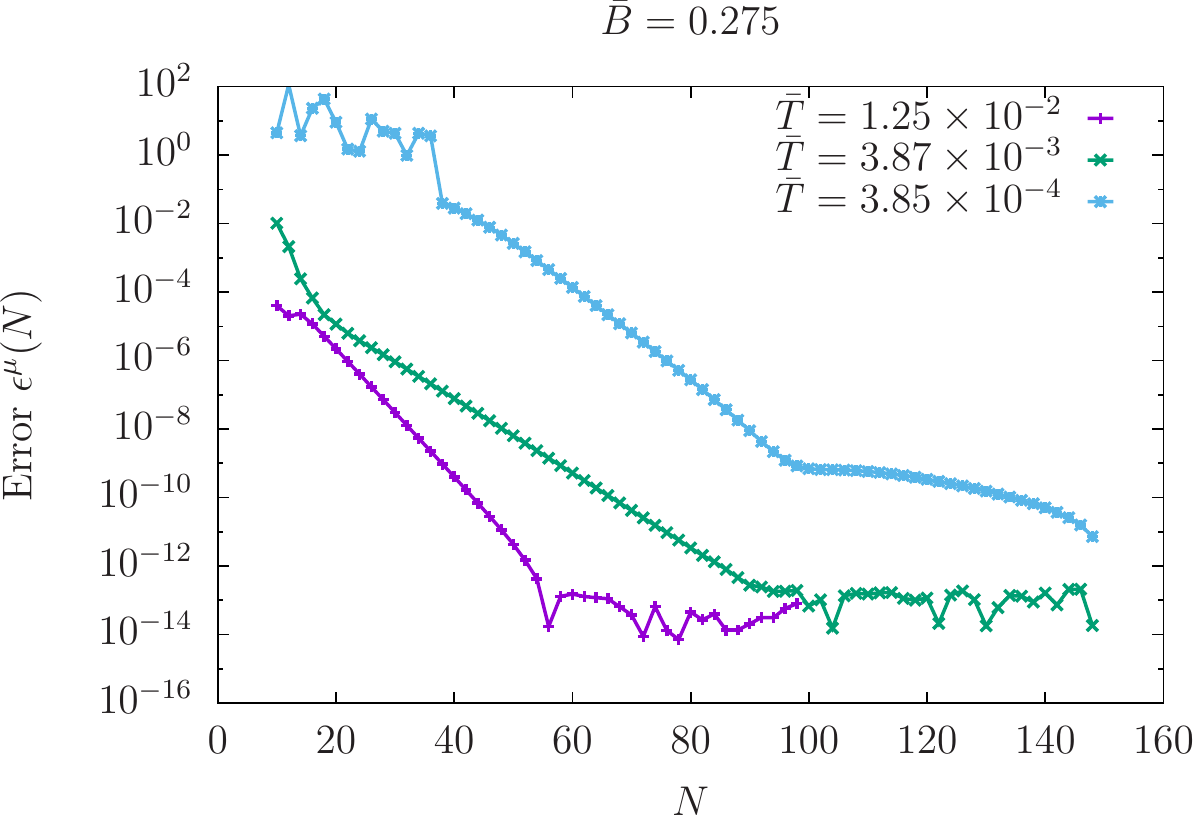}
\end{center}
\caption{Numerical error for low temperatures. Despite the exponential convergence rate, the method becomes less efficient as the temperature decreases due to the presence of strong gradients near $z=1$ (see figure \ref{fig:LowTemp_gamma17_funcQ}). A reliable highly accurate solution requires a massive increase in the numerical resolution.}
\label{fig:gamma15_Error_TempDep}
\end{figure}

\begin{figure}[t!]
\begin{center}
\includegraphics[width=7.2cm]{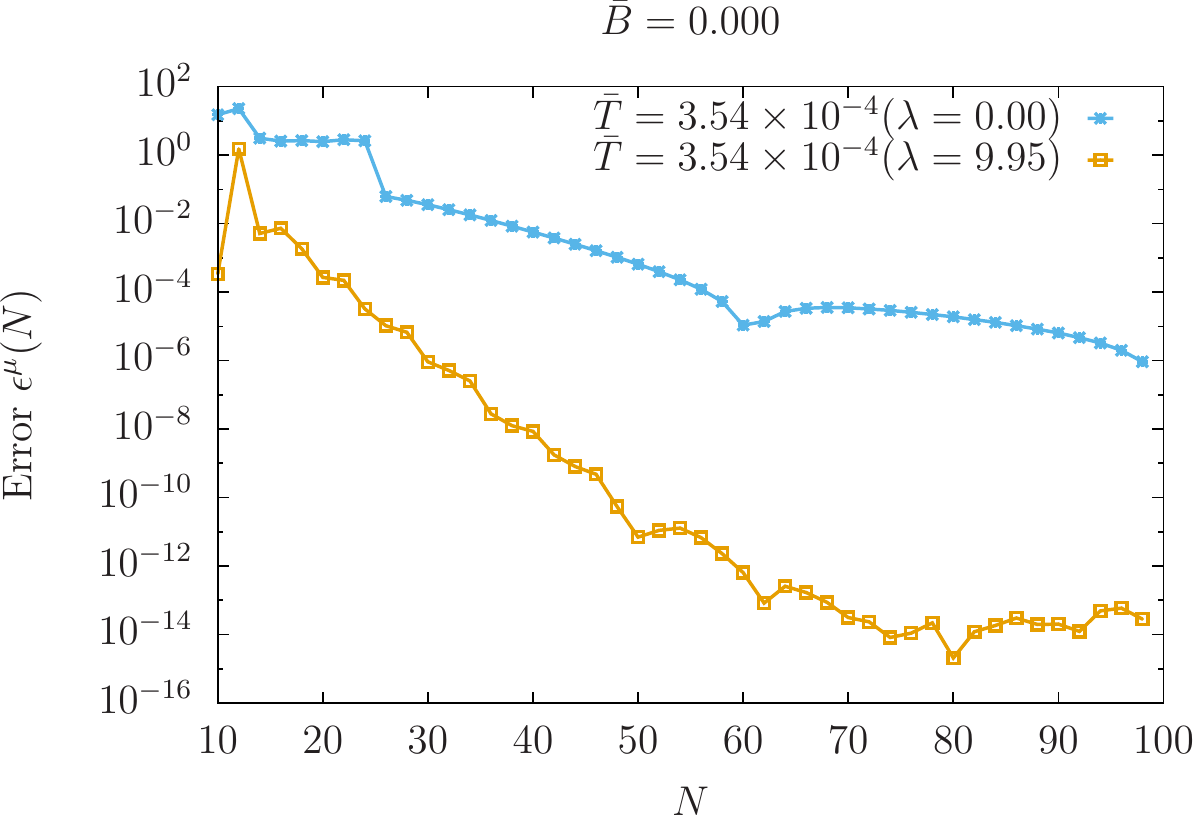}
\includegraphics[width=7.2cm]{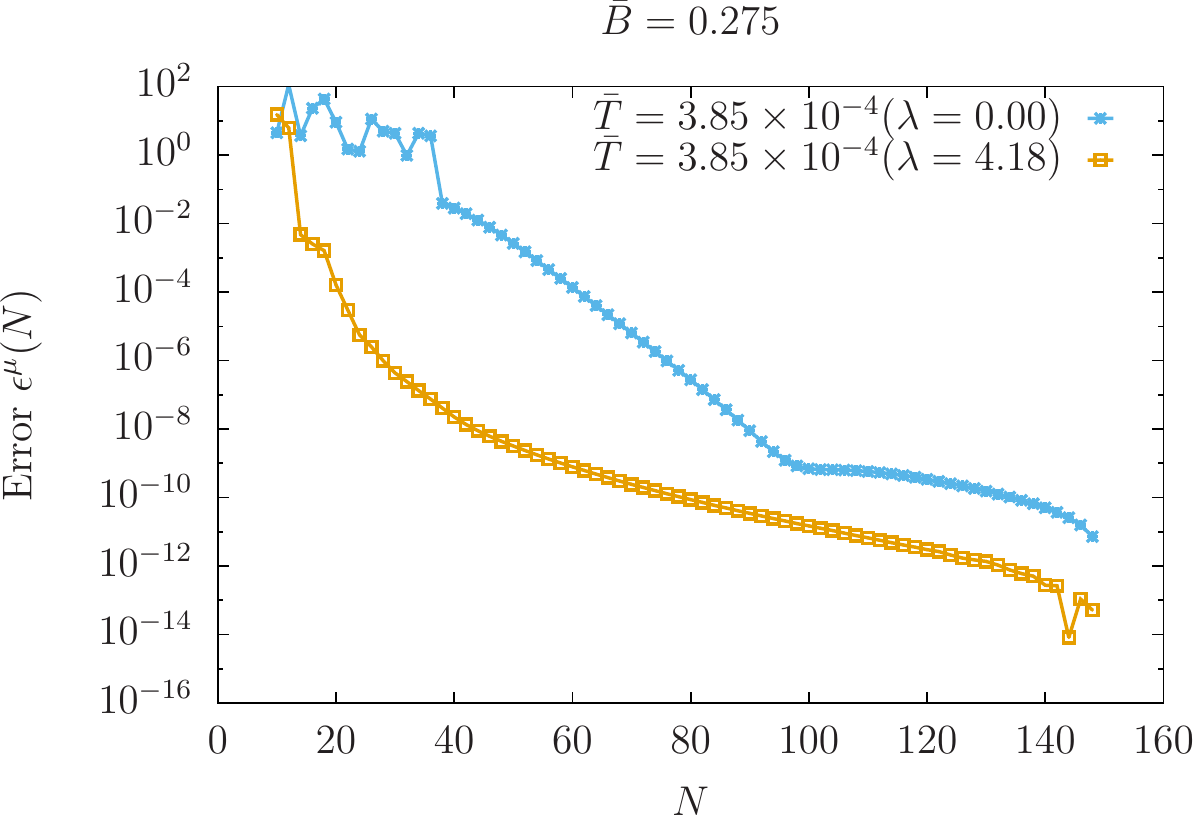}
\end{center}
\caption{Numerical error for low temperatures with the analytical mesh-refinement. Left panel: for $\bar{B}=0$ the convergence rate is significantly improved and we achieve the machine precision with a moderate number of grid points. Right panel: for $\bar{B} \neq 0$ we still obtain a better convergence rate, though the method worsens the algebraic decay due to the logarithmic terms.}
\label{fig:gamma15_Error_TempDep_MeshRef}
\end{figure}

Even though the method provides a high accuracy solution for moderate values of $\bar{T}$, we note that the small temperature regime requires a massive increase in resolution. This feature becomes evident in figure \ref{fig:gamma15_Error_TempDep}, where we compare the convergence rate (for instance for $\mu$) at different temperatures. As already mentioned in appendix \ref{sec:DonosGauntlett} and illustrated in figure \ref{fig:LowTemp_gamma17_funcQ}, the main reason is the presence of strong gradients around the horizon $z=1$.
One technique to deal with such strong gradients is the so called analytical mesh-refinement, described in \cite{Meinel:2008, Macedo2014}. It consists of mapping the coordinate $z\in[0,1]$ into $\zeta \in [0,1]$ via
\be
z =  1- \frac{\sinh[ \lambda (1-\zeta)]}{\sinh{\lambda}}.
\ee
By choosing an adequate parameter $\lambda$, the mapping increases the number of grid points around $z=1$ and smoothen out the solution. In our case we set $\lambda = \left | b_2  \ln(\bar{T})  \right|$. In figure \ref{fig:gamma15_Error_TempDep_MeshRef} one sees the significant improvement of the convergence rate, specially in the case $\bar{B}=0$. For $\bar{B} \neq 0$, the method is still effective at low temperatures, but it also intensifies the algebraic decay rate introduced by logarithmic terms. Even though the analytical mesh-refinement provides the necessary tools to study the low temperature limit within the scope of this work, we note that there are still possibilities for enhancing the accuracy of the $\bar{B}\neq 0$ case.\footnote{An option is to use a multi-domain code with another coordinate map to remove the logarithmic terms \cite{Kalisch:2015via}.}

\newpage
\bibliographystyle{JHEP}

\bibliography{ref}

\providecommand{\href}[2]{#2}\begingroup\raggedright\begin{thebibliography}{10}

\bibitem{Maldacena:1997re}
J.~M. Maldacena, \emph{{The large N limit of superconformal field theories and
  supergravity}}, {\emph{Adv. Theor. Math. Phys.} {\bf 2} (1998) 231--252},
  [\href{http://arxiv.org/abs/hep-th/9711200}{{\tt hep-th/9711200}}].

\bibitem{Ammon:2015wua}
M.~Ammon and J.~Erdmenger, \emph{{Gauge/gravity duality}}.
\newblock Cambridge Univ. Pr., Cambridge, UK, 2015.

\bibitem{Nastase}
H.~Nastase, \emph{{Introduction to the AdS/CFT Correspondence}}.
\newblock Cambridge Univ. Pr., Cambridge, UK, 2015.

\bibitem{Schalm}
J.~Zaanen, Y.~Liu, Y.-W. Sun and K.~Schalm, \emph{{Holographic Duality in
  Condensed Matter Physics}}.
\newblock Cambridge Univ. Pr., Cambridge, UK, 2016.

\bibitem{Hartnoll:2009sz}
S.~A. Hartnoll, \emph{{Lectures on holographic methods for condensed matter
  physics}},
  \href{http://dx.doi.org/10.1088/0264-9381/26/22/224002}{\emph{Class. Quant.
  Grav.} {\bf 26} (2009) 224002}, [\href{http://arxiv.org/abs/0903.3246}{{\tt
  0903.3246}}].

\bibitem{Herzog:2009xv}
C.~P. Herzog, \emph{{Lectures on Holographic Superfluidity and
  Superconductivity}},
  \href{http://dx.doi.org/10.1088/1751-8113/42/34/343001}{\emph{J. Phys.} {\bf
  A42} (2009) 343001}, [\href{http://arxiv.org/abs/0904.1975}{{\tt
  0904.1975}}].

\bibitem{McGreevy:2009xe}
J.~McGreevy, \emph{{Holographic duality with a view toward many-body physics}},
  \href{http://dx.doi.org/10.1155/2010/723105}{\emph{Adv. High Energy Phys.}
  {\bf 2010} (2010) 723105}, [\href{http://arxiv.org/abs/0909.0518}{{\tt
  0909.0518}}].

\bibitem{CasalderreySolana:2011us}
J.~Casalderrey-Solana, H.~Liu, D.~Mateos, K.~Rajagopal and U.~A. Wiedemann,
  \emph{{Gauge/String Duality, Hot QCD and Heavy Ion Collisions}},
  \href{http://arxiv.org/abs/1101.0618}{{\tt 1101.0618}}.

\bibitem{D'Hoker:2009mm}
E.~D'Hoker and P.~Kraus, \emph{{Magnetic Brane Solutions in AdS}},
  \href{http://dx.doi.org/10.1088/1126-6708/2009/10/088}{\emph{JHEP} {\bf 10}
  (2009) 088}, [\href{http://arxiv.org/abs/0908.3875}{{\tt 0908.3875}}].

\bibitem{D'Hoker:2009bc}
E.~D'Hoker and P.~Kraus, \emph{{Charged Magnetic Brane Solutions in AdS (5) and
  the fate of the third law of thermodynamics}},
  \href{http://dx.doi.org/10.1007/JHEP03(2010)095}{\emph{JHEP} {\bf 03} (2010)
  095}, [\href{http://arxiv.org/abs/0911.4518}{{\tt 0911.4518}}].

\bibitem{D'Hoker:2010rz}
E.~D'Hoker and P.~Kraus, \emph{{Holographic Metamagnetism, Quantum Criticality,
  and Crossover Behavior}},
  \href{http://dx.doi.org/10.1007/JHEP05(2010)083}{\emph{JHEP} {\bf 05} (2010)
  083}, [\href{http://arxiv.org/abs/1003.1302}{{\tt 1003.1302}}].

\bibitem{D'Hoker:2010ij}
E.~D'Hoker and P.~Kraus, \emph{{Magnetic Field Induced Quantum Criticality via
  new Asymptotically AdS5 Solutions}},
  \href{http://dx.doi.org/10.1088/0264-9381/27/21/215022}{\emph{Class. Quant.
  Grav.} {\bf 27} (2010) 215022}, [\href{http://arxiv.org/abs/1006.2573}{{\tt
  1006.2573}}].

\bibitem{D'Hoker:2012ej}
E.~D'Hoker and P.~Kraus, \emph{{Charge Expulsion from Black Brane Horizons, and
  Holographic Quantum Criticality in the Plane}},
  \href{http://dx.doi.org/10.1007/JHEP09(2012)105}{\emph{JHEP} {\bf 09} (2012)
  105}, [\href{http://arxiv.org/abs/1202.2085}{{\tt 1202.2085}}].

\bibitem{D'Hoker:2012ih}
E.~D'Hoker and P.~Kraus, \emph{{Quantum Criticality via Magnetic Branes}},
  \href{http://dx.doi.org/10.1007/978-3-642-37305-3_18}{\emph{Lect. Notes
  Phys.} {\bf 871} (2013) 469--502},
  [\href{http://arxiv.org/abs/1208.1925}{{\tt 1208.1925}}].

\bibitem{Hartnoll:2008vx}
S.~A. Hartnoll, C.~P. Herzog and G.~T. Horowitz, \emph{{Building a Holographic
  Superconductor}},
  \href{http://dx.doi.org/10.1103/PhysRevLett.101.031601}{\emph{Phys. Rev.
  Lett.} {\bf 101} (2008) 031601}, [\href{http://arxiv.org/abs/0803.3295}{{\tt
  0803.3295}}].

\bibitem{Hartnoll:2008kx}
S.~A. Hartnoll, C.~P. Herzog and G.~T. Horowitz, \emph{{Holographic
  Superconductors}},
  \href{http://dx.doi.org/10.1088/1126-6708/2008/12/015}{\emph{JHEP} {\bf 12}
  (2008) 015}, [\href{http://arxiv.org/abs/0810.1563}{{\tt 0810.1563}}].

\bibitem{Gubser:2008wv}
S.~S. Gubser and S.~S. Pufu, \emph{{The gravity dual of a p-wave
  superconductor}},
  \href{http://dx.doi.org/10.1088/1126-6708/2008/11/033}{\emph{JHEP} {\bf 11}
  (2008) 033}, [\href{http://arxiv.org/abs/0805.2960}{{\tt 0805.2960}}].

\bibitem{Ammon:2009xh}
M.~Ammon, J.~Erdmenger, V.~Grass, P.~Kerner and A.~O'Bannon, \emph{{On
  Holographic p-wave Superfluids with Back-reaction}},
  \href{http://dx.doi.org/10.1016/j.physletb.2010.02.021}{\emph{Phys. Lett.}
  {\bf B686} (2010) 192--198}, [\href{http://arxiv.org/abs/0912.3515}{{\tt
  0912.3515}}].

\bibitem{Nakamura:2009tf}
S.~Nakamura, H.~Ooguri and C.-S. Park, \emph{{Gravity Dual of Spatially
  Modulated Phase}},
  \href{http://dx.doi.org/10.1103/PhysRevD.81.044018}{\emph{Phys. Rev.} {\bf
  D81} (2010) 044018}, [\href{http://arxiv.org/abs/0911.0679}{{\tt
  0911.0679}}].

\bibitem{Ooguri:2010kt}
H.~Ooguri and C.-S. Park, \emph{{Holographic End-Point of Spatially Modulated
  Phase Transition}},
  \href{http://dx.doi.org/10.1103/PhysRevD.82.126001}{\emph{Phys. Rev.} {\bf
  D82} (2010) 126001}, [\href{http://arxiv.org/abs/1007.3737}{{\tt
  1007.3737}}].

\bibitem{Donos:2012wi}
A.~Donos and J.~P. Gauntlett, \emph{{Black holes dual to helical current
  phases}}, \href{http://dx.doi.org/10.1103/PhysRevD.86.064010}{\emph{Phys.
  Rev.} {\bf D86} (2012) 064010}, [\href{http://arxiv.org/abs/1204.1734}{{\tt
  1204.1734}}].

\bibitem{Erdmenger:2015qqa}
J.~Erdmenger, B.~Herwerth, S.~Klug, R.~Meyer and K.~Schalm, \emph{{S-Wave
  Superconductivity in Anisotropic Holographic Insulators}},
  \href{http://dx.doi.org/10.1007/JHEP05(2015)094}{\emph{JHEP} {\bf 05} (2015)
  094}, [\href{http://arxiv.org/abs/1501.07615}{{\tt 1501.07615}}].

\bibitem{Domenech:2010nf}
O.~Domenech, M.~Montull, A.~Pomarol, A.~Salvio and P.~J. Silva, \emph{{Emergent
  Gauge Fields in Holographic Superconductors}},
  \href{http://dx.doi.org/10.1007/JHEP08(2010)033}{\emph{JHEP} {\bf 08} (2010)
  033}, [\href{http://arxiv.org/abs/1005.1776}{{\tt 1005.1776}}].

\bibitem{Bolognesi:2010nb}
S.~Bolognesi and D.~Tong, \emph{{Monopoles and Holography}},
  \href{http://dx.doi.org/10.1007/JHEP01(2011)153}{\emph{JHEP} {\bf 01} (2011)
  153}, [\href{http://arxiv.org/abs/1010.4178}{{\tt 1010.4178}}].

\bibitem{Ammon:2011je}
M.~Ammon, J.~Erdmenger, P.~Kerner and M.~Strydom, \emph{{Black Hole Instability
  Induced by a Magnetic Field}},
  \href{http://dx.doi.org/10.1016/j.physletb.2011.10.067}{\emph{Phys. Lett.}
  {\bf B706} (2011) 94--99}, [\href{http://arxiv.org/abs/1106.4551}{{\tt
  1106.4551}}].

\bibitem{Almuhairi:2011ws}
A.~Almuhairi and J.~Polchinski, \emph{{Magnetic AdS x R2: Supersymmetry and
  stability}},  \href{http://arxiv.org/abs/1108.1213}{{\tt 1108.1213}}.

\bibitem{Bu:2012mq}
Y.-Y. Bu, J.~Erdmenger, J.~P. Shock and M.~Strydom, \emph{{Magnetic field
  induced lattice ground states from holography}},
  \href{http://dx.doi.org/10.1007/JHEP03(2013)165}{\emph{JHEP} {\bf 03} (2013)
  165}, [\href{http://arxiv.org/abs/1210.6669}{{\tt 1210.6669}}].

\bibitem{Cremonini:2012ir}
S.~Cremonini and A.~Sinkovics, \emph{{Spatially Modulated Instabilities of
  Geometries with Hyperscaling Violation}},
  \href{http://dx.doi.org/10.1007/JHEP01(2014)099}{\emph{JHEP} {\bf 01} (2014)
  099}, [\href{http://arxiv.org/abs/1212.4172}{{\tt 1212.4172}}].

\bibitem{Montull:2012fy}
M.~Montull, O.~Pujolas, A.~Salvio and P.~J. Silva, \emph{{Magnetic Response in
  the Holographic Insulator/Superconductor Transition}},
  \href{http://dx.doi.org/10.1007/JHEP04(2012)135}{\emph{JHEP} {\bf 04} (2012)
  135}, [\href{http://arxiv.org/abs/1202.0006}{{\tt 1202.0006}}].

\bibitem{Salvio:2012at}
A.~Salvio, \emph{{Holographic Superfluids and Superconductors in
  Dilaton-Gravity}},
  \href{http://dx.doi.org/10.1007/JHEP09(2012)134}{\emph{JHEP} {\bf 09} (2012)
  134}, [\href{http://arxiv.org/abs/1207.3800}{{\tt 1207.3800}}].

\bibitem{Salvio:2013jia}
A.~Salvio, \emph{{Transitions in Dilaton Holography with Global or Local
  Symmetries}}, \href{http://dx.doi.org/10.1007/JHEP03(2013)136}{\emph{JHEP}
  {\bf 03} (2013) 136}, [\href{http://arxiv.org/abs/1302.4898}{{\tt
  1302.4898}}].

\bibitem{Bao:2013fda}
N.~Bao, S.~Harrison, S.~Kachru and S.~Sachdev, \emph{{Vortex Lattices and
  Crystalline Geometries}},
  \href{http://dx.doi.org/10.1103/PhysRevD.88.026002}{\emph{Phys. Rev.} {\bf
  D88} (2013) 026002}, [\href{http://arxiv.org/abs/1303.4390}{{\tt
  1303.4390}}].

\bibitem{Jokela:2014dba}
N.~Jokela, M.~Jarvinen and M.~Lippert, \emph{{Gravity dual of spin and charge
  density waves}}, \href{http://dx.doi.org/10.1007/JHEP12(2014)083}{\emph{JHEP}
  {\bf 12} (2014) 083}, [\href{http://arxiv.org/abs/1408.1397}{{\tt
  1408.1397}}].

\bibitem{Donos:2015eew}
A.~Donos and J.~P. Gauntlett, \emph{{Minimally packed phases in holography}},
  \href{http://arxiv.org/abs/1512.06861}{{\tt 1512.06861}}.

\bibitem{Basar:2010zd}
G.~Basar, G.~V. Dunne and D.~E. Kharzeev, \emph{{Chiral Magnetic Spiral}},
  \href{http://dx.doi.org/10.1103/PhysRevLett.104.232301}{\emph{Phys. Rev.
  Lett.} {\bf 104} (2010) 232301}, [\href{http://arxiv.org/abs/1003.3464}{{\tt
  1003.3464}}].

\bibitem{Kim:2010pu}
K.-Y. Kim, B.~Sahoo and H.-U. Yee, \emph{{Holographic chiral magnetic spiral}},
  \href{http://dx.doi.org/10.1007/JHEP10(2010)005}{\emph{JHEP} {\bf 10} (2010)
  005}, [\href{http://arxiv.org/abs/1007.1985}{{\tt 1007.1985}}].

\bibitem{BallonBayona:2012wx}
A.~Ballon-Bayona, K.~Peeters and M.~Zamaklar, \emph{{A chiral magnetic spiral
  in the holographic Sakai-Sugimoto model}},
  \href{http://dx.doi.org/10.1007/JHEP11(2012)164}{\emph{JHEP} {\bf 11} (2012)
  164}, [\href{http://arxiv.org/abs/1209.1953}{{\tt 1209.1953}}].

\bibitem{2010ARCM}
E.~{Fradkin}, S.~A. {Kivelson}, M.~J. {Lawler}, J.~P. {Eisenstein} and A.~P.
  {MacKenzie}, \emph{{Nematic Fermi Fluids in Condensed Matter Physics}},
  \href{http://dx.doi.org/10.1146/annurev-conmatphys-070909-103925}{\emph{Annual
  Review of Condensed Matter Physics} {\bf 1} (Apr., 2010) 153--178},
  [\href{http://arxiv.org/abs/0910.4166}{{\tt 0910.4166}}].

\bibitem{Buchel:2006gb}
A.~Buchel and J.~T. Liu, \emph{{Gauged supergravity from type IIB string theory
  on Y**p,q manifolds}},
  \href{http://dx.doi.org/10.1016/j.nuclphysb.2007.03.001}{\emph{Nucl. Phys.}
  {\bf B771} (2007) 93--112}, [\href{http://arxiv.org/abs/hep-th/0608002}{{\tt
  hep-th/0608002}}].

\bibitem{Gauntlett:2006ai}
J.~P. Gauntlett, E.~O~Colgain and O.~Varela, \emph{{Properties of some
  conformal field theories with M-theory duals}},
  \href{http://dx.doi.org/10.1088/1126-6708/2007/02/049}{\emph{JHEP} {\bf 02}
  (2007) 049}, [\href{http://arxiv.org/abs/hep-th/0611219}{{\tt
  hep-th/0611219}}].

\bibitem{Gauntlett:2007ma}
J.~P. Gauntlett and O.~Varela, \emph{{Consistent Kaluza-Klein reductions for
  general supersymmetric AdS solutions}},
  \href{http://dx.doi.org/10.1103/PhysRevD.76.126007}{\emph{Phys. Rev.} {\bf
  D76} (2007) 126007}, [\href{http://arxiv.org/abs/0707.2315}{{\tt
  0707.2315}}].

\bibitem{Colgain:2014pha}
E.~O. Colg\'{a}in, M.~M. Sheikh-Jabbari, J.~F. V\'{a}zquez-Poritz,
  H.~Yavartanoo and Z.~Zhang, \emph{{Warped Ricci-flat reductions}},
  \href{http://dx.doi.org/10.1103/PhysRevD.90.045013}{\emph{Phys. Rev.} {\bf
  D90} (2014) 045013}, [\href{http://arxiv.org/abs/1406.6354}{{\tt
  1406.6354}}].

\bibitem{Henningson:1998gx}
M.~Henningson and K.~Skenderis, \emph{{The Holographic Weyl anomaly}},
  \href{http://dx.doi.org/10.1088/1126-6708/1998/07/023}{\emph{JHEP} {\bf 07}
  (1998) 023}, [\href{http://arxiv.org/abs/hep-th/9806087}{{\tt
  hep-th/9806087}}].

\bibitem{Balasubramanian:1999re}
V.~Balasubramanian and P.~Kraus, \emph{{A Stress tensor for Anti-de Sitter
  gravity}}, \href{http://dx.doi.org/10.1007/s002200050764}{\emph{Commun. Math.
  Phys.} {\bf 208} (1999) 413--428},
  [\href{http://arxiv.org/abs/hep-th/9902121}{{\tt hep-th/9902121}}].

\bibitem{Taylor:2000xw}
M.~Taylor, \emph{{More on counterterms in the gravitational action and
  anomalies}},  \href{http://arxiv.org/abs/hep-th/0002125}{{\tt
  hep-th/0002125}}.

\bibitem{Amado:2011zx}
I.~Amado, K.~Landsteiner and F.~Pena-Benitez, \emph{{Anomalous transport
  coefficients from Kubo formulas in Holography}},
  \href{http://dx.doi.org/10.1007/JHEP05(2011)081}{\emph{JHEP} {\bf 05} (2011)
  081}, [\href{http://arxiv.org/abs/1102.4577}{{\tt 1102.4577}}].

\bibitem{Janiszewski:2015ura}
S.~Janiszewski and M.~Kaminski, \emph{{Quasinormal modes of magnetic and
  electric black branes versus far from equilibrium anisotropic fluids}},
  \href{http://arxiv.org/abs/1508.06993}{{\tt 1508.06993}}.

\bibitem{Bhattacharya:2011eea}
J.~Bhattacharya, S.~Bhattacharyya and S.~Minwalla, \emph{{Dissipative
  Superfluid dynamics from gravity}},
  \href{http://dx.doi.org/10.1007/JHEP04(2011)125}{\emph{JHEP} {\bf 04} (2011)
  125}, [\href{http://arxiv.org/abs/1101.3332}{{\tt 1101.3332}}].

\bibitem{Donos:2013woa}
A.~Donos, J.~P. Gauntlett and C.~Pantelidou, \emph{{Competing p-wave orders}},
  \href{http://dx.doi.org/10.1088/0264-9381/31/5/055007}{\emph{Class. Quant.
  Grav.} {\bf 31} (2014) 055007}, [\href{http://arxiv.org/abs/1310.5741}{{\tt
  1310.5741}}].

\bibitem{Rogatko:2007pv}
M.~Rogatko, \emph{{First Law of Black Rings Thermodynamics in Higher
  Dimensional Chern-Simons Gravity}},
  \href{http://dx.doi.org/10.1103/PhysRevD.75.024008}{\emph{Phys. Rev.} {\bf
  D75} (2007) 024008}, [\href{http://arxiv.org/abs/hep-th/0611260}{{\tt
  hep-th/0611260}}].

\bibitem{Suryanarayana:2007rk}
N.~V. Suryanarayana and M.~C. Wapler, \emph{{Charges from Attractors}},
  \href{http://dx.doi.org/10.1088/0264-9381/24/20/009}{\emph{Class. Quant.
  Grav.} {\bf 24} (2007) 5047--5072},
  [\href{http://arxiv.org/abs/0704.0955}{{\tt 0704.0955}}].

\bibitem{Meinel:2008}
R.~{Meinel}, M.~{Ansorg}, A.~{Kleinw{\"a}chter}, G.~{Neugebauer} and
  D.~{Petroff}, \emph{Relativistic figures of equilibrium}.
\newblock Cambridge University Press, June, 30th, 2008.

\bibitem{Macedo2014}
R.~P. Macedo and M.~Ansorg, \emph{{Axisymmetric fully spectral code for
  hyperbolic equations}},
  \href{http://dx.doi.org/10.1016/j.jcp.2014.07.040}{\emph{J. Comput. Phys.}
  {\bf 276} (2014) 357--379}, [\href{http://arxiv.org/abs/1402.7343}{{\tt
  1402.7343}}].

\bibitem{Boyd00}
J.~P. Boyd, \emph{Chebyshev and Fourier Spectral Methods (Second Edition,
  Revised)}.
\newblock Dover Publications, New York, 2001.

\bibitem{canuto_2006_smf}
C.~Canuto, M.~Hussaini, A.~Quarteroni and T.~Zang, \emph{Spectral Methods:
  Fundamentals in Single Domains}.
\newblock Springer, Berlin, 2006.

\bibitem{Chesler:2013lia}
P.~M. Chesler and L.~G. Yaffe, \emph{{Numerical solution of gravitational
  dynamics in asymptotically anti-de Sitter spacetimes}},
  \href{http://dx.doi.org/10.1007/JHEP07(2014)086}{\emph{JHEP} {\bf 07} (2014)
  086}, [\href{http://arxiv.org/abs/1309.1439}{{\tt 1309.1439}}].

\bibitem{Dias:2015nua}
O.~J.~C. Dias, J.~E. Santos and B.~Way, \emph{{Numerical Methods for Finding
  Stationary Gravitational Solutions}},
  \href{http://arxiv.org/abs/1510.02804}{{\tt 1510.02804}}.

\bibitem{Kalisch:2015via}
M.~Kalisch and M.~Ansorg, \emph{{Highly Deformed Non-uniform Black Strings in
  Six Dimensions}},  in \emph{{14th Marcel Grossmann Meeting}}, 2015.
\newblock \href{http://arxiv.org/abs/1509.03083}{{\tt 1509.03083}}.

\end{thebibliography}\endgroup

\end{document}